\documentclass{statsoc}

\usepackage{filecontents}
\usepackage{geometry}
\usepackage[OT1]{fontenc}
\usepackage{natbib}
\usepackage{float}
\usepackage{graphicx}
\graphicspath{{figures/}}

\usepackage[colorlinks,allcolors=blue]{hyperref}

\usepackage{comment}
\usepackage{mathtools}
\mathtoolsset{showonlyrefs}
\usepackage{amsmath}

\newtheorem{theorem}{Theorem}

\newtheorem{lemma}[theorem]{Lemma}
\newtheorem{remark}{Remark}

\usepackage{amsfonts}
\usepackage{color}
\usepackage{subcaption}
\usepackage{url}

\usepackage{acro}
\acsetup{
  tooltip = true,
  long-format = \it,
  short-format = \sc
}
\DeclareAcronym{ode}{short=ode, long=ordinary differential equation}
\DeclareAcronym{pde}{short=pde, long=partial differential equation}
\DeclareAcronym{mcmc}{short=mcmc, long=Markov chain Monte Carlo}
\DeclareAcronym{hmc}{short=hmc, long=Hamiltonian Monte Carlo}
\DeclareAcronym{rmhmc}{short=rm-hmc, long=Riemannian-manifold Hamiltonian Monte Carlo}
\DeclareAcronym{chmc}{short=c-hmc, long=constrained Hamiltonian Monte Carlo}
\DeclareAcronym{rwm}{short=rwm, long=random-walk Metropolis}
\DeclareAcronym{mala}{short=mala, long=Metropolis-adjusted Langevin algorithm}
\DeclareAcronym{ess}{short=ess, long=effective sample size}
\DeclareAcronym{nuts}{short=nuts, long=No-U-Turn Sampler}
\DeclareAcronym{ad}{short=ad, long=algorithmic differentation}
\DeclareAcronym{svd}{short=svd, long=singular value decomposition}
\DeclareAcronym{fem}{short=fem, long=finite element method}
\DeclareAcronym{ssm}{short=ssm, long=state space model} 
\usepackage{algorithm}
\usepackage[noend]{algpseudocode}

\usepackage{xr}
\algrenewcommand{\algorithmicrequire}{\textbf{Inputs:}}
\algrenewcommand{\algorithmicensure}{\textbf{Outputs:}}

\newcommand{\vct}[1]{\boldsymbol{#1}}
\newcommand{\mtx}[1]{\mathit{#1}}

\newcommand{\BK}[1]{ {\left( #1 \right)} }

\newcommand{\curBK}[1]{ {\left\{ #1 \right\}} }

\newcommand{\cO}{\mathcal{O}}
\newcommand{\cN}{\mathcal{N}}

\newcommand{\cS}{\mathcal{S}}
\newcommand{\cT}{\mathcal{T}}
\newcommand{\cX}{\mathcal{X}}
\newcommand{\cY}{\mathcal{Y}}

\newcommand{\bR}{\mathbb{R}}

\newcommand{\EE}{\ensuremath{\operatorname{E}}}

\newcommand{\gbar}[1]{\bar g_\text{#1}}

\renewcommand\epsilon{\varepsilon}
\newcommand{\manifold}{\mathcal{M}}
\newcommand{\forw}{F}
\newcommand{\constr}{C}
\newcommand{\jac}{\mathbf{D}}
\newcommand{\hauss}{\mathcal{H}}

\newcommand{\infnorm}[1]{\| #1 \|_{\infty}}
\newcommand{\id}{\mtx{I}}
\newcommand{\gram}[2][]{\mtx{G}^{#1}(#2)}

\newcommand{\tr}{^\mathsf{T}}
\newcommand{\jacob}{\jac}
\newcommand{\grad}{\nabla}
\newcommand{\idmtx}{\mathbb{I}}

\newcommand{\dr}{\mathrm{d}}

\newcommand{\normal}{\mathsf{Normal}}
\newcommand{\halfnormal}{\mathsf{HalfNormal}}
\newcommand{\lognormal}{\mathsf{LogNormal}}
\newcommand{\uniform}{\mathsf{Uniform}}

\title{Manifold lifting: scaling Markov chain Monte Carlo to the vanishing noise regime}
\author{Khai Xiang Au}
\address{National University of Singapore, Singapore}
\author{Matthew M. Graham}
\address{University College London, UK}
\author[K.X. Au, M.M. Graham and A.H. Thiery]{Alexandre H. Thiery}
\address{National University of Singapore, Singapore}
\email{{khai@u.nus.edu}, {m.graham@ucl.ac.uk} \& {a.h.thiery@nus.edu.sg}}

\begin{document}

\begin{abstract}
Standard Markov chain Monte Carlo methods struggle to explore distributions that concentrate in the neighbourhood of low-dimensional structures. This pathology naturally occurs in Bayesian inference settings when there is a high signal-to-noise ratio in the observational data but the model is inherently over-parametrized or non-identifiable. In this paper, we propose a strategy that transforms the original sampling problem into the task of exploring a distribution supported on a manifold embedded in a higher-dimensional space; in contrast to the original posterior this lifted distribution remains diffuse in the limit of vanishing observation noise. We employ a constrained Hamiltonian Monte Carlo method, which exploits the geometry of this lifted distribution, to perform efficient approximate inference. We demonstrate in numerical experiments that, contrarily to competing approaches, the sampling efficiency of our proposed methodology does not degenerate as the target distribution to be explored concentrates near low-dimensional structures.
\end{abstract}


\section{Introduction}
Under a Bayesian framework, inference corresponds to deducing the posterior distribution over the configurations a generative model for observations given observed data and a prior distribution on the model configurations. While we generally expect conditioning on observed data to reduce the uncertainty about the possible model configurations, in most real-world situations we also expect to retain some uncertainty in our posterior beliefs. Accurately quantifying this uncertainty is important in downstream tasks such as using a model to make predictions or comparing how well competing models explain the data.

An obvious source of uncertainty in our posterior beliefs is measurement noise, where observations are imperfectly measured due to, for instance, instrument limitations or human error. Uncertainty also naturally arises in ill-posed or underdetermined problems, that is when the number of degrees of freedom in the unknown variables in a model exceeds the number of degrees of freedom constrained by the observations. Even in the case of plentiful noiseless observations, uncertainty may still arise due to inherent non-identifiabilities in the model.

In this paper, we consider inference of a vector of \emph{unknown variables} $\vct{\theta} \in \Theta = \bR^{d_\Theta}$ given noisy \emph{observations} $\vct{y} \in \cY = \bR^{d_{\cY}}$. For ease of exposition, we first consider the special case where the model has additive, isotropic, Gaussian noise of a known variance. The observation model is then defined as
\begin{equation} \label{eq.model}
  \vct{y} = \forw(\vct{\theta}) + \sigma \vct{\eta}
\end{equation}
for a vector of \emph{noise variables} $\vct{\eta} \sim \normal(\vct{0}, \idmtx_{d_\cY})$, \emph{noise scale} $\sigma > 0$ and a \emph{forward function} $\forw: \Theta \to \cY$, which is generally non-linear and computationally expensive to evaluate. The methodology described in this work can be applied more generally, and we will relax the assumptions of known $\sigma$ and Gaussian noise in Section \ref{sec.state_space_augmentation}.

We adopt a Bayesian approach and specify a prior distribution on $\Theta$ with unnormalised density $\exp(-\Phi_{\vct{\theta}}(\vct{\theta}))$ with respect to the Lebesgue measure. The negative logarithm of the posterior density $\pi^{\sigma}: \Theta \to \bR_{\geq 0}$ then reads
\begin{equation} \label{eq.neg_log_posterior}
  -\log \pi^{\sigma}(\vct{\theta}) =
  \Phi_{\vct{\theta}}(\vct{\theta}) +
  \frac{1}{2 \sigma^2} \left\| \vct{y} - \forw(\vct{\theta}) \right\|^2 +
  d_{\cY} \log\sigma +
  \mathrm{constant}.
\end{equation}
In the case where the forward function $\forw$ is linear and the prior distribution Gaussian, the posterior distribution is also Gaussian. In general, however, computations involving the posterior distribution are intractable and it is necessary to employ approximate inference methods.

There are natural situations where the dimension of the observations $d_{\mathcal{Y}}$ is lower, and often much lower, than the dimension of the unknown variables $d_\Theta$. In these settings, i.e. $d_\cY \ll d_\Theta$, even as the noise scale $\sigma$ decreases to zero, exact reconstruction of the unknown variables is typically not possible: the system is underdetermined. In \emph{Bayesian inverse problems} \citep{stuart2010inverse, knapik2011bayesian, petra2014computational}, a particularly representative class of models where this type of scenario naturally occurs, the unknown variables $\vct{\theta}$ typically represents a spatially extended field. The forward function $\forw$ describes the process of collecting a (typically small) set of measurements derived from $\vct{\theta}$. For example, in Section \ref{sec.poisson} we consider the problem of reconstructing a thermal conductivity field from a discrete set of noisy measurements of a temperature field; the temperature is obtained from the conductivity as the solution to an elliptic \ac{pde}.

Conversely, even when there are ample observations compared to the number of unknowns, i.e. $d_\cY \gg d_\Theta$, structural non-identifiabilities in the model formulation --- that multiple values of $\vct{\theta}$ give the same conditional distribution on observations $\vct{y}$ \citep{rothenberg1971identification} --- can also mean that the posterior distribution on $\vct{\theta}$ does not collapse to a point as the observation noise vanishes ($\sigma \to 0$). As an example, in Sections \ref{sec.fhn_ode} and \ref{sec.hh} we consider the problem of inferring the parameters of a \ac{ode} models of neuronal dynamics given a sequence of noisy observations. In some cases, such as the FitzHugh--Nagumo model in Section \ref{sec.fhn_ode}, while the natural parametrisation of a model is such that a subset of parameters are non-identifiable, reparametrisations can be used to remove non-identifiabilities \citep{dasgupta2007non}. However, such parametrisations may not always be possible to find, even for relatively simple models \citep{ballnus2017comprehensive, hines2014determination}. Working with more complex models can exacerbate  the challenge of removing non-identifiabilities, as exemplified in the Hodgkin--Huxley \ac{ode} model discussed in Section \ref{sec.hh}. In practice, it is also not easy to know if non-identifiabilities are present  prior to performing inference \citep{raue2013joining}.

Although in the above-mentioned cases it is not possible to exactly reconstruct the quantity of interest as the observation noise vanishes $\sigma \to 0$, the posterior distribution will concentrate on a subset $\cS \subset \Theta$ of the latent space given by
\begin{equation*}
  \label{eq.limiting_manifold}
  \cS =  \curBK{\vct{\theta} \in \Theta :  \forw(\vct{\theta}) = \vct{y}}.
\end{equation*}
Under mild regularity assumptions on the forward operator $\forw$, the subset $\cS$ is a submanifold of dimension $d_\cS = d_\Theta - d_\cY$ embedded in $\Theta$ and the bulk of the posterior mass is distributed in a neighbourhood of radius $\cO(\sigma)$ around $\cS$.

The key computation involving the posterior distribution in downstream tasks is the evaluation of posterior expectations of the form $\int_\Theta \varphi(\vct{\theta}) \pi^{\sigma}(\vct{\theta})\,\dr \vct{\theta}$ for a test function $\varphi: \Theta \to \bR$. We focus in this text on \ac{mcmc} methods for approximating such posterior expectations. We assume throughout that derivatives of the forward function $\forw$ and prior negative log-density $\Phi_{\vct{\theta}}$ can be computed and develop \ac{mcmc} methods which exploit this derivative information.

Many of the most widely-used \ac{mcmc} methods for general target distributions on $\bR^{d_\Theta}$ such as \ac{rwm} with Gaussian proposals, the \ac{mala} \citep{besag1993comments}, and \ac{hmc} \citep{duane1987hybrid}, require the setting of a step size parameter which controls the scale of the proposed moves. The efficiency of these methods is highly dependent on an appropriate choice of this step size parameter, with overly large steps leading to a high probability of proposed moves being rejected, while overly small steps leading to slow exploration.

In the past decades, much analysis has been done to identify optimal scaling of the step size of various \ac{mcmc} algorithms as the target distribution dimension becomes large, i.e. the $d_\Theta \to \infty$ asymptotic \citep{gupta1990acceptance, roberts1997weak, roberts2001optimal, beskos2013optimal}. In contrast, comparably little attention has been devoted to the vanishing noise asymptotic, i.e. $\sigma \to 0$. As the posterior concentrates increasingly close to the manifold $\cS$, this induces a strong anisotropy in the scaling of the posterior distribution in different directions, with large changes in posterior density in directions normal to $\cS$ and much smaller changes in direction tangential to $\cS$. Importantly, for non-linear manifolds $\cS$, the tangential and normal directions vary across the manifold so that a simple global rescaling will not be sufficient to counteract the anisotropy.

In \citet{beskos2018asymptotic} the authors analyse the performance of \ac{rwm} algorithms in this vanishing noise regime in the specific case where the manifold $\cS$ is a linear subspace of $\Theta$. They prove that the step size needs to be scaled linearly with $\sigma$ to avoid the acceptance probability decreasing to zero when $\sigma \to 0$. As we will describe in the following section, this limitation also applies to gradient-based \ac{mcmc} methods such as \ac{mala} and \ac{hmc}. Although \citet{beskos2018asymptotic} concentrate on the linear case, they acknowledge that the more practically relevant case is of a non-linear limiting manifold $\cS$, and state that an important area for future work is the `study of \ac{mcmc} algorithms that better exploit the manifold structure of the support of the target distribution' where the `manifold can be of smaller dimension than the general space'. The main contribution of this paper is precisely a practical methodology for this setting: we propose an \ac{mcmc} algorithm which exploits the manifold structure of the target posterior distribution in order to remain computationally efficient in the vanishing noise regime.

The remainder of the paper is structured as follows. We begin in Section \ref{sec.vanishing_noise} by illustrating how standard \ac{mcmc} methods breakdown in the limit of $\sigma \to 0$ in an illustrative two-dimensional example. In Section \ref{sec.state_space_augmentation} we introduce an augmented state space formulation which lifts the target posterior distribution on to a manifold embedded in a higher-dimensional space, with this lifted target distribution not suffering the degenerate anisotropic scaling exhibited in the original space as $\sigma \to 0$. In Section \ref{sec.proposed_method} we describe our proposed approach of using a constrained \ac{hmc} algorithm to generate samples according to this manifold-supported lifted target distribution. In Section \ref{sec.theory} we provide some theoretical justification for the approach. Finally, in Section \ref{sec.experiments} we present the results of numerical experiments which empirically demonstrate that, in contrast to existing \ac{mcmc} methods, the proposed methodology is robust to low observation noise.

\section{Vanishing noise asymptotic regime}
\label{sec.vanishing_noise}

\begin{figure}[t]
  \centering
  \begin{subfigure}[b]{0.3\linewidth}%
    \includegraphics[width=\linewidth]{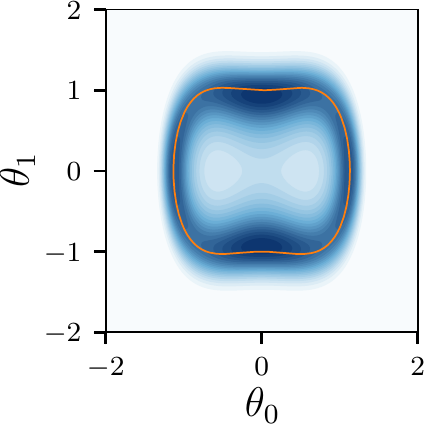}%
    \label{sfig.shrinking_loop_prob_mass_sigma_5e-1}%
  \end{subfigure}%
  ~
  \begin{subfigure}[b]{0.3\linewidth}%
    \includegraphics[width=\linewidth]{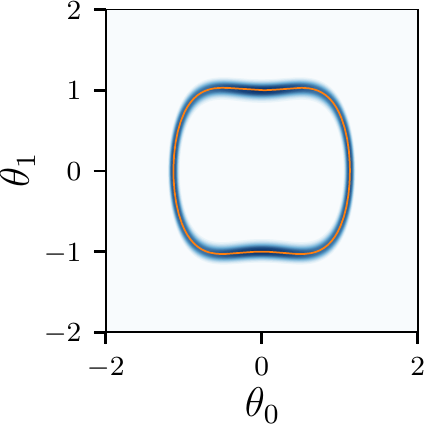}%
    \label{sfig.shrinking_loop_prob_mass_sigma_1e-1}%
  \end{subfigure}%
  ~
  \begin{subfigure}[b]{0.3\linewidth}%
    \includegraphics[width=\linewidth]{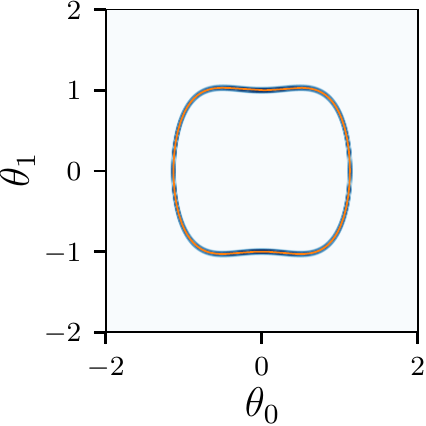}%
    \label{sfig.shrinking_loop_prob_mass_sigma_2e-2}%
  \end{subfigure}
\caption{Posterior distribution for noise scales $\sigma \in \lbrace 0.5, 0.1, 0.02\rbrace$ (left, centre, right) in toy example with forward operator $\forw(\vct{\theta}) =  \theta_1^2 + \theta_0^2 \BK{\theta_0^2 - \tfrac{1}{2} }$. The blue heatmaps show the posterior density $\pi^\sigma$ with darker colours indicating higher density. The orange curves show the limiting manifold $\cS$.}
\label{fig.shrinking_loop_prob_mass}
\end{figure}

In this section we numerically illustrate in a toy example the behaviour of standard \ac{mcmc} methods in the vanishing noise asymptotic $\sigma \to 0$. Consider a model of the form defined in \eqref{eq.model} with $d_\Theta=2$ and $d_\cY = 1$ and a forward function $\forw: \bR^2 \to \bR$
\begin{equation}
	\label{eq.running_example}
	\forw(\vct{\theta}) =  \theta_1^2 + \theta_0^2 \BK{\theta_0^2 - \tfrac{1}{2} }.
\end{equation}
We assume we observe $y = 1$ and place a standard centred Gaussian prior on the unknown parameter, i.e. $\vct{\theta} \sim \normal(\vct{0}, \idmtx_2)$. The posterior distribution is depicted in Figure \ref{fig.shrinking_loop_prob_mass} for three noise scales $\sigma \in \lbrace 0.5, 0.1, 0.02\rbrace$. As $\sigma \to 0$, the posterior distribution can be seen to concentrate in the neighbourhood of the manifold $\cS$ implicitly defined by the set of solutions to the equation $\forw(\vct{\theta}) = \vct{y}$.
\begin{figure}[t]
  \centering
  \includegraphics[width=0.93\linewidth]{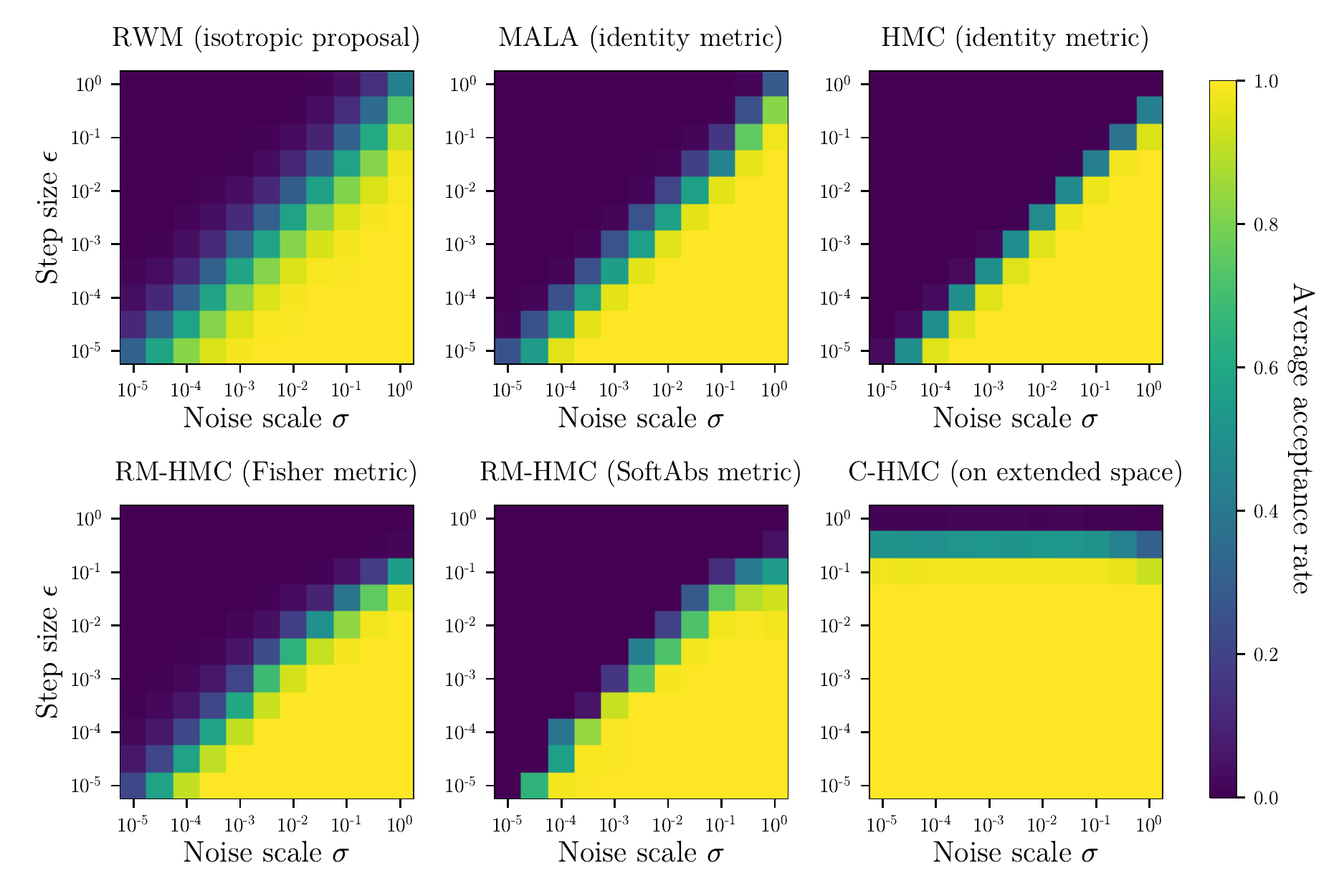}
  \caption{Average acceptance rate for chains simulated using different \ac{mcmc} methods targeting the posterior $\pi^\sigma$ of the toy example for varying noise scales $\sigma$ and step sizes $\epsilon$. For the \ac{hmc} methods $N=10$ integrator steps were used per sample. The values displayed are means of the acceptance probabilities over four chains of 1000 samples initialised at equispaced points around the limiting manifold $\cS$.
  }
  \label{fig.loop_experimental_results}
\end{figure}

We consider first the \ac{rwm} algorithm with isotropic Gaussian proposals with standard deviation (step size) $\epsilon > 0$. For the proposal to have a non-vanishing acceptance rate, the step size $\epsilon$ needs to be of order $\cO(\sigma)$ (see top-left panel in Figure \ref{fig.loop_experimental_results}). As \ac{rwm} chains evolve on a diffusive scale, we require $\cO(\epsilon^{-2}) = \cO(\sigma^{-2})$ steps to move a $\cO(1)$ distance in $\Theta$ \citep{beskos2018asymptotic}. To informally describe this phenomenon, we state that \ac{rwm} degenerates at rate $\sigma^{-2}$ as $\sigma \to 0$.

Maybe surprisingly, despite the use of gradient information, the performance of \ac{mala} chains is typically not better than \ac{rwm} (see top-centre panel in Figure \ref{fig.loop_experimental_results}). Indeed, since in the neighbourhood of $\cS$ the gradient of the (logarithm of the) posterior density is approximately orthogonal to the tangent plane to $\cS$, the gradient information by itself does not help \ac{mala} chains to explore the directions in the parameter space with greatest variation, that is, the directions that are tangential to $\cS$. The use of gradient information in settings were the target distribution is highly heterogeneous can in fact often have the negative effect of making performance highly sensitive to the choice of step size \citep{livingstone2019barker}, with this effect evident in the example here in the sharper transition in the acceptance rate for the \ac{mala} panel in Figure \ref{fig.loop_experimental_results} compared to \ac{rwm}. Further, as the exploration of the parameter  space remains diffusive, \ac{mala} also degenerates at rate $\sigma^{-2}$.

\ac{hmc} methods with a fixed metric (mass matrix) enjoy a slightly better scaling. The step size parameter $\epsilon>0$ of the leapfrog integrator used to generate proposals, like the \ac{rwm} step size, still needs to be chosen of order $\cO(\sigma)$ to maintain a non-zero acceptance rate (see top-right panel in Figure \ref{fig.loop_experimental_results}). However for proposals generated by simulating $\cO(\epsilon^{-1}) = \cO(\sigma^{-1})$ leapfrog steps, the \ac{hmc} method can move a distance $\cO(1)$ in the parameter space. Therefore, \ac{hmc} degenerates at rate $\sigma^{-1}$ as $\sigma \to 0$.

\acf{rmhmc} \citep{girolami2011riemann} and related position-dependent \ac{mala} \citep{roberts2002langevin, xifara2014langevin} and \ac{rwm} \citep{livingstone2015geometric} methods use a locally-varying metric to rescale the target distribution or, equivalently, locally rescale the size of the proposed moves in different directions. For an appropriate choice of metric, which differentially rescales proposed moves tangentially and orthogonally to the limiting manifold, these approaches may seem to offer a potential solution to the performance degeneration of standard \ac{rwm}, \ac{mala} and \ac{hmc} algorithms as $\sigma \to 0$.

In practice however these approaches still require the use of a vanishing step size as $\sigma \to 0$ in order to maintain a non-zero acceptance rate, as shown in the bottom-left and bottom-centre panels in Figure \ref{fig.loop_experimental_results}. These two panels show results for \ac{rmhmc} using two different metric functions: the expected Fisher information matrix plus prior covariance, as recommended by \citet{girolami2011riemann} and a `SoftAbs' regularisation of the Hessian of the negative log posterior density as recommended by \citet{betancourt2013general}. Both metrics have the desired behaviour of rescaling the steps to make small moves normal to, and large moves tangential to, the limiting manifold. Despite this, the \ac{rmhmc} chains show the same pattern as \ac{hmc} with a fixed metric of requiring $\epsilon \propto \sigma$ to maintain a non-vanishing acceptance rate as $\sigma \to 0$. In Section \ref{app.position-dependent-methods-vanishing-noise} in the Supplementary Material we show that simpler \ac{mcmc} schemes with position-dependent proposals also exhibit the same behaviour.

At a high-level, these phenomena can be attributed to the fact that, although the corresponding continuous time dynamics underlying \ac{rmhmc} and related approaches `follow' the curvature of the limiting manifold, for a finite discretisation step size the simulated trajectories only approximately replicate this idealised dynamic. In particular, as $\sigma \to 0$, smaller steps are needed for the dynamics to remain adequately close to the limiting manifold for the acceptance probability to remain non-zero and to ensure the iterative solvers used in the (implicit) generalised leapfrog integrator required by the \ac{rmhmc} algorithm do not diverge.

In contrast to the aforementioned standard \ac{hmc} and \ac{rmhmc} approaches, the \acf{chmc} method proposed in this paper is able to maintain a constant acceptance rate when using a fixed integrator step size $\epsilon$ as $\sigma \to 0$, as shown for the toy example in the bottom-right panel of Figure \ref{fig.loop_experimental_results}. When combined with the coherent (non-diffusive) exploration of the state space afforded by the Hamiltonian dynamics based proposals, this means the methodology we propose allows efficient \ac{mcmc} based estimates in the vanishing noise regime.

\section{Lifting the posterior distribution onto a manifold}
\label{sec.state_space_augmentation}

In this section, we describe our approach for constructing an \ac{mcmc} method which allows us to efficiently approximate expectations with respect to the posterior distribution in situations where the noise scale $\sigma$ may be small compared to the scale of the observed data $\vct{y}$. Here, we generalise our discussion from the statistical model \eqref{eq.model} to models with $\sigma$ as a function of $\vct{\theta}$,
\begin{equation}\label{eq.general_model}
\vct{y} = \forw(\vct{\theta}) + \sigma(\vct{\theta}) \, \vct{\eta},
\end{equation}
for a $\cY$-valued noise random variable $\vct{\eta}$ with negative log-density $\Phi_{\vct{\eta}}$.
Our approach relies on considering an extended state-space $\cX = \Theta \times \cY = \bR^{d_\cX}$, with dimension $d_\cX = d_\Theta + d_\cY$, of pairs $\vct{q}=(\vct{\theta}, \vct{\eta}) \in \cX$ and an extended auxiliary distribution $\overline{\pi}(\dr \vct{q})=\overline{\pi}(\dr\vct{\theta}, \dr\vct{\eta})$ that has the posterior distribution $\pi(\dr\vct{\theta})$ as $\vct{\theta}$-marginal. We then construct a Markov chain which leaves this extended distribution invariant.

Before describing the construction of the extended distribution $\overline{\pi}(\dr \vct{q})$ our methodology is based upon, we recall a few standard results on conditioning of random variables. Consider a random variable $\vct{Q}$ with density $\mu : \cX \to \bR_{\geq 0}$ with respect to the Lebesgue measure on $\cX$, as well as a continuously differentiable function $\constr: \cX \to \cY$ whose Jacobian matrix $\jac \constr$ has full-rank $\mu$-almost everywhere. A consequence of the co-area formula \citep{rousset2010free,diaconis2013sampling,graham2017asymptotically} gives that, for any test function $\varphi:\cX \to \bR$,
\begin{equation}\label{eq.coarea}
  \EE[\varphi(\vct{Q}) \, | \, \constr(\vct{Q})=\vct{0}] = \frac{1}{Z} \int_{\manifold}
    \varphi(\vct{q}) \det \gram{\vct{q}}^{-\frac{1}{2}} \mu(\vct{q})
  \, \hauss(\dr \vct{q})
\end{equation}
where $Z = \int_{\manifold} \det\gram{\vct{q}}^{-\frac{1}{2}} \mu(\vct{q}) \,\hauss(\dr \vct{q})$ is a normalization constant,
$\hauss(\dr \vct{q})$ is the $d_{\manifold} = d_{\cX} - d_{\cY} = d_{\Theta}$  dimensional \emph{Hausdorff measure} on the manifold $\manifold  =  \constr^{-1}(\vct{0})$, and
\(
  \gram{\vct{q}} =
  \jac\constr(\vct{q}) \jac\constr(\vct{q})\tr \in \bR^{\dr_\cY \times \dr_\cY}
\)
is the \emph{Gram matrix} of $\constr$. The Gram matrix is positive-definite, and hence non-singular, for any $\sigma > 0$.
Informally, this means that the posterior density is equal to the prior density $\mu$, restricted to the manifold $\manifold$, and
with a multiplicative volume term $\det \gram{\vct{q}}^{-\frac{1}{2}}$ that accounts for the change of measure due to the non-linearity of the constraint function $C$.

The model \eqref{eq.general_model} with a-priori
$\vct{\eta} \sim \exp(-\Phi_{\vct{\eta}}(\cdot))$ and $\vct{\theta} \sim \exp(-\Phi_{\vct{\theta}}(\cdot))$ corresponds to the case with constraint function $\constr$ and resulting manifold $\manifold$
\begin{align}
\label{eq.constraint}
 \constr(\vct{q}) \equiv \constr(\vct{\theta},\vct{\eta}) &=
  \forw(\vct{\theta}) + \sigma(\vct{\theta}) \vct{\eta} - \vct{y}\\
  \label{eq.manifold}
  \manifold &=
  \big\lbrace
    (\vct{\theta}, \vct{\eta}) \in \cX  :
    \vct{y} = \forw(\vct{\theta}) + \sigma(\vct{\theta}) \vct{\eta}
  \big\rbrace
\end{align}
while the negative log-density of the pair $\vct{q} = (\vct{\theta}, \vct{\eta})$ on $\cX$ then reads
\begin{align*}
-\log \mu(\vct{q})
\; \equiv\;
-\log \mu(\vct{\theta}, \vct{\eta})
\; = \;
\Phi_{\vct{\theta}}(\vct{\theta}) +
\Phi_{\vct{\eta}}(\vct{\eta}) +
\textrm{constant}.
\end{align*}
Consequently, Equation \eqref{eq.coarea} implies that the posterior distribution on the pair $\vct{q}=(\vct{\theta},\vct{\eta})$ has a density $\overline{\pi}(\vct{\theta}, \vct{\eta})$ with respect to the Hausdorff measure $\hauss(\dr \vct{q})$ on the manifold $\manifold = \constr^{-1}(\vct{0})$ given by
\begin{equation}
\label{eq.potential_energy}
  -\log\overline{\pi}(\vct{\theta}, \vct{\eta})
  = \Phi_{\vct{\theta}}(\vct{\theta})
  +  \Phi_{\vct{\eta}}(\vct{\eta})
  + \tfrac{1}{2}  \log\det\gram[]{\vct{\theta}, \vct{\eta}} + \textrm{constant}.
\end{equation}
\begin{figure}[t]
  \centering
  \includegraphics[width=0.3\linewidth]{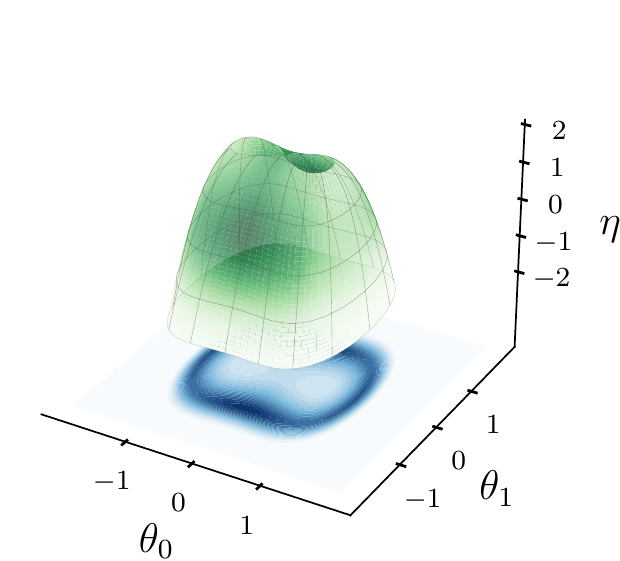}%
  \includegraphics[width=0.3\linewidth]{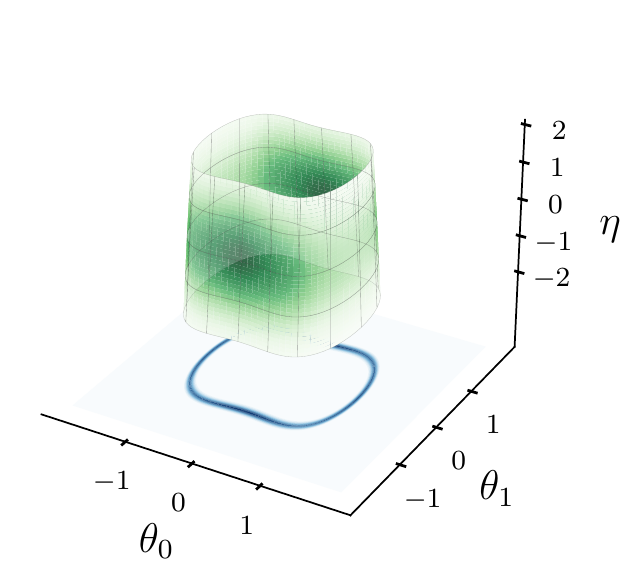}%
  \includegraphics[width=0.3\linewidth]{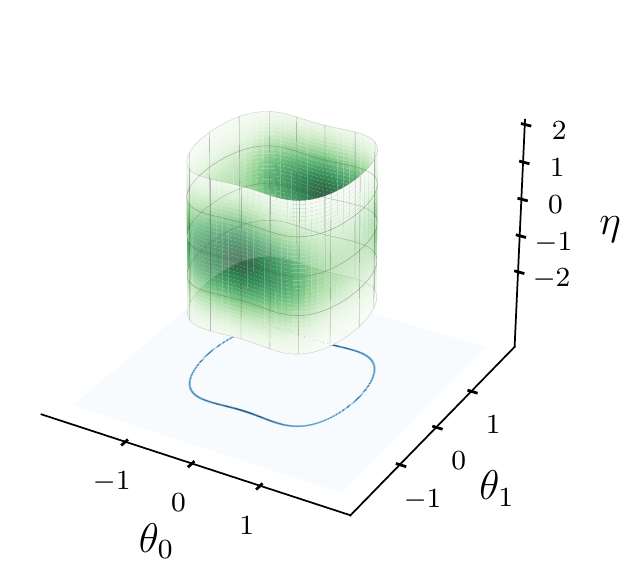}
  \caption{Manifold $\manifold$ and density $\overline{\pi}$ (green) in the extended space $\cX$ for $\sigma \in \lbrace 0.5, 0.1, 0.02 \rbrace$ (left-to-right). Only subset of $\manifold$ with $|\eta| \leq 2$ is shown. The density $\pi$ on $\Theta$ (blue) is shown for comparison.}
  \label{fig.manifold}
\end{figure}
In the running example \eqref{eq.running_example}, the original parameter space $\Theta$ is two-dimensional and, as $\sigma \to 0$, the posterior distribution concentrates in a neighbourhood of the one-dimensional manifold $\cS$. Figure \ref{fig.manifold} shows the two-dimensional manifold $\manifold$ embedded in the three-dimensional extended space $\cX$ and the lifted posterior density $\overline{\pi}$ on the manifold for $\sigma \in \lbrace 0.5, 0.1, 0.02\rbrace$. It can be seen that though the geometry of the manifold does change as $\sigma \to 0$, unlike the posterior distribution $\pi(\dr\vct{\theta})$ in the original space $\Theta$, the lifted posterior distribution $\overline{\pi}(\dr\vct{q})$ does not concentrate and remains diffuse. Furthermore, note that in the limit $\sigma \to 0$ the manifold $\manifold$ tends to the product manifold $\{(\vct{\theta}, \vct{\eta}) \in \Theta \times  \cY : \vct{y} = \forw(\vct{\theta})\} = \cS \times \cY$, and $\vct{\theta}$ and $\vct{\theta}$ become independent under the lifted posterior distribution. This phenomenon partly motivates why our proposed methodology does not degenerate as $\sigma \to 0$.

In summary, our methodology consists in lifting the posterior distribution $\pi(\dr\vct{\theta})$ that is concentrated in the neighbourhood of a set $\cS \subset \Theta$ onto a manifold $\manifold$ in a higher-dimensional space $\cX = \Theta \times \cY$. The task is then to sample from the lifted distribution $\overline{\pi}(\dr\vct{q})$, which is supported on $\manifold$ and importantly remains diffuse in the vanishing noise limit $\sigma \to 0$. We propose to use a constrained \ac{hmc} algorithm to efficiently sample from the lifted distribution $\overline{\pi}(\dr\vct{q})$ using first- and second-order derivative information. As will be numerically demonstrated in Section \ref{sec.experiments}, the resulting sampler does not degenerate as $\sigma \to 0$.
\section{Exploring the manifold-supported extended distribution}
\label{sec.proposed_method}

In this section, we describe the constrained \ac{hmc} method we use to construct an ergodic Markov chain leaving the lifted distribution $\overline{\pi}(\dr\vct{q})$ invariant. The use of simulated constrained Hamiltonian dynamics within a \ac{hmc} scheme has been proposed multiple times before, see for example \citet{hartmann2005constrained,rousset2010free,brubaker2012family}. Our formulation most closely follows that described in \citet{graham2017asymptotically}, which along with \citet{lelievre2019hybrid}, applies an idea proposed in \citet{zappa2018monte} to ensure the overall reversibility of the \ac{chmc} transitions despite the use of a numerical integrator with implicit steps that may fail to converge to a unique solution.

For conceptual clarity, we proceed by first giving a very high-level description of the standard \ac{hmc} methodology \citep{duane1987hybrid,neal2012mcmc} in order to draw distinctions with the \ac{chmc} approach. In order to explore a $\cX$-valued distribution with negative log-density $U(\vct{q})$, the \ac{hmc} algorithm introduces an auxiliary, and independent, momentum variable $\vct{p} \in \cX$ with standard Gaussian distribution. This defines an extended distribution on $\cX \times \cX$ with negative log-density equal, up to an irrelevant additive constant, to the Hamiltonian $H(\vct{q},\vct{p}) = U(\vct{q}) + \tfrac{1}{2} \|\vct{p}\|^2$. Proposed moves in the extended state space $\cX \times \cX$ are computed by numerically integrating the Hamiltonian dynamics $(\dot{\vct{q}}, \dot{\vct{p}}) = (\nabla_2, -\nabla_1) H(\vct{q},\vct{p})$ forward in time and then accepting or rejecting the final state in a Rosenbluth--Teller--Metropolis acceptance step \citep{metropolis1953equation}. To ensure ergodicity these dynamics based updates are interleaved with resampling of the momentum variable. The Hamiltonian dynamics is simulated using a reversible and volume preserving integrator, a common choice being the standard \emph{leapfrog} or St\"ormer--Verlet integrator with a fixed step size $\epsilon>0$ (see Algorithm \ref{alg.leapfrog_integrator}). More sophisticated integrators \citep{leimkuhler2004simulating,blanes2014numerical} are also possible.

\begin{algorithm}[t]
  \small
  \caption{Unconstrained leapfrog integrator}
  \label{alg.leapfrog_integrator}
  \algorithmicrequire{ $\vct{p}_0$ current momentum, $\vct{q}_0$ current position, $\epsilon$ step size.}\\
  \algorithmicensure{ $\vct{p}_1$ updated momentum, $\vct{q}_1$ updated position.}
  \begin{algorithmic} %
  \Function{LeapfrogStep}{$\vct{p}_0,\vct{q}_0,\epsilon$}
  \State $\vct{p}_{1/2} \gets \vct{p}_0 - \frac{\epsilon}{2} \nabla U(\vct{q}_0) $  \Comment{Momentum half-step update}
  \State $\vct{q}_{1} \gets \vct{q}_0 + \epsilon \vct{p}_{1/2} $ \Comment{Position full-step update}
  \State $\vct{p}_{1} \gets \vct{p}_{1/2} - \frac{\epsilon}{2} \nabla U(\vct{q}_1) $  \Comment{Momentum half-step update}
  \State \textbf{return} $\vct{p}_{1} , \vct{q}_{1} $
  \EndFunction
  \end{algorithmic}
\end{algorithm}

The constrained \ac{hmc} algorithm allows one to instead explore a distribution supported on an embedded manifold $\manifold  = \constr^{-1}(\vct{0}) \subset \cX$ implicitly defined by a constraint function $\constr: \cX \to \cY$.
At any point $\vct{q} \in \manifold$ we can define two orthogonal linear subspaces: the \emph{tangent space}
\[
  \cT_\manifold (\vct{q}) = \curBK{ \vct{p} \in \bR^{d_\cX} : \jac\constr(\vct{q}) \vct{p} = \vct{0} },
\]
which is the space of directions tangential to manifold at $\vct{q}$ and the \emph{normal space}
\[
  \cN_\manifold(\vct{q}) = \lbrace \jac \constr(\vct{q})\tr \vct{\lambda} : \vct{\lambda} \in \bR^{d_\cY}\rbrace,
\]
which is the space of directions normal to the manifold. Differentiating the constraint equation with respect to time gives us that $\jac\constr(\vct{q})\dot{\vct{q}} = \jac\constr(\vct{q}) \vct{p} = \vct{0}$, i.e. that the momentum $\vct{p}$ (under an assumption of an identity metric equivalent to the velocity) is in the tangent space $\cT_\manifold(\vct{q})$ at $\vct{q} \in \manifold$ at all time points.

Constrained \ac{hmc} operates on the same principles as standard \ac{hmc}, however additional projection steps are necessary in the integrator used to ensure that the position $\vct{q}$ remains on the manifold $\manifold$ and that the momentum $\vct{p}$ remains in the tangent space $\cT_\manifold(\vct{q})$ at all time steps. The resulting \emph{constrained leapfrog integrator} \citep{leimkuhler2016efficient}, summarised in Algorithm \ref{alg.constrained_integrator}, %
is a variant of the \emph{RATTLE integrator} \citep{andersen1983rattle} with an additional intermediate momentum projection, and, subject to an appropriate projection function, is symplectic and so volume-preserving \citep{leimkuhler1994symplectic,reich1996symplectic}. As will be described in Section \ref{sec.reversibility_check}, additional \emph{reversibility checks} ensure the reversibility of the method.
\begin{algorithm}[t]
  \small
  \algorithmicrequire{
    $\lbrace \vct{p}_0, \vct{q}_0, \epsilon \rbrace $ see Algorithm \ref{alg.leapfrog_integrator}, $\tau$ convergence tolerance and $M$ maximum iterations for position projection.
  }\\
  \algorithmicensure{
    $\lbrace \vct{p}_1, \vct{q}_1 \rbrace $ see Algorithm \ref{alg.leapfrog_integrator}, \textsc{failed} flag for failed position projection.
  }
  \caption{Constrained leapfrog integrator}\label{alg.constrained_integrator}
  \begin{algorithmic} %
    \Function{ConstrainedStep}{$\vct{p}_0,\vct{q}_0,\epsilon, \tau, M$}
      \State $\tilde{\vct{p}}_{1/2} \gets \vct{p}_0 - \frac{\epsilon}{2} \nabla U(\vct{q}_0) $  \Comment{Unconstrained momentum half-step update}
      \State $\bar{\vct{p}}_{1/2} \gets \Call{projectMomentum}{\tilde{\vct{p}}_{1/2}, \vct{q}_0} $  \Comment{Project $\tilde{\vct{p}}_{1/2}$ onto $\cT _\manifold (\vct{q}_0)$ (see \S\ref{sec.momentum_projection})}
      \State $\tilde{\vct{q}}_{1} \gets \vct{q}_0 + \epsilon \bar{\vct{p}}_{1/2} $ \Comment{Unconstrained position full-step update}
      \State $\vct{q}_{1}, \textsc{failed} \gets \Call{projectPosition}{\tilde{\vct{q}}_{1}, \vct{q}_0, \tau, M} $  \Comment{Project $\tilde{\vct{q}}_1$ onto $\manifold$ (see \S\ref{sec.position_projection})}
      \State $\vct{p}_{1/2} = \frac{1}{\epsilon} (\vct{q}_1 - \vct{q}_0)$ \Comment{Compute $\vct{p}_{1/2}$ such that $\vct{q}_1 = \vct{q}_0 + \epsilon \vct{p}_{1/2}$}
      \State $\tilde{\vct{p}}_{1} \gets \vct{p}_{1/2} - \frac{\epsilon}{2} \nabla U(\vct{q}_1) $  \Comment{Unconstrained momentum half-step update}
      \State $\vct{p}_{1} \gets \Call{projectMomentum}{\tilde{\vct{p}}_{1}, \vct{q}_{1}} $  \Comment{Project $\tilde{\vct{p}}_{1}$ onto $\cT_\manifold (\vct{q}_1) $}
      \State \textbf{return} $\vct{p}_{1} , \vct{q}_{1}, \textsc{failed}$
    \EndFunction
  \end{algorithmic}
\end{algorithm}

\subsection{Momentum projection}
\label{sec.momentum_projection}

The constrained integrator in Algorithm \ref{alg.constrained_integrator}, uses a function \textsc{projectMomentum}, which orthogonally projects a momentum $\tilde{\vct{p}} \in \bR^{d_\cX}$ onto $\cT_\manifold (\vct{q})$ at a position $\vct{q} \in \manifold$. The projected momentum $\vct{p}$ can be expressed as $\vct{p} = \tilde{\vct{p}} - \jac\constr(\vct{q})\tr \vct{\lambda}$ for a vector $\vct{\lambda} \in \bR^{d_\cY}$ as the rows of $\jac\constr(\vct{q})$ span $\cN_\manifold(\vct{q})$. Solving $\jac\constr(\vct{q}) (\tilde{\vct{p}} - \jac\constr(\vct{q})\tr \vct{\lambda}) = \vct{0}$ for $\vct{\lambda}$ gives that
\begin{equation} \label{eq.momentum_projection}
  \vct{p} = \Call{projectMomentum}{\tilde{\vct{p}}, \vct{q}} =
  \big(
    \id_{d_\cX} - \jac\constr(\vct{q})\tr \gram{\vct{q}}^{-1}  \jac\constr(\vct{q})
  \big)\tilde{\vct{p}}.
\end{equation}
As the Gram matrix $\gram{\vct{q}}$ is non-singular, the projection in \eqref{eq.momentum_projection} is well-defined.

\subsection{Position projection}
\label{sec.position_projection}

For a momentum $\vct{p} \in \cT_\manifold(\vct{q})$ an update to the position $\tilde{\vct{q}} \gets \vct{q} + \epsilon \vct{p}$ will move tangentially to the manifold. However, for a non-linear manifold $\manifold$ the updated position $\tilde{\vct{q}}$ in general will not lie on the manifold. To map $\tilde{\vct{q}}$ to a point on the manifold we project $\tilde{\vct{q}} \in \cX$ along $\cN_\manifold(\vct{q})$, i.e. the normal space at the previous point $\vct{q} \in \manifold$. In other words, we need to solve the non-linear equation in $\vct{\lambda} \in \bR^{d_{\cY}}$
\begin{equation} \label{eq.position_projection}
  \chi(\vct{\lambda}) =
  \constr\big(\tilde{\vct{q}} - \jac\constr(\vct{q})\tr \vct{\lambda}\big) = \vct{0}.
\end{equation}
A range of numerical methods can potentially be used for solving Equation \eqref{eq.position_projection}.
 Applying Newton's method gives the iterative update,
\begin{equation} \label{eq.newton}
  \vct{\lambda}_{m+1} =
  \vct{\lambda}_m + \big(
    \jac \constr(\vct{q}_m) \jac \constr(\vct{q})\tr
  \big)^{-1} \chi(\vct{\lambda}_m),
  \quad
  \vct{q}_m = \tilde{\vct{q}} - \jac \constr(\vct{q})\tr \vct{\lambda}_m.
\end{equation}
This requires evaluating the constraint function and its Jacobian on each iteration, and solving a linear system in a $d_\mathcal{Y}\times d_\mathcal{Y}$ matrix.

An alternative quasi-Newton approach, proposed by \citet{barth1995algorithms} and termed the \emph{symmetric Newton method}, iterates the update
\begin{equation} \label{eq.quasi_newton}
  \vct{\lambda}_{m+1} = \vct{\lambda}_m + \gram{\vct{q}}^{-1} \chi(\vct{\lambda}_{m}).
\end{equation}
In contrast to the full Newton's update in \eqref{eq.newton} the matrix $\gram{\vct{q}}$ solved for here is fixed over all iterations. This is motivated by the observation that $\jac \constr(\vct{q}_m) $ often stays largely constant over the Newton iterations \citep{barth1995algorithms}. As a single decomposition of the Gram matrix can be used for all iterations and the constraint Jacobian does not need to be recomputed on each iteration, this approach can have significantly lower per-iteration computational costs. However, if the constraint Jacobian does vary significantly, the symmetric Newton method can take more iterations to converge than Newton's method or fail to converge at all.

In practice, we stop the non-linear solvers as soon as the norm $\|\chi(\vct{\lambda}_m)\|_\infty = \|\constr(\vct{q}_m)\|_\infty$ falls below a tolerance threshold $\tau$, or a maximum number $M$ of iterations is reached. We suggest setting $\tau$ such that the projection error is comparable to the numerical error already incurred from using floating-point arithmetic. Furthermore, as mentioned in \citet{zappa2018monte}, we also recommend setting the maximum number of iterations $M$ relatively small. In doing so, we avoid wasting computational effort on proposals whose position projections take abnormally long to converge. In our work, we set $\tau = 10^{-9}$ and $M = 50$, and the average number of iterations to convergence in practice was around 5 to 10 in our experiments. In the case when the numerical solver does not converge, which does happen for a non-negligible fraction of the proposals, the proposal is rejected.

We make an early remark that in most of our experiments using the full Newton's method yielded better sampling performance than the quasi-Newton approach, even after taking into account the relatively higher cost per iteration. We discuss this in more details in Section \ref{sec.experiments}. As explained in the next section, the possible non-convergence of the numerical solver, as well as the possible existence of multiple solutions to the non-linear equation \eqref{eq.position_projection}, require us to perform some additional checks to avoid compromising the correctness of the overall method.

\subsection{Reversibility check}
\label{sec.reversibility_check}
The correctness of the standard \ac{hmc} method is ensured by the reversibility and volume preservation of the leapfrog integrator. Although the constrained integrator is also volume-preserving, the reversibility property needs to be checked more carefully. This is because the non-linear equation \eqref{eq.position_projection} that needs to be solved to implement the \textsc{projectPosition} function may not in general have a unique solution, and the numerical method used to find such a solution may not converge within the prescribed number of iterations.

To enforce reversibility we follow the scheme proposed by \citet{zappa2018monte} (for a constrained \ac{rwm} algorithm) by manually checking if each constrained step constitutes a reversible map. For each forward step $ (\vct{q},\vct{p}) \rightarrow (\vct{q}',\vct{p}')$, we run the step in reverse (by negating the time step) $ (\vct{q}',\vct{p}')  \rightarrow (\vct{q}^*,\vct{p}^*) $. We proceed with the next step only if the current step is reversible, i.e., $(\vct{q},\vct{p}) = (\vct{q}^*, \vct{p}^*)$ to within a numerical tolerance $\rho$, \emph{and} neither of the iterative solves in the position projections in the forward nor reverse steps failed to converge (indicated by the \textsc{failed} flag returned by the \textsc{projectPosition} function in Algorithm \ref{alg.constrained_integrator} being \textsc{true}); otherwise the trajectory terminates early and we reject. By construction this reversibility check guarantees the simulated trajectory defines a reversible map.

\subsection{Complete algorithm}
\label{sec.complete_algorithm}

Recall that the lifted target distribution $\overline{\pi}(\dr\vct{q})$ is supported on the manifold $\manifold$ and has a density given by
\begin{equation*}
  \frac{\dr \overline{\pi}}{\dr \hauss}(\vct{q}) =
  \frac{d \overline{\pi}}{\dr \hauss}(\vct{\theta}, \vct{\eta}) \propto
  \exp\Big(
    -\Phi_{\vct{\theta}}(\vct{\theta})
    -  \Phi_{\vct{\eta}}(\vct{\eta})
    -\tfrac{1}{2}  \log \det\gram[]{\vct{\theta}, \vct{\eta}}
  \Big).
\end{equation*}
Consequently, for \ac{chmc} with an identity metric, the Hamiltonian reads
\begin{equation*}
  H(\vct{q},\vct{p}) = H\big((\vct{\theta}, \vct{\eta}), \vct{p}\big) =
  \Phi_{\vct{\theta}}(\vct{\theta})
  +  \Phi_{\vct{\eta}}(\vct{\eta}) +
  \tfrac{1}{2} \log\det\gram[]{\vct{\theta}, \vct{\eta}} +
  \tfrac{1}{2} \| \vct{p} \|^2.
\end{equation*}
For initializing the \ac{chmc} algorithm, one can choose an arbitrary initial $\theta_0 \in \Theta$ (e.g. by sampling from the prior) and map to the corresponding unique extended state $\vct{q}_0 = (\vct{\theta}_0, (\vct{y} - \forw(\vct{\theta}_0))/\sigma(\vct{\theta}_0)) \in \manifold$. Algorithm \ref{alg.constrained_hmc_full} provides a complete description of a single chain transition of the \ac{chmc} method.

In general, the linear algebra operations involved in each constrained leapfrog integrator step will have a $\mathcal{O}(\min(d_\Theta d_\cY^2, d_\cY d_\Theta^2))$ complexity and will also require evaluation of the constraint Jacobian at $\mathcal{O}(\min(d_\Theta, d_{\mathcal{Y}}))$ times the cost of evaluating the forward function $\forw$ (see Section \ref{app.computational_cost} in the Supplementary Material for details). While these costs are significant compared to, for example, the cost of each leapfrog integrator step in the standard \ac{hmc} algorithm, in the numerical experiments in Section \ref{sec.experiments} we will see that the ability to use a much larger integrator size and so take fewer integrator steps will often outweigh the increased per-step costs.

\begin{algorithm}[t]
\small
\caption{\ac{chmc} transition}\label{alg.constrained_hmc_full}
\algorithmicrequire{ $\vct{q}$ current position, $N$ number of integrator steps per transition, $\rho$ reversibility check tolerance, $\lbrace \epsilon, \tau, M \rbrace$ see Algorithm \ref{alg.constrained_integrator}}.\\
\algorithmicensure{ $\vct{q}'$ updated position such that $\vct{q} \sim \overline{\pi}(\cdot) \implies \vct{q}' \sim \overline{\pi}(\cdot)$.}
\begin{algorithmic} %
\Function{ConstrainedHMC}{$\vct{q}, \epsilon, \tau, \rho, N, M$}
\State $\vct{q}_0 \gets \vct{q}$
\State $\tilde{\vct{p}}_0 \gets \normal(\vct{0},\id_{d_\cX})$ \Comment{Sample momentum from standard Gaussian}
\State $\vct{p}_0 \gets \Call{projectMomentum}{\tilde{\vct{p}}_0, \vct{q}_0}$ \Comment{Project initial momentum onto $\cT_{\manifold}(\vct{q}_0)$}
\For{$n \in \lbrace 1, \dots, N \rbrace$}
\State $\vct{q}_{n}, \vct{p}_{n}, \textsc{fwFailed}  \gets \Call{ConstrainedStep}{\vct{q}_{n-1}, \vct{p}_{n-1}, \epsilon, \tau, M} $ \Comment{Forward step}
\State $\vct{q}_*, \vct{p}_*, \textsc{rvFailed} \gets \Call{ConstrainedStep}{\vct{q}_{n}, \vct{p}_{n},  -\epsilon, \tau, M}$ \Comment{Reverse step}
\If{$\textsc{fwFailed} ~\mathbf{or}~ \textsc{rvFailed} ~\mathbf{or}~ \infnorm{\vct{q}_* - \vct{q}_{n-1}} \geq \rho $} \Comment{Reversibility check}
\State return $\vct{q} $ \Comment{Reject proposal if non-reversible}
\EndIf
\EndFor
\vspace{-2pt}
\If { $\text{uniform}(0,1) < \exp\big(H(\vct{q}_0, \vct{p}_0) -  H(\vct{q}_n, \vct{p}_n) \big)$ } \Comment{Acceptance step}
\State return $ \vct{q}_{N}$ \Comment{Accept proposal}
\Else
\State  return $\vct{q}$ \Comment{Reject proposal}
\EndIf
\EndFunction
\end{algorithmic}
\end{algorithm}

\section{Lifted Hamiltonian dynamic in original latent space}
\label{sec.theory}

To provide some theoretical insight to our approach, we now show that the lifted Hamiltonian dynamic is equivalent to a Hamiltonian dynamic on the original latent space under a particular choice of metric (mass matrix). For linear-Gaussian models the lifted dynamic corresponds to an `optimal' choice of setting the metric to the posterior precision matrix. For general models, the lifted dynamic corresponds to a Hamiltonian dynamic with a position-dependent metric that accounts for the locally varying scale and curvature of the distribution, analogously to \ac{rmhmc}. %

\subsection{Linear-Gaussian models}

We first consider the special case of a linear-Gaussian model with fixed $\sigma > 0$ of the form
\begin{equation*}
  \vct{y} = \mtx{F}\vct{\theta} + \vct{f} + \sigma \vct{\eta},
  \quad \vct{\theta} \sim \normal(\vct{0}, \idmtx),
  \quad \vct{\eta} \sim \normal(\vct{0}, \idmtx).
\end{equation*}
\begin{remark} \label{rmk.reparameterise}
Any linear-Gaussian model of the more general form
\begin{equation*}
  \vct{y} = \mtx{F}'\vct{\theta}' + \vct{f}' + \sigma \vct{\eta},
  \quad \vct{\theta}' \sim \normal(\vct{m}, \mtx{C}),
  \quad \vct{\eta} \sim \normal(\vct{0}, \idmtx)
\end{equation*}
can be reparametrized into the form above by defining $\vct{\theta} = \mtx{L}^{-1}(\vct{\theta}' - \vct{m})$, $\mtx{F} = \mtx{F}'\mtx{L}$ and $\vct{f} = \vct{f}' - \mtx{F}'\vct{m}$ for any $\mtx{L}$ such that $\mtx{L} \mtx{L}
\tr = \mtx{C}$.
\end{remark}
The posterior distribution in this case can be derived analytically as
\begin{equation*}
  \vct{\theta} \,|\, \vct{y} \sim \normal(\vct{\mu}, \mtx{\Sigma}),
  \quad \mtx{\Sigma} = (\idmtx + \tfrac{1}{\sigma^2}\mtx{F}\tr\mtx{F})^{-1},
  \quad \vct{\mu} = \tfrac{1}{\sigma^2} \mtx{\Sigma} \mtx{F}\tr (\vct{y} - \vct{f}).
\end{equation*}
We can define a Hamiltonian dynamic on $\Theta$, the flow map of which marginally preserves the posterior, by first introducing a momentum variable $\vct{p} \in \mathbb{R}^{d_\Theta}$ which is independent of $\vct{\theta}$ and with marginal distribution $\normal(\vct{0}, \mtx{M})$ for some positive definite matrix $\mtx{M}$. The \emph{Hamiltonian} is then equal to the negative logarithm of the joint density on $(\vct{\theta}, \vct{p})$,
\begin{equation*}
  H(\vct{\theta}, \vct{p}) =
  \tfrac{1}{2}(\vct{\theta} - \vct{\mu})\tr \mtx{\Sigma}^{-1}(\vct{\theta} -\vct{\mu}) + \tfrac{1}{2}\vct{p}\tr \mtx{M}^{-1}\vct{p},
\end{equation*}
with corresponding Hamiltonian dynamics described by the system of \acp{ode}
\begin{equation}\label{eq:linear-gaussian-dynamic}
\dot{\vct{\theta}} = \mtx{M}^{-1} \vct{p}, \qquad \dot{\vct{p}} = -\mtx{\Sigma}^{-1}(\vct{\theta} - \vct{\mu}).
\end{equation}
Alternatively, if we lift the posterior distribution on to the manifold
\begin{equation*}
\mathcal{M} = \lbrace \vct{\theta} \in \mathbb{R}^{d_\Theta}, \vct{\eta} \in \mathbb{R}^{d_{\mathcal{Y}}} : C(\vct{\theta}, \vct{\eta}) = \mtx{F}\vct{\theta} + \vct{f} + \sigma \vct{\eta} - \vct{y} = 0 \rbrace
\end{equation*}
then lifted posterior distribution $\bar{\pi}$ has density
\begin{equation*}
\bar{\pi}(\vct{\theta},\vct{\eta}) \propto \exp\left( -\tfrac{1}{2}\| \vct{\theta} \|^2 -\tfrac{1}{2}\| \vct{\eta} \|^2 \right) \left\vert \mtx{F}\mtx{F}\tr + \sigma^2 \idmtx\right\vert^{-\tfrac{1}{2}}
\end{equation*}
with respect to the $d_\Theta$-dimensional Hausdorff measure on $\mathcal{M}$.
Introducing momenta $(\vct{v}_{\vct{\theta}}, \vct{v}_{\vct{\eta}}) \in \mathcal{T}_\mathcal{M}(\vct{\theta},\vct{\eta})$ with density
\(
  \exp(-\tfrac{1}{2}\| \vct{v}_{\vct{\theta}}\|^2 -\tfrac{1}{2}\| \vct{v}_{\vct{\eta}}\|^2)
\)
with respect to the $d_\Theta$-dimensional Hausdorff measure on $\mathcal{T}_\mathcal{M}(\vct{\theta},\vct{\eta})$ and a vector of Lagrange multipliers $\vct{\lambda} \in \mathbb{R}^{d_\mathcal{Y}}$, implicitly defined by the constraint $C(\vct{\theta}, \vct{\eta}) = 0$, the Hamiltonian for the resulting constrained system is
\begin{equation*}
  \bar{H}((\vct{\theta}, \vct{\eta}), (\vct{v}_{\vct{\theta}}, \vct{v}_{\vct{\eta}})) =
  \tfrac{1}{2}\| \vct{\theta} \|^2 + \tfrac{1}{2}\| \vct{\eta} \|^2 + \tfrac{1}{2}\| \vct{v}_{\vct{\theta}}\|^2 + \| \vct{v}_{\vct{\eta}}\|^2 + (\mtx{F}\vct{\theta} + \vct{f} + \sigma\vct{\eta} - \vct{y})\tr\vct{\lambda}
\end{equation*}
with corresponding constrained Hamiltonian dynamics described by the system of differential algebraic equations
\begin{equation}\label{eq:linear-gaussian-constrained-dynamic}
  \begin{aligned}
  \dot{\vct{\theta}} = \vct{v}_{\vct{\theta}},
  \quad
  \dot{\vct{\eta}} = \vct{v}_{\vct{\eta}},
  \quad
  \dot{\vct{v}}_{\vct{\theta}} = -\vct{\theta} - \mtx{F}\tr\vct{\lambda},
  \quad
  \dot{\vct{v}}_{\vct{\eta}} = -\vct{\eta} - \sigma \vct{\lambda} \\
  \text{subject to } \quad
  \mtx{F}\vct{\theta} + \vct{f} + \sigma\vct{\eta} - \vct{y} = \vct{0}
  \; \text{ and } \;
  \mtx{F}\vct{v}_{\vct{\theta}} + \sigma \vct{v}_{\vct{\eta}} = 0.
  \end{aligned}
\end{equation}
The flow map corresponding to these dynamics marginally preserves the lifted distribution $\bar{\pi}$ on $(\vct{\theta}, \vct{\eta})$ and so the original posterior $\pi$ on $\vct{\theta}$.
We are now in a position to state our first result (proof in Appendix \ref{sec.thm.equivalent_dynamics}).
\begin{theorem} \label{thm.equivalent_dynamics}
The continuous-time constrained Hamiltonian dynamics of $\vct{\theta}$ in  \eqref{eq:linear-gaussian-constrained-dynamic} are equivalent to the Hamiltonian dynamics of $\vct{\theta}$ in \eqref{eq:linear-gaussian-dynamic} with the metric set to the posterior precision matrix of $\vct{\theta}$ under $\pi$, i.e. $\mtx{M} = \mtx{\Sigma}^{-1}$.
\end{theorem}

This result illustrates that for linear-Gaussian models, the lifting approach results in continuous-time Hamiltonian dynamics corresponding to those produced by the `optimal' choice of metric $\mtx{M} = \mtx{\Sigma}^{-1}$, that is resulting in dynamics corresponding to decoupled simple harmonic motions of unit angular frequency in each component of $\vct{\theta}$. Importantly this result is dependent on the model being parametrized such that $\vct{\theta}$ has a standard normal prior; for models of the more general form in Remark \ref{rmk.reparameterise} with a non-standard normal prior, the result in Theorem \ref{thm.equivalent_dynamics} does not hold without reparametrization. This motivates our advice to use a standard normal prior parametrization where possible when using the manifold lifting approach. For models with Gaussian priors this can be equivalently achieved by using a non-identity metric in the constrained Hamiltonian dynamics however to avoid introducing unnecessary additional notation we assume an identity metric for the constrained dynamics here.

\subsection{More general models}

We now consider a more general model of the form
\begin{equation*}
\vct{y} = F(\vct{\theta})+ \sigma(\vct{\theta}) \vct{\eta}, \quad \vct{\theta} \sim \normal(\vct{0}, \idmtx), \quad \vct{\eta} \sim \normal(\vct{0}, \idmtx),
\end{equation*}
where $F$ and $\sigma$ are both potentially non-linear functions of latent variables $\vct{\theta}$.
If we lift the posterior distribution on to the manifold
\begin{equation*}
\mathcal{M} = \lbrace \vct{\theta} \in \mathbb{R}^{d_\Theta}, \vct{\eta} \in \mathbb{R}^{d_{\mathcal{Y}}} : C(\vct{\theta}, \vct{\eta}) = F(\vct{\theta}) + \sigma(\vct{\theta}) \vct{\eta} - \vct{y} = \vct{0} \rbrace
\end{equation*}
then the lifted posterior distribution $\bar{\pi}$ has density
\begin{equation*}
\bar{\pi}(\vct{\theta},\vct{\eta}) \propto \exp\left( -\tfrac{1}{2}\| \vct{\theta} \|^2 -\tfrac{1}{2}\| \vct{\eta} \|^2 \right) \left\vert G(\vct{\theta}, \vct{\eta}) \right\vert^{-\tfrac{1}{2}}
\end{equation*}
with respect to the $d_\Theta$-dimensional Hausdorff measure on $\mathcal{M}$, where
\begin{equation*}
  G(\vct{\theta}, \vct{\eta}) = (\jacob F(\vct{\theta}) + \vct{\eta} \jacob\sigma(\vct{\theta}))(\jacob F(\vct{\theta})+ \vct{\eta} \jacob\sigma(\vct{\theta}))\tr + \sigma(\vct{\theta})^2\idmtx.
\end{equation*}
Introducing momenta $(\vct{v}_{\vct{\theta}}, \vct{v}_{\vct{\eta}}) \in \mathcal{T}_{\mathcal{M}}(\vct{\theta},\vct{\eta})$ with density $\exp(-\tfrac{1}{2}\| \vct{v}_{\vct{\theta}}\|^2 -\tfrac{1}{2}\| \vct{v}_{\vct{\eta}}\|^2)$ with respect to the $d_\Theta$-dimensional Hausdorff measure on $\mathcal{T}_{\mathcal{M}}(\vct{\theta},\vct{\eta})$, the Hamiltonian function for the constrained system is then
\begin{equation*}
  \bar{H}((\vct{\theta},\vct{\eta}), (\vct{v}_{\vct{\theta}}, \vct{v}_{\vct{\eta}})) =
  \tfrac{1}{2}\| \vct{\theta} \|^2 + \tfrac{1}{2}\| \vct{\eta} \|^2 + \tfrac{1}{2}\log|G(\vct{\theta}, \vct{\eta})| + \tfrac{1}{2}\| \vct{v}_{\vct{\theta}}\|^2 + \tfrac{1}{2}\| \vct{v}_{\vct{\eta}}\|^2 + C(\vct{\theta}, \vct{\eta})\tr\vct{\lambda}.
\end{equation*}
Alternatively, working with the posterior distribution $\pi$ which has Lebesgue density
\begin{equation*}
  \pi(\vct{\theta}) \propto
  \exp\left(
    -\tfrac{1}{2\sigma(\vct{\theta})^2}(\vct{y}-F(\vct{\theta}))\tr(\vct{y}-F(\vct{\theta})) -d_{\mathcal{Y}}\log\sigma(\vct{\theta})-\tfrac{1}{2}\| \vct{\theta} \|^2
  \right),
\end{equation*}
if we introduce a momentum $\vct{p} \in \mathbb{R}^{d_\Theta}$ with conditional distribution  $\normal(0, M(\vct{\theta}))$ given $\vct{\theta}$ for a positive-definite matrix valued function $M$, then we can define a Hamiltonian equal to the negative logarithm of the joint density on $\vct{\theta}$ and $\vct{p}$
\begin{equation*}
  H(\vct{\theta}, \vct{p}) =
  \tfrac{1}{2\sigma(\vct{\theta})^2}\|\vct{y}-F(\vct{\theta})\|^2 + d_{\mathcal{Y}}\log\sigma(\vct{\theta}) + \tfrac{1}{2}\| \vct{\theta} \|^2 \\+ \tfrac{1}{2}\vct{p} \mtx{M}(\vct{\theta})^{-1}\vct{p} + \tfrac{1}{2}\log|\mtx{M}(\vct{\theta})|.
\end{equation*}
We then have that the constrained Hamiltonian dynamics on $\mathcal{M}$ are equivalent to the Hamiltonian dynamics on $\Theta$ with a particular \emph{position-dependent} choice of $\mtx{M}$ (proof in Appendix \ref{sec.thm.equivalent_dyna.general}).
\begin{theorem} \label{thm.equivalent_dyna.general}
The continuous-time constrained Hamiltonian dynamics of $\vct{\theta}$ in the lifted system are equivalent to the Hamiltonian dynamics of $\vct{\theta}$ in the original system with the position-dependent metric (mass matrix)
\begin{equation*}
  M(\vct{\theta}) = \idmtx + \tfrac{1}{\sigma(\vct{\theta})^2} (\jacob F(\vct{\theta}) +\vct{\eta}(\vct{\theta})\jacob\sigma(\vct{\theta}))\tr(\jacob F(\vct{\theta}) +\vct{\eta}(\vct{\theta})\jacob\sigma(\vct{\theta})),
  ~~
  \vct{\eta}(\vct{\theta}) = \tfrac{1}{\sigma(\vct{\theta})}(\vct{y} - F(\vct{\theta})).
\end{equation*}
\end{theorem}
For the special case where $\sigma(\vct{\theta}) = \sigma$ is a constant function (that is the observation noise standard deviation is assumed to be known and fixed), the metric function simplifies to
\(
  M(\vct{\theta}) =  \idmtx + \tfrac{1}{\sigma^2} \jacob F(\vct{\theta})\tr\jacob F(\vct{\theta}),
\)
corresponding to the expected Fisher information matrix $\tfrac{1}{\sigma^2} \jacob F(\vct{\theta})\tr\jacob F(\vct{\theta})$  plus the negative Hessian of the log prior $\idmtx$ for this model. This specific choice of metric is recommended by \citet{girolami2011riemann}, and used in their numerical experiments.

From this equivalence of the continuous-time dynamics, the observed performance differences between our proposed \ac{chmc} approach and \ac{rmhmc} (with a specific choice of metric) can be seen as a direct consequence of the improved performance of the constrained leapfrog integrator used in \ac{chmc} compared to the generalised leapfrog integrator used in \ac{rmhmc}. Although both involve implicit steps, as described in Section \ref{sec.position_projection}, for each constrained leapfrog integrator step we can use Newton's method to efficiently solve the non-linear system of equations during the position projection.

In contrast in each generalised leapfrog step, two separate systems of non-linear equations must be solved, with existing implementations generally using a simple direct fixed-point iteration \citep{girolami2011riemann}. Empirically we observe that these simple fixed-point iterations have a much higher tendency to diverge or converge slowly compared to the Newton iteration used for the constrained integrator. We also stress that as the fixed-point solvers used in the generalised leapfrog integrator may fail to converge, or converge to a different solution in a time-reversed step, a similar reversibility check as used for the constrained leapfrog integrator (and originally proposed in a \ac{rwm} setting by \citet{zappa2018monte}), is required to ensure the correctness of the \ac{rmhmc} transitions. The \ac{rmhmc} implementation we used in our experiments includes such checks.

\section{Numerical experiments}
\label{sec.experiments}

In this section, we evaluate the performance of our proposed methodology on artificial and real-world inference problems. We test variants of the \ac{chmc} algorithm using both a symmetric Newton and full Newton solver for the projection step. In each example, to provide a representative benchmark for existing \ac{mcmc} methods we compare to a standard \ac{hmc} algorithm with either a diagonal or dense metric. We also compare to \ac{rmhmc} on the first (toy) example, however due to its poor convergence, even with long chains, we do not include results on the other examples.

For each method (standard \ac{hmc}, \ac{chmc}, and \ac{rmhmc}), we use the updated version of the \ac{nuts} algorithm \citep{hoffman2014no, betancourt2017conceptual} currently underlying the probabilistic programming framework \emph{Stan} \citep{carpenter2017stan} to automate setting the number of integrator steps in each proposal's trajectory. During an initial warm-up stage of each chain, the integrator step size $\epsilon > 0$ was tuned with a dual-averaging algorithm to achieve an average acceptance rate of 0.9 as described in \citet{hoffman2014no}. For the standard \ac{hmc} chains, the diagonal or dense metric was additionally tuned during this warm-up stage using an equivalent scheme to that employed by \emph{Stan}. All experiments were performed using the \ac{mcmc} sampler implementations in the Python package \emph{Mici} \citep{graham2019mici}, which defines a high-level interface that abstracts away the details of Algorithm \ref{alg.constrained_hmc_full} and requires only specification of the model functions and optionally, their derivatives.

We report the minimum estimated \ac{ess} over the quantities of interest, normalised by the total chain run time, as our measure of sampling efficiency. We compute the \ac{ess} estimates using the \emph{ArviZ} \citep{arviz2019} implementation of the \emph{bulk \ac{ess}} statistic \citep{vehtari2019rank}. We also report, where informative, the maximum potential scale reduction $\hat{R}$ convergence diagnostic across the quantities of interest, using the \emph{ArviZ} implementation of the rank-normalized split-$\hat{R}$ statistic proposed by \citet{vehtari2019rank}, with values greater than one indicative of non-convergence. For each of the experiments we ran three sets of four chains (twelve in total) for each method tested, with 1000 adaptive warm-up iterations and 2500 main iterations per-chain. For the standard \ac{hmc} chains, the metric adaptation was shared across each set of four chains, otherwise all chains were independent and with initial states independently sampled from the prior. Only the non-warm-up samples were used in computing the \ac{ess} estimates.

All plots were produced using the Python package \emph{Matplotlib} \citep{hunter2007matplotlib} and the Python packages \emph{NumPy} \citep{harris2020array} and \emph{SciPy} \citep{scipy2020scipy} were used for array and linear-algebra operations. In all cases unless otherwise mentioned, the forward models were implemented using the numerical computing framework \emph{JAX} \citep{jax2018github} to allow efficient calculation of the required derivatives using automatic differentiation.

\subsection{Toy example}
\label{sec.toy}

\begin{figure}[t]
  \centering
  \includegraphics[width=0.9\textwidth]{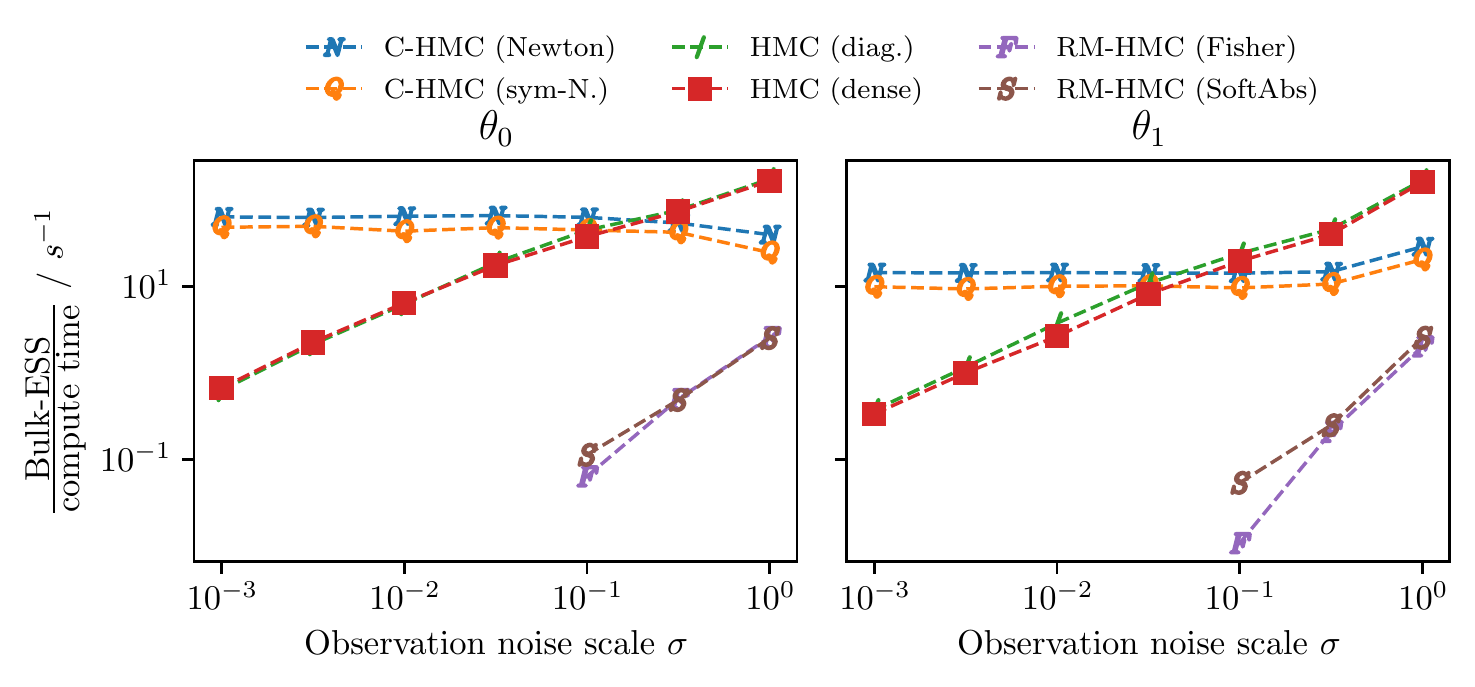}
  \caption{Toy two-dimensional model. Sampling efficiency for varying $\sigma$.}
  \label{fig.loop_ess_results}
\end{figure}

We first return to our running toy example considered in Section \ref{sec.vanishing_noise}. Figure \ref{fig.loop_ess_results} shows the sampling efficiencies for standard \ac{hmc}, \ac{rmhmc} and our proposed approach of \ac{chmc} on the extended space. For \ac{rmhmc} the same Fisher information based and \emph{SoftAbs} metrics are used as described previously. It can be immediately seen that both tested variants of \ac{hmc} and \ac{rmhmc} degenerate as $\sigma \to 0$ while the efficiency of \ac{chmc} with both projection solvers is unaffected by the vanishing noise asymptotic $\sigma \to 0$. Figure \ref{fig.loop_rhat_step_size_results} in the Supplementary Material shows the $\hat{R}$ convergence statistics and integrator step sizes $\epsilon$ for varying $\sigma$. For all but the largest $\sigma=1$ value the \ac{rmhmc} chains show signs of non-convergence. To maintain a non-zero average acceptance rate, both standard \ac{hmc} and \ac{rmhmc} chains require an integrator step size which decreases with $\sigma$; the \ac{chmc} chains on the other hand maintain a stable acceptance rate with an approximately constant step size.

\subsection{State space model}
\label{sec.statespace}

We apply our proposed methodology to a \ac{ssm} to illustrate that the approach can also give gains in models with high-dimensional latent spaces. A sequence of $T$ latent states $x_{1{:}T}$ are generated from a linear-Gaussian dynamic
\begin{equation}
  x_{1} \sim \normal(\mu, \tfrac{\gamma^2}{1 - \rho^2}),
  \quad
  x_{t} = \mu + \rho(x_{t-1} - \mu) + \gamma  \nu_{t},
  \quad
  \nu_t \sim \normal(0, 1)
  \quad
  \forall t \in 2{:}T
\end{equation}
for  $T = 1000$ and parameter values $\mu = -0.5$, $\gamma = 0.4$ and $\rho = 0.9$. From this latent sequence $x_{1{:}T}$,  we generate $T$ observations according to
\begin{equation}
  y_{t} = \exp(x_{t}) + \sigma \eta_t,
  \quad
  \eta_t \sim \normal(0,1)
  \quad
  ~\forall t \in 1{:}T,
\end{equation}
for each observation noise standard deviation $\sigma \in \lbrace 10^{-3},10^{-2},10^{-1},10^0,10^1\rbrace$.
Then, we jointly infer the parameters $\mu, \rho, \gamma, \sigma$ and latent state sequence $x_{1{:}T}$ given the observations $y_{1{:}T}$ for each of the true $\sigma$ values.

We assign weakly informative priors on the parameters: $\mu \sim \normal(0, 1)$, $\gamma \sim \halfnormal(1)$, $\rho \sim \uniform(-1, 1)$ and $\sigma \sim \halfnormal(3)$. We use a non-centred parametrization of the model in terms of the a-priori independent noise sequence $\nu_{1{:}T}$ rather than the state sequence $x_{1{:}T}$, and express the parameters with bounded support ($\gamma$, $\rho$ and $\sigma$) as smooth transforms of unbounded variables.

As the number of observations and latent variables both scale with $T$, a direct application of the constrained \ac{hmc} algorithm using the lifting described in Section \ref{sec.state_space_augmentation} would result in an $\mathcal{O}(T^3)$ complexity per constrained integrator step, due to the need to compute, evaluate the determinant of and solve linear systems in a $\mathcal{O}(T) \times \mathcal{O}(T)$ Gram matrix at each step. By exploiting the invertibility of the observation function here and Markovian structure of the model, the cost can be reduced to $\mathcal{O}(T)$. In brief, by using an alternative definition of the constraint function the resulting Gram matrix corresponds to the sum of a block tridiagonal matrix and low rank product, allowing efficient $\mathcal{O}(T)$ operations. More details are given in Section \ref{sec.exploiting-markovian-structure-in-ssms} in the Supplementary Material.

\begin{figure}
  \centering
  \includegraphics[width=0.9\textwidth]{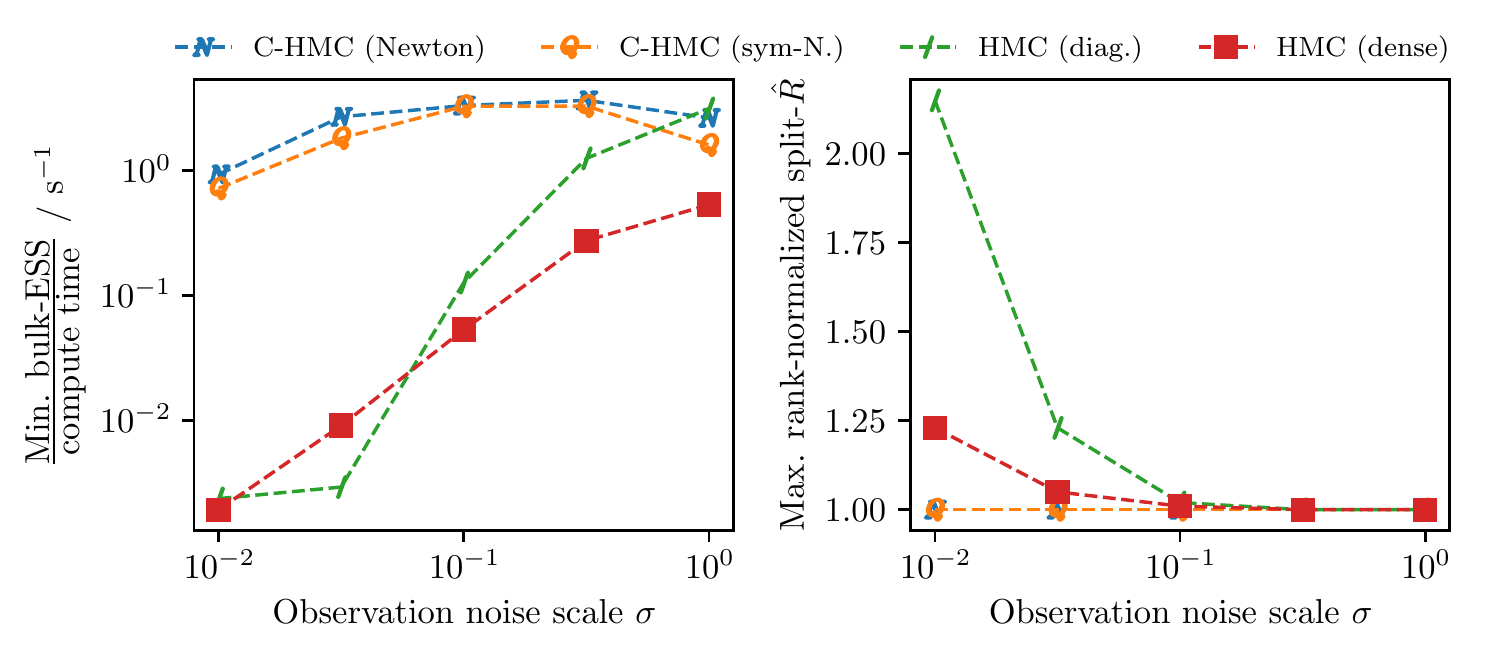}
  \caption{Nonlinear \ac{ssm} model. Left: minimum $\tfrac{\textsc{ess}}{\text{compute time}}$ for varying $\sigma$; right: maximum $\hat{R}$ for varying $\sigma$. Min./max. are across the model parameters $\mu, \rho, \gamma, \sigma$.}
  \label{fig.nonlinear_ssm_results}
\end{figure}

The left panel in Figure \ref{fig.nonlinear_ssm_results} shows the measured computational efficiencies (in terms of estimated bulk-\ac{ess} per computation time) versus the true observation noise standard deviation $\sigma$ for this \ac{ssm} model. We see that the efficiency of our \ac{chmc} approach (with both projection solvers) varies only minimally across all $\sigma$ values, while, in contrast, the standard \ac{hmc} methods both show a sustained degradation in efficiency as $\sigma$ becomes smaller. The $\hat{R}$ convergence statistics values show in the right panel in Figure \ref{fig.nonlinear_ssm_results} are also increasingly indicative of non-convergence for the \ac{hmc} chains as $\sigma$ becomes smaller. To give a sense of anistropy being induced in the posterior as $\sigma$ decreases, Figure \ref{fig.nonlinear_ssm_hessian_evals} in the Supplementary Material shows the Hessian of the negative log-density of the posterior (evaluated at a \emph{maximum a posteriori} estimate) becoming increasingly ill-conditioned as $\sigma \to 0$.

Due to our ability to exploit additional model structure in this case, the constrained \ac{hmc} method also remains competitive for even relatively large $\sigma$ values despite the large latent and observation dimensions ($d_\Theta = 1004$ and $d_{\mathcal{Y}} = 1000$). The $\mathcal{O}(T)$ scaling with respect to the number of observation times of our approach is the same as standard \ac{hmc} in this case.

\subsection{FitzHugh--Nagumo ordinary differential equation model}
\label{sec.fhn_ode}

Next, we consider parameter inference in non-linear dynamical system with noisy observations. The FitzHugh--Nagumo model \citep{fitzhugh1961impulses,nagumo1962active}, a simplified description of the dynamics of action potential generation within an neuron, can be described by the following system of \acp{ode}
\begin{equation}\label{eq.fhn}
  \begin{aligned}
  \partial_1 x_0(t, \vct{\theta}) &= \alpha x_1(t, \vct{\theta}) - \beta x_0(t, \vct{\theta})^3 + \gamma x_1(t, \vct{\theta}),
  \\
  \partial_1 x_1(t, \vct{\theta}) &= -\delta x_0(t, \vct{\theta}) - \epsilon x_1(t, \vct{\theta}) + \zeta.
  \end{aligned}
\end{equation}
We use a fourth-order Runge--Kutta method to discretize the system in time. The ground truth parameters used were $\alpha = 3$, $\beta = 1$, $\gamma = 3$, $\delta = 1/3$, $\varepsilon = 1/15$, and $\zeta = 1/15$, and initial state $\vct{x}(0, \cdot) = (1, -1)$. We simulated noisy observations $y_k = x_0(t_k, \vct{\theta}) + \sigma \eta_k$, with $\eta_k \sim \normal(0,1)$, of the first state coordinate at $d_\cY=200$ equispaced times $0 = t_1 < \ldots < t_{d_\cY}=20$.

The parametrisation of the Fitzhugh--Nagumo system of \acp{ode} in Equation \eqref{eq.fhn} is non-identifiable; the dynamics remains unchanged under the transformation
\begin{equation*}
  (x_0(0, \cdot), x_1(0, \cdot), \alpha, \beta, \gamma, \delta, \varepsilon, \zeta, \sigma) \mapsto
  (x_0(0, \cdot), s x_1(0, \cdot), \alpha, \beta, s^{-1} \gamma, s \delta, \varepsilon, s \zeta, \sigma)
\end{equation*}
for any scaling factor $s > 0$. This means that $x_2(0, \cdot)$, $\gamma$, $\delta$ and $\zeta$ are only identifiable up to a common scaling factor $s$. Therefore, the posterior distribution on these variables concentrates around a one-dimensional limiting manifold as $\sigma \to 0$.

We use prior distributions $\alpha  \sim \lognormal(0, 1)$, $\beta  \sim \lognormal(0, 1)$, $\gamma  \sim \lognormal(0, 1)$, $\delta  \sim \lognormal(-1, 1)$, $\varepsilon \sim \lognormal(-2, 1)$, $\zeta  \sim \lognormal(-2, 1)$, $x_0(0) \sim \normal(0, 1)$, and $x_1(0) \sim \normal(0, 1)$, using a logarithmic parametrisation for $(\beta, \gamma, \delta, \varepsilon, \zeta)$ to enforce non-negativity constraints.

\begin{figure}
    \centering
    \includegraphics[width=0.9\textwidth]{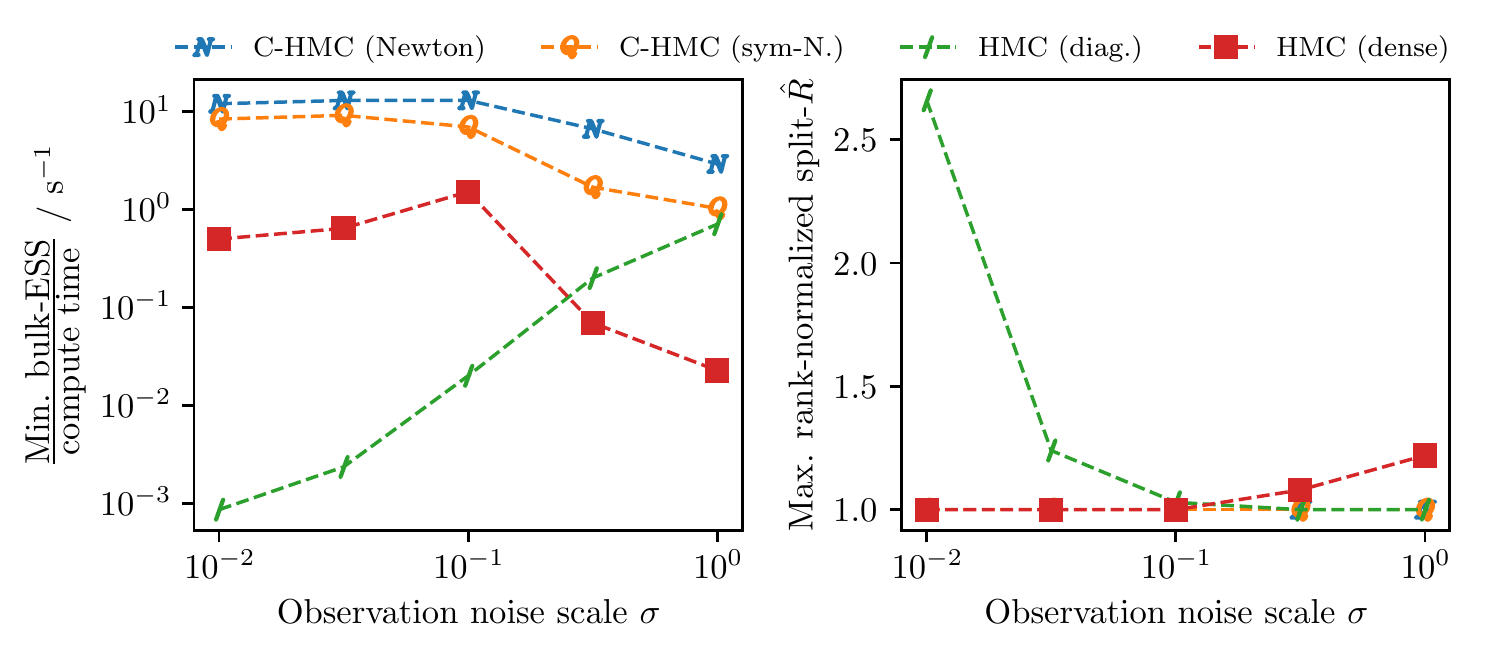}
    \caption{FitzHugh--Nagumo \ac{ode} model. Left: minimum $\tfrac{\textsc{ess}}{\text{compute time}}$ for varying $\sigma$; right: maximum $\hat{R}$ for varying $\sigma$. Min./max. are across the 9 model parameters.}
    \label{fig.fhn_results}
\end{figure}

Figure \ref{fig.fhn_results} shows the estimated \ac{ess} and $\hat{R}$ convergence statistics for varying $\sigma$ for this \ac{ode} model. The \ac{chmc} chains using both the full Newton and symmetric Newton solver show a similar pattern of performance, with near constant \ac{ess} per compute time for smaller $\sigma$ values and a slight drop-off in efficiency as $\sigma$ becomes larger; as for the previous \ac{ssm} model we speculate that this is due to a greater variation in curvature in the posterior for the more diffuse large $\sigma$ settings, requiring a smaller integrator step size $\epsilon$ to maintain the stability of the projection solver and in turn a drop in sampling efficiency. In support of this hypothesis, Figure \ref{fig.simulated_data_model_step_size_results} in the Supplementary Material shows that the adapted integrator step sizes show a small decreasing trend for large $\sigma$ for the \ac{chmc} chains. The Newton solver appears to give a small but consistent improvement in efficiency across all $\sigma$ values compared to the symmetric Newton solver.

The results for the \ac{hmc} chains with diagonal metric show the same decreasing sampling efficiency with $\sigma$ as encountered previously, with the $\hat{R}$ values indicative of non-convergence for the smaller observation noise scales $\sigma$. The results for the \ac{hmc} chains with dense metric on the other hand seem to indicate a quite different behaviour, with the sampling efficiency initially seeming to \emph{increase} as the observation noise scale $\sigma$ is \emph{decreased} before beginning to decrease again for $\sigma < 0.1$, though at a slower rate than for the \ac{hmc} chains with diagonal metric. The adapted integrator step size $\epsilon$ for the dense metric \ac{hmc} chains (Figure \ref{fig.simulated_data_model_step_size_results} in the Supplementary Material) also show a similar non-monotonic relationship with $\sigma$. The split-$\hat{R}$ values also only indicate non-convergence for larger $\sigma$ values for the \ac{hmc} chains with dense metric, with values close to one for the smallest $\sigma$ values.

We believe these results are due a pathological behaviour of the adaptation of the \ac{hmc} chains with dense metric for small $\sigma$ in this model, causing the chains to become effectively non-ergodic. It appears that for small $\sigma$ the step size adaptation typically produces step sizes which give a vanishingly low probability of the \ac{hmc} trajectories being able to enter the `narrowest' regions of the posterior in the tails. Figure \ref{fig.fhn_pair_plots_sigma} in the Supplementary Material shows pairwise scatter plots for samples of the non-identifiable parameters $(x_{1}(0), \gamma, \delta, \zeta)$ for both \ac{hmc} (with dense metric) and \ac{chmc} (with Newton solver) chains for $\sigma = 0.01$ and $\sigma = 0.1$. It is immediately evident that the \ac{hmc} chains have explored only a subset of the posterior mass covered by the \ac{chmc} chain samples, with the \ac{hmc} chains seeming to have particularly failed to explore the narrow regions of the posterior in the tails along the directions spanned by the variables in which the limiting manifold is non-linear.

The generated \ac{hmc} trajectories rarely approach these narrow posterior regions and when they do, as the integrator step size is too large for the local curvature, they typically diverge leading to a rejection; however, as the proportion of the posterior mass in these regions is small (for $\sigma=0.01$ just over two percent of \ac{hmc} with dense metric chain transitions reject due to integrator divergence), these rejections do not significantly affect the average acceptance rate and so fail to cause the step size to be adapted to a more appropriate smaller value. Eventually for long enough runs the chains will enter the narrower regions and will typically then get `stuck' and reject for many successive iterations due to the inappropriately large step size for the local curvature - trace plots for a \ac{hmc} (with dense metric) chain showing such a sticking event are shown in Figure \ref{fig.fitzhugh_nagumo_hmc_sticking_example_trace} in the Supplementary Material. However, for finite length chains, such sticking events are rare and do not appear to have adversely effected the $\hat{R}$ convergence diagnostic.

This issue seems to particularly effect the \ac{hmc} chains with dense metric as the equivalent linear transformation effected by the adapted dense metric, while leading to a beneficial rescaling and decorrelation of the variables in a local region of the posterior where the limiting manifold $\mathcal{S}$ is close to linear, conversely has a detrimental effect in the tail regions of the posterior with the linear transformation meaning an even smaller step size is needed for stability in these regions. This pathology and the difficult in diagnosing it with standard heuristics, provide a clear illustration of the benefits of using an \ac{mcmc} method, such as our proposed \ac{chmc} based approach, which is robust to a range of posterior scales.

\subsection{Hodgkin--Huxley ordinary differential equation model}
\label{sec.hh}

While the parameter non-identifiability in the previous example (Section \ref{sec.fhn_ode}) could be removed by reparametrization, we now consider a non-linear dynamical model with  structural non-identifiabilities \citep{daly2015hodgkin} that, to the best of the authors' knowledge, does not admit such a reparametrisation.

Conductance-based or \emph{Hodgkin--Huxley} models \citep{hodgkin1952quantitative}, are a family of non-linear dynamical system that describe the time-varying membrane voltage of neurons in terms of the ionic currents associated with voltage-gated ion channels. This class of mechanistic models is widely used in computational neuroscience, and inference in the vanishing noise regime is of particular relevance as modern intracellular recording technologies allow high-fidelity recording of cell membrane voltages with high signal-to-noise ratios \citep{weckstrom2010intracellular}.

Here, we specifically consider a conductance-based model using three channel types, with the model behaviour determined by a `minimal' set of 6 free parameters $(\gbar{Na}, \gbar{K}, \gbar{M}, g_\text{L}, V_T, k_{\tau_p})$ which nonetheless allow good fits to experimental data from a broad range of cell types with differing characteristic behaviours \citep{pospischil2008minimal}. We consider the case of a cell subject to a square-wave stimulus current and with the cell membrane voltage noisily observed at $d_{\mathcal{Y}} = 1000$ equispaced time points. We infer the $d_{\Theta} = 7$ model parameters $\vct{\theta} = (\gbar{Na}, \gbar{K}, \gbar{M}, g_\text{L}, V_T, k_{\tau_p}, \sigma)$ for observations generated for each of the true observation noise standard deviations $\sigma \in \lbrace 10^{-2}, 10^{-1.5}, 10^{-1}, 10^0\rbrace$. A full description of the model used including the priors and time discretisation is given in Section \ref{sec.hh_full} in the Supplementary Material.
\begin{figure}
    \includegraphics[width=0.9\textwidth]{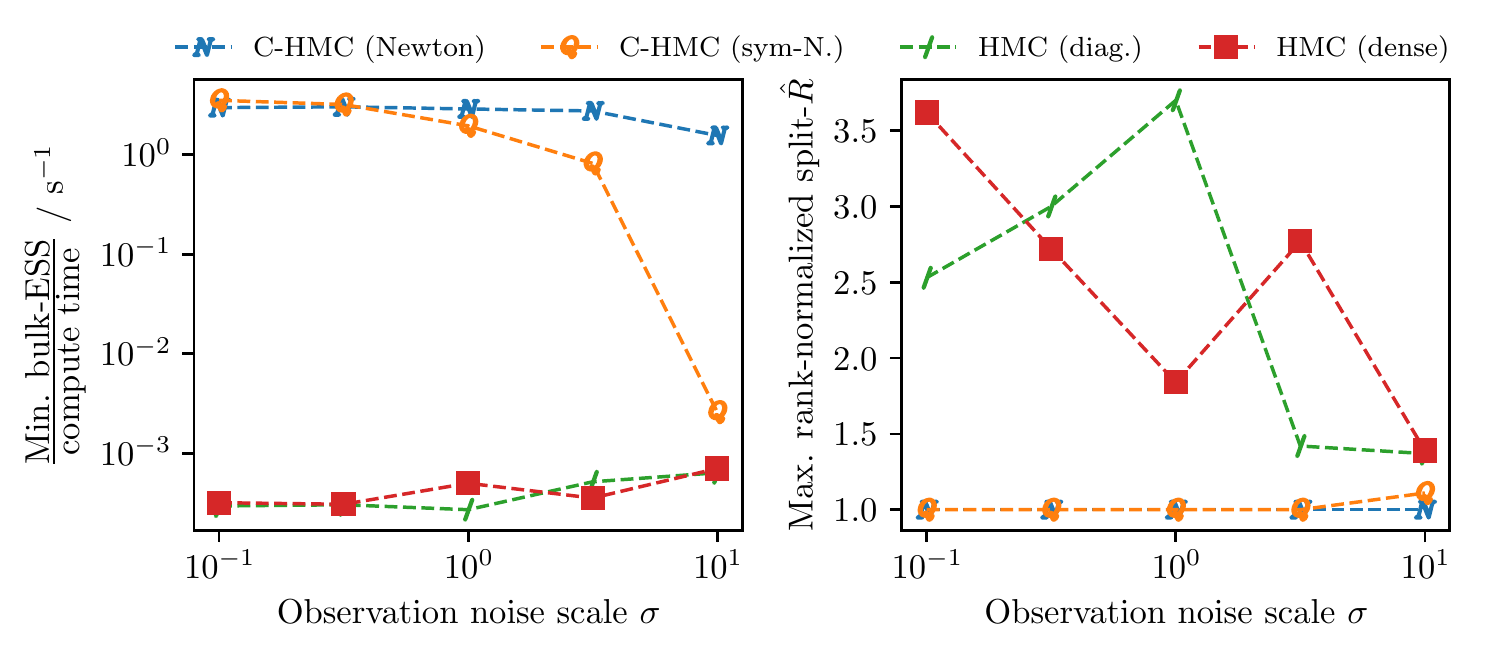}
    \caption{Hodgkin--Huxley \ac{ode} model. Left: minimum $\tfrac{\textsc{ess}}{\text{compute time}}$ for varying $\sigma$; right: maximum $\hat{R}$ for varying $\sigma$. Min./max. are across the 7 model parameters.}
    \label{fig.hh_results}
\end{figure}

Figure \ref{fig.hh_results} shows the estimated sampling efficiencies and convergence diagnostics for varying true observation noise scales $\sigma$. \ac{hmc} with both diagonal and dense metrics fails to converge for all $\sigma$ values with $\hat{R}$ values much greater than one. For small $\sigma$ values \ac{chmc} with both projection solvers show similarly good efficiency and $\hat{R}$ values indicative of convergence, however for the larger $\sigma$ values tested there is a noticeable drop off in efficiency for the \ac{chmc} chains with symmetric-Newton projection solver (and some indication of non-convergence for $\sigma = 1$). This appears to be due to the larger variation in curvatures in this more diffuse setting causing the symmetric-Newton solver to become unstable in some regions of the state space, potentially due to the constant Jacobian approximation becoming increasingly poor in such regions. The rejections due to non-convergence of the projection solver cause the step size to be adapted to a smaller value to maintain the acceptance rate, this in turn leading to a reduction in sampling efficiency. While there is also a small drop-off in efficiency when using the Newton solver, it is much less marked suggesting it is continues to remain stable for larger step sizes.

\subsection{Poisson partial differential equation model}
\label{sec.poisson}

\begin{figure}[t]
    \centering
    \includegraphics[width=0.9\textwidth]{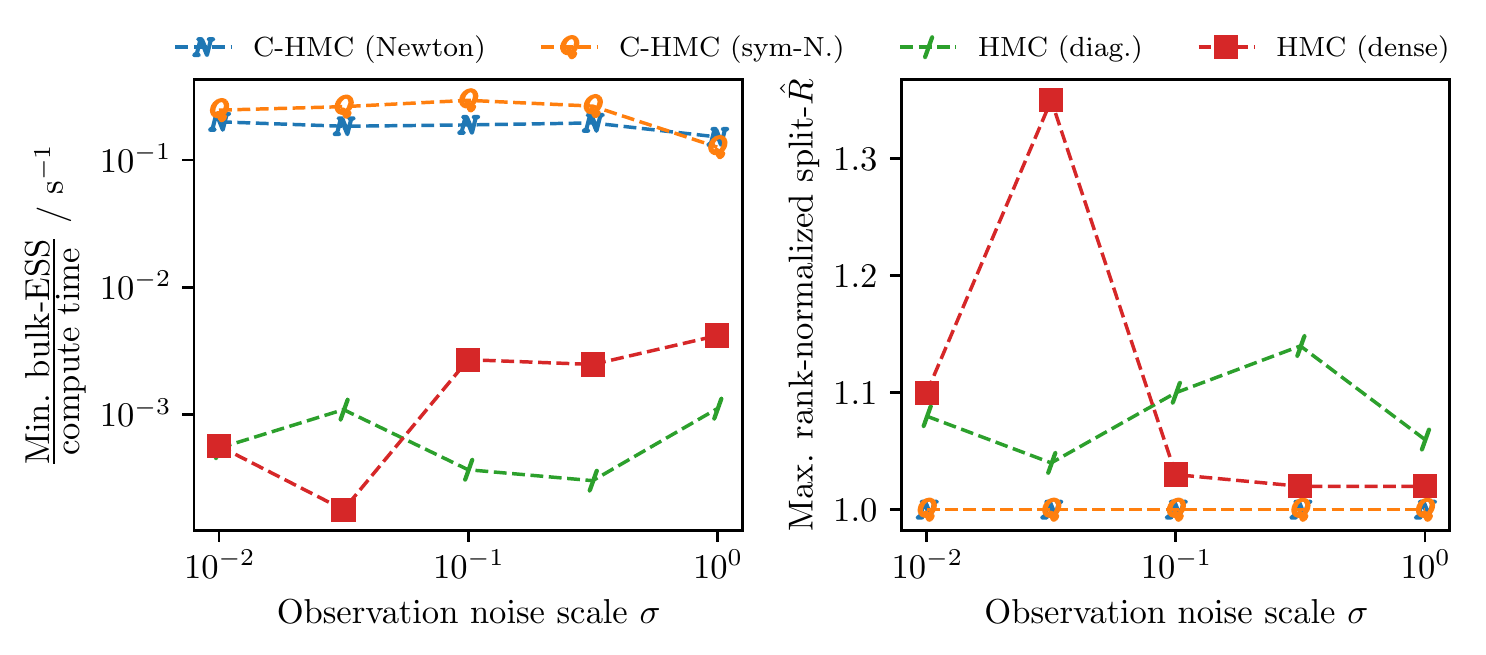}
    \caption{Poisson \ac{pde} model. Left: minimum $\tfrac{\textsc{ess}}{\text{compute time}}$ for varying $\sigma$; right: maximum $\smash{\hat{R}}$ for varying $\sigma$. Min./max. are across $\text{mean}(\vct{z})$, $\text{std}(\vct{z})$ and $\sigma$.}
    \label{fig.pde_results}
\end{figure}

To illustrate the applicability of our model to inference in \ac{pde} constrained Bayesian inverse problems, we consider a two-dimensional domain $\Omega \subset \bR^2$ and the problem of inferring a log-conductivity field $\phi:\Omega\to \bR$ from a discrete set of noisy observations of a temperature field $u: \Omega \to \bR$ and a known source term $f:\Omega \to \bR$. These quantities satisfy the following Poisson \ac{pde} on the spatial domain $\Omega$,
\begin{equation}
\label{eq.pde}
  -\nabla \cdot \big( \exp(\phi)  \nabla u \big) = f,
\end{equation}
with vanishing Dirichlet boundary conditions $u_{|\partial \Omega} = 0$.
Noisy observations of the temperature field $y_k = u(x_k) + \sigma \eta_k ~~ \forall k \in \lbrace 1, \dots, d_{\cY}\rbrace$, are available at $d_\cY=10$ distinct locations $x_k \in \Omega$ (right panel of Figure \ref{fig.pde_coefficient_and_solution_fields}  in the Supplementary Material). The library \emph{FEniCS} \citep{AlnaesBlechta2015a} was used to numerically solved the \ac{pde} \eqref{eq.pde} with a first order \ac{fem} on a triangular mesh discretization of $\Omega$ with dimension $K = 469$ (left panel of Figure \ref{fig.pde_coefficient_and_solution_fields}  in the Supplementary Material).

Specifically the discretised log-conductivity field can be expressed as $\phi(x) = \sum_{k=1}^{K} z_k e_k(x)$ for piecewise linear basis functions $e_k: \Omega \to \bR$ and coefficients $\vct{z} = (z_1, \ldots, z_{K}) \in \bR^{K}$.

We use a Gaussian random field with a Mat{\'e}rn covariance \citep{lindgren2011explicit} as a prior model for $\phi$. This can be realised as a \ac{fem} discretization of the Poisson equation $(\kappa^2 - \Delta) \phi = w$, with scaling parameter $\kappa = 0.1$, for a spatial Gaussian white noise $w: \Omega \to \bR$ with unit variance. This induces a zero-mean Gaussian prior on the coefficients $\vct{z}$ with covariance matrix $\mtx{\Gamma} \in \bR^{K\times K}$.

Our objective is then to approximate the posterior on the $d_\Theta = K + 1$ vector $\vct{\theta} = (z_1, \ldots, z_{K}, \sigma)$ of coefficients and observation noise scale, given the observations $\vct{y} = (y_1,\ldots, y_{d_\cY})$. Note that each evaluation of the posterior density involves solving the Poisson equation in \eqref{eq.pde}. The automatic differentiation functionality of the \emph{unified form language} \citep{alnaes2014unified} library in \emph{FEniCS} was used to calculate the required derivatives of the forward operator implicitly defined by solving \eqref{eq.pde}.

Figure \ref{fig.pde_results} shows estimates of how the sampling efficiency and convergence to stationarity varies with $\sigma$ for the different methods tested. As for the Hodgkin--Huxley \ac{ode} model in the previous section, the standard \ac{hmc} chains with both dense and diagonal metrics, appear to have not converged for all $\sigma$ values tested, with $\hat{R}$ values greater than one. While the non-convergence mean the \ac{ess} estimates for the \ac{hmc} chains are unreliable, the adapted integrator step sizes $\epsilon$, shown in Figure \ref{fig.simulated_data_model_step_size_results} in the Supplementary Material, suggest a similar requirement of $\epsilon$ scaling proportionally to $\sigma$ for the \ac{hmc} chains to maintain a stable acceptance rate, which would be expected to lead to a similar drop in efficiency with $\sigma$. The \ac{chmc} chains with both projection solvers do not show any signs of non-convergence across all $\sigma$ values and the \ac{ess} estimates are close to constant aside from a small drop-off in efficiency when using the symmetric Newton solver for the largest $\sigma$ values. For this model the symmetric Newton solver however in general appears to be slightly more performant than the Newton solver, with the large latent dimension $d_\Theta$ compared to observation dimension $d_{\mathcal{Y}}$ here meaning there is a greater gain in reducing the number of constraint Jacobian evaluations and associated linear algebra operations.

\subsection{Ordinary differential equation models fitted to real data}
\label{sec.ode_real_data}

So far, the data in our experiments have been simulated using the underlying generative models. This gave direct control over the observation noise scale $\sigma$ (and so the signal-to-noise ratio in the model) which allowed us to show the robustness of the performance of our proposed methodology to differing signal-to-noise ratios.

In this section, we demonstrate the competitive performance of our proposed methodology in approximating the posteriors of models fitted to real data:
\begin{enumerate}
\item \label{ode.lotka_volterra} A Lotka--Volterra predator-prey model \citep{lotka1925elements,volterra1926fluctuations} fitted to a dataset of hare and lynx populations measured by the Hudson Bay Company in the years 1900--1920 \citep{hewitt1921conservation, howard2009modeling}.
\item \label{ode.carbon} A two-pool model for carbon flow within soil fitted to data from two soil incubation experiments taken from the \emph{R} package \emph{SoilR} \citep{sierra2014modeling}. The two datasets, \emph{HN-T35} and \emph{AK-T25} use soil samples from two different sites and incubated at different temperatures.
\item \label{ode.hh_real_data} A conductance-based model for action potential generation in the giant squid axon proposed by Hodgkin and Huxley \citep{hodgkin1952quantitative} fitted to a digitization of the original voltage clamp experimental data from \cite{daly2015hodgkin}. Posteriors are separately fitted to the sodium (\emph{Na}) and potassium (\emph{K}) ionic conductance data. Compared to the conductance-based model fitted in Section \ref{sec.hh}, here the membrane voltage is assumed to be clamped resulting in a set of decoupled linear \acp{ode} for each of the channel gating variables.
\end{enumerate}
Further details of all three models are given in Section \ref{sec.real-data-model-details} in the Supplementary Material. Approximate samples from each posterior were generated using: (i) our proposed approach of a constrained \ac{hmc} algorithm targeting the lifted posterior (using a Newton projection solver), (ii) a standard \ac{hmc} algorithm with diagonal metric, (iii) a standard \ac{hmc} algorithm with dense metric. %

Table \ref{tab:real-data-results} shows the parameter dimension $d_\Theta$ and observation dimension $d_{\mathcal{Y}}$ for each of the posteriors as well as an estimate of the effective sample size per compute time in seconds, with the values shown corresponding to the median (across the three independent chain sets) of the minimum per-parameter time normalised effective sample size (across the $d_\Theta$ model parameters). The constrained \ac{hmc} algorithm outperforms standard \ac{hmc} with a diagonal metric in all cases, and standard \ac{hmc} with a dense metric in all but one case.

In general, while the cost of each constrained integrator step in the constrained \ac{hmc} algorithm is higher compared to the leapfrog integrator step cost in the standard \ac{hmc} algorithm, typically the adaptive tuning of the integrator step size results in much higher step sizes for a given target acceptance rate for the constrained \ac{hmc} algorithm, resulting in fewer integrator steps per sample and so often an overall reduced computational cost per sample.

When the posterior is close to Gaussian, however, standard \ac{hmc} with an appropriately tuned metric can also support larger integrator step sizes. In this case, due to its lower per-step computational cost standard \ac{hmc} will typically outperform constrained \ac{hmc}. This is seen in the Lotka-Volterra results where the posterior in the unconstrained space is well approximated by a Gaussian unlike the other posteriors (see pair plots in Section \ref{sec.real-data-posterior-pair-plots} in the  Supplementary Material).

\begin{table}[!t]
\centering
\begin{tabular}{cccccc}
\hline
& & & \multicolumn{3}{c}{Min. bulk-\ac{ess} per compute time / s$^{-1}$}
\\
Posterior &  $d_\Theta$ & $d_{\mathcal{Y}}$ & \ac{chmc} & \ac{hmc} (diag.) & \ac{hmc}
(dense)
\\
\hline
Lotka--Volterra & 8 & 42 & 35 & 15 & \textbf{300}
\\
Soil carbon (HN-T35) & 7 & 25 & \textbf{6.6} & 2.1 & 3.4
\\
Soil carbon (AK-T25) & 7 & 25 & \textbf{17} & 4.4 & 2.8
\\
Hodgkin-Huxley (K) & 7 & 136 & \textbf{52} & 23 & 29
\\
Hodgkin-Huxley (Na) & 11 & 142  & \textbf{3.9} & 1.6 & 3.0
\\
\hline
\end{tabular}
\caption{Problem dimensions and measured minimum bulk-\ac{ess} per compute time for real-data examples. Bold values indicate best performing method for each posterior.}
\label{tab:real-data-results}
\end{table}

\section{Conclusion}

While a large literature has been devoted to designing \ac{mcmc} samplers with good mixing properties when used in high-dimensional settings, comparatively little attention has been devoted to the problem of exploring distributions concentrating near low-dimensional manifolds or exhibiting multi-scale structures  \citep{beskos2018asymptotic,livingstone2019barker,schillings2019convergence}. To the best of our knowledge, the sampling efficiency of all currently existing general-purpose \ac{mcmc} samplers degenerates in the vanishing noise $\sigma \to 0$ asymptotic studied in this text. By reformulating the original problem into the task of exploring a distribution supported on a manifold embedded in a higher-dimensional space, our proposed methodology is able to maintain high sampling efficiency in the $\sigma \to 0$ regime. Our work opens up a number of avenues for future work in this area. While we concentrated in this work on additive observation noise, there are natural possible extensions to more general observation processes. Furthermore, an important open question is whether it is possible to define variants of the proposed approach that improve scalability in the regime where both latent and observation dimensions, $d_\Theta$ and $d_\cY$, are high, with the current scheme having a $\cO(\min(d_\Theta d_\cY^2, d_\cY d_\Theta^2))$ computational complexity for general problems. The approach used to achieve a linear scaling in the number of observations for the \ac{ssm} in Section \ref{sec.statespace}, illustrates some initial progress along this line by exploiting structure in specific model classes.

\bigskip
{\bf Acknowledgement:}
The authors acknowledge support from a National University of Singapore (NUS) Young Investigator Award Grant (R-155-000-180-133) and a Singapore Ministry of Education Academic Research Funds Tier 2 (MOE2016-T2-2-135).

\bibliographystyle{rss}

\begin{thebibliography}{62}
\expandafter\ifx\csname natexlab\endcsname\relax\def\natexlab#1{#1}\fi
\expandafter\ifx\csname url\endcsname\relax
  \def\url#1{\texttt{#1}}\fi
\expandafter\ifx\csname urlprefix\endcsname\relax\def\urlprefix{URL: }\fi

\bibitem[{Aln{\ae}s et~al.(2015)Aln{\ae}s, Blechta, Hake, Johansson, Kehlet,
  Logg, Richardson, Ring, Rognes and Wells}]{AlnaesBlechta2015a}
Aln{\ae}s, M.~S., Blechta, J., Hake, J., Johansson, A., Kehlet, B., Logg, A.,
  Richardson, C., Ring, J., Rognes, M.~E. and Wells, G.~N. (2015) The {FEniCS}
  project version 1.5.
\newblock \textit{Archive of Numerical Software}, \textbf{3}.

\bibitem[{Aln{\ae}s et~al.(2014)Aln{\ae}s, Logg, {\O}lgaard, Rognes and
  Wells}]{alnaes2014unified}
Aln{\ae}s, M.~S., Logg, A., {\O}lgaard, K.~B., Rognes, M.~E. and Wells, G.~N.
  (2014) Unified form language: A domain-specific language for weak
  formulations of partial differential equations.
\newblock \textit{ACM Transactions on Mathematical Software (TOMS)},
  \textbf{40}, 1--37.

\bibitem[{Andersen(1983)}]{andersen1983rattle}
Andersen, H.~C. (1983) {RATTLE}: A `velocity' version of the {SHAKE} algorithm
  for molecular dynamics calculations.
\newblock \textit{Journal of Computational Physics}, \textbf{52}, 24--34.

\bibitem[{Ballnus et~al.(2017)Ballnus, Hug, Hatz, G{\"o}rlitz, Hasenauer and
  Theis}]{ballnus2017comprehensive}
Ballnus, B., Hug, S., Hatz, K., G{\"o}rlitz, L., Hasenauer, J. and Theis, F.~J.
  (2017) Comprehensive benchmarking of markov chain monte carlo methods for
  dynamical systems.
\newblock \textit{BMC systems biology}, \textbf{11}, 1--18.

\bibitem[{Barth et~al.(1995)Barth, Kuczera, Leimkuhler and
  Skeel}]{barth1995algorithms}
Barth, E., Kuczera, K., Leimkuhler, B. and Skeel, R.~D. (1995) Algorithms for
  constrained molecular dynamics.
\newblock \textit{Journal of Computational Chemistry}, \textbf{16}, 1192--1209.

\bibitem[{Besag(1994)}]{besag1993comments}
Besag, J. (1994) Comments on `{R}epresentations of knowledge in complex
  systems' by {G}renander and {M}iller.
\newblock \textit{Journal of the Royal Statistical Society, Series B},
  \textbf{56}, 591--592.

\bibitem[{Beskos et~al.(2013)Beskos, Pillai, Roberts, Sanz-Serna and
  Stuart}]{beskos2013optimal}
Beskos, A., Pillai, N., Roberts, G., Sanz-Serna, J.-M. and Stuart, A. (2013)
  Optimal tuning of the hybrid {M}onte {C}arlo algorithm.
\newblock \textit{Bernoulli}, \textbf{19}, 1501--1534.

\bibitem[{Beskos et~al.(2018)Beskos, Roberts, Thiery and
  Pillai}]{beskos2018asymptotic}
Beskos, A., Roberts, G., Thiery, A. and Pillai, N. (2018) Asymptotic analysis
  of the random walk {M}etropolis algorithm on ridged densities.
\newblock \textit{The Annals of Applied Probability}, \textbf{28}, 2966--3001.

\bibitem[{Betancourt(2013)}]{betancourt2013general}
Betancourt, M. (2013) A general metric for riemannian manifold hamiltonian
  monte carlo.
\newblock In \textit{International Conference on Geometric Science of
  Information}, 327--334. Springer.

\bibitem[{{Betancourt}(2017)}]{betancourt2017conceptual}
{Betancourt}, M. (2017) A conceptual introduction to {H}amiltonian {M}onte
  {C}arlo.
\newblock \textit{arXiv e-prints}.

\bibitem[{Blanes et~al.(2014)Blanes, Casas and
  Sanz-Serna}]{blanes2014numerical}
Blanes, S., Casas, F. and Sanz-Serna, J. (2014) Numerical integrators for the
  {H}ybrid {M}onte {C}arlo method.
\newblock \textit{SIAM Journal on Scientific Computing}, \textbf{36},
  A1556--A1580.

\bibitem[{Bradbury et~al.(2018)Bradbury, Frostig, Hawkins, Johnson, Leary,
  Maclaurin and Wanderman-Milne}]{jax2018github}
Bradbury, J., Frostig, R., Hawkins, P., Johnson, M.~J., Leary, C., Maclaurin,
  D. and Wanderman-Milne, S. (2018) {JAX}: composable transformations of
  {P}ython+{N}um{P}y programs.
\newblock \urlprefix\url{http://github.com/google/jax}.

\bibitem[{Brubaker et~al.(2012)Brubaker, Salzmann and
  Urtasun}]{brubaker2012family}
Brubaker, M., Salzmann, M. and Urtasun, R. (2012) A family of {MCMC} methods on
  implicitly defined manifolds.
\newblock In \textit{Proceedings of the Fifteenth International Conference on
  Artificial Intelligence and Statistics (AISTATS)}, 161--172. Proceedings of
  Machine Learning Research.

\bibitem[{Carpenter et~al.(2017)Carpenter, Gelman, Hoffman, Lee, Goodrich,
  Betancourt, Brubaker, Guo, Li and Riddell}]{carpenter2017stan}
Carpenter, B., Gelman, A., Hoffman, M.~D., Lee, D., Goodrich, B., Betancourt,
  M., Brubaker, M., Guo, J., Li, P. and Riddell, A. (2017) {Stan}: A
  probabilistic programming language.
\newblock \textit{Journal of Statistical Software}, \textbf{76}, 1.

\bibitem[{Daly et~al.(2015)Daly, Gavaghan, Holmes and Cooper}]{daly2015hodgkin}
Daly, A.~C., Gavaghan, D.~J., Holmes, C. and Cooper, J. (2015)
  {H}odgkin--{H}uxley revisited: reparametrization and identifiability analysis
  of the classic action potential model with approximate {B}ayesian methods.
\newblock \textit{Royal Society open science}, \textbf{2}, 150499.

\bibitem[{Dasgupta et~al.(2007)Dasgupta, Self and Gupta}]{dasgupta2007non}
Dasgupta, A., Self, S.~G. and Gupta, S.~D. (2007) Non-identifiable parametric
  probability models and reparametrization.
\newblock \textit{Journal of Statistical Planning and Inference}, \textbf{137},
  3380--3393.

\bibitem[{Diaconis et~al.(2013)Diaconis, Holmes and
  Shahshahani}]{diaconis2013sampling}
Diaconis, P., Holmes, S. and Shahshahani, M. (2013) \textit{Sampling from a
  Manifold}, vol. Volume 10 of \textit{Collections}, 102--125.
\newblock Beachwood, Ohio, USA: Institute of Mathematical Statistics.

\bibitem[{Duane et~al.(1987)Duane, Kennedy, Pendleton and
  Roweth}]{duane1987hybrid}
Duane, S., Kennedy, A.~D., Pendleton, B.~J. and Roweth, D. (1987) Hybrid
  {M}onte {C}arlo.
\newblock \textit{Physics letters B}, \textbf{195}, 216--222.

\bibitem[{FitzHugh(1961)}]{fitzhugh1961impulses}
FitzHugh, R. (1961) Impulses and physiological states in theoretical models of
  nerve membrane.
\newblock \textit{Biophysical Journal}, \textbf{1}, 445--466.

\bibitem[{Girolami and Calderhead(2011)}]{girolami2011riemann}
Girolami, M. and Calderhead, B. (2011) Riemann manifold {L}angevin and
  {H}amiltonian {M}onte {C}arlo methods.
\newblock \textit{Journal of the Royal Statistical Society: Series B
  (Statistical Methodology)}, \textbf{73}, 123--214.

\bibitem[{Graham(2019)}]{graham2019mici}
Graham, M.~M. (2019) Mici: manifold {MCMC} methods in {P}ython.
\newblock \urlprefix\url{https://git.io/mici.py}.

\bibitem[{Graham and Storkey(2017)}]{graham2017asymptotically}
Graham, M.~M. and Storkey, A.~J. (2017) Asymptotically exact inference in
  differentiable generative models.
\newblock \textit{Electronic Journal of Statistics}, \textbf{11}, 5105--5164.

\bibitem[{Gupta et~al.(1990)Gupta, Irb{\"a}ck, Karsch and
  Petersson}]{gupta1990acceptance}
Gupta, S., Irb{\"a}ck, A., Karsch, F. and Petersson, B. (1990) The acceptance
  probability in the hybrid {M}onte {C}arlo method.
\newblock \textit{Physics Letters, B}, \textbf{242}.

\bibitem[{Harris et~al.(2020)Harris, Millman, van~der Walt, Gommers, Virtanen,
  Cournapeau, Wieser, Taylor, Berg, Smith, Kern, Picus, Hoyer, van Kerkwijk,
  Brett, Haldane, del R{\'{i}}o, Wiebe, Peterson, G{\'{e}}rard-Marchant,
  Sheppard, Reddy, Weckesser, Abbasi, Gohlke and Oliphant}]{harris2020array}
Harris, C.~R., Millman, K.~J., van~der Walt, S.~J., Gommers, R., Virtanen, P.,
  Cournapeau, D., Wieser, E., Taylor, J., Berg, S., Smith, N.~J., Kern, R.,
  Picus, M., Hoyer, S., van Kerkwijk, M.~H., Brett, M., Haldane, A., del
  R{\'{i}}o, J.~F., Wiebe, M., Peterson, P., G{\'{e}}rard-Marchant, P.,
  Sheppard, K., Reddy, T., Weckesser, W., Abbasi, H., Gohlke, C. and Oliphant,
  T.~E. (2020) Array programming with {NumPy}.
\newblock \textit{Nature}, \textbf{585}, 357--362.
\newblock \urlprefix\url{https://doi.org/10.1038/s41586-020-2649-2}.

\bibitem[{Hartmann and Sch{\"u}tte(2005)}]{hartmann2005constrained}
Hartmann, C. and Sch{\"u}tte, C. (2005) A constrained hybrid {M}onte-{C}arlo
  algorithm and the problem of calculating the free energy in several
  variables.
\newblock \textit{ZAMM-Journal of Applied Mathematics and Mechanics/Zeitschrift
  f{\"u}r Angewandte Mathematik und Mechanik: Applied Mathematics and
  Mechanics}, \textbf{85}, 700--710.

\bibitem[{Hewitt(1921)}]{hewitt1921conservation}
Hewitt, C. (1921) \textit{The Conservation of thc Wild Life of {C}anada}.
\newblock Charles Scribner's Sons.

\bibitem[{Hines et~al.(2014)Hines, Middendorf and
  Aldrich}]{hines2014determination}
Hines, K.~E., Middendorf, T.~R. and Aldrich, R.~W. (2014) Determination of
  parameter identifiability in nonlinear biophysical models: A bayesian
  approach.
\newblock \textit{Journal of General Physiology}, \textbf{143}, 401--416.

\bibitem[{Hodgkin and Huxley(1952)}]{hodgkin1952quantitative}
Hodgkin, A.~L. and Huxley, A.~F. (1952) A quantitative description of membrane
  current and its application to conduction and excitation in nerve.
\newblock \textit{The Journal of physiology}, \textbf{117}, 500.

\bibitem[{Hoffman and Gelman(2014)}]{hoffman2014no}
Hoffman, M.~D. and Gelman, A. (2014) The {No-U-Turn} sampler: adaptively
  setting path lengths in {Hamiltonian Monte Carlo}.
\newblock \textit{Journal of Machine Learning Research}, \textbf{15},
  1593--1623.

\bibitem[{Howard(2009)}]{howard2009modeling}
Howard, P. (2009) Modeling basics, lecture notes for {M}ath 442.
\newblock Texas A\&M University.
\newblock \urlprefix\url{http://www.math.tamu.edu/~phoward/m442/modbasics.pdf}.

\bibitem[{Hunter(2007)}]{hunter2007matplotlib}
Hunter, J.~D. (2007) Matplotlib: A 2d graphics environment.
\newblock \textit{Computing in Science \& Engineering}, \textbf{9}, 90--95.

\bibitem[{Knapik et~al.(2011)Knapik, van~der Vaart and van
  Zanten}]{knapik2011bayesian}
Knapik, B.~T., van~der Vaart, A.~W. and van Zanten, J.~H. (2011) Bayesian
  inverse problems with {G}aussian priors.
\newblock \textit{The Annals of Statistics}, \textbf{39}, 2626--2657.

\bibitem[{Kumar et~al.(2019)Kumar, Carroll, Hartikainen and Martin}]{arviz2019}
Kumar, R., Carroll, C., Hartikainen, A. and Martin, O.~A. (2019) {ArviZ} a
  unified library for exploratory analysis of {Bayesian} models in {Python}.
\newblock \textit{The Journal of Open Source Software}.

\bibitem[{Leimkuhler and Matthews(2016)}]{leimkuhler2016efficient}
Leimkuhler, B. and Matthews, C. (2016) Efficient molecular dynamics using
  geodesic integration and solvent--solute splitting.
\newblock \textit{Proceedings of the Royal Society A: Mathematical, Physical
  and Engineering Sciences}, \textbf{472}, 20160138.

\bibitem[{Leimkuhler and Reich(2004)}]{leimkuhler2004simulating}
Leimkuhler, B. and Reich, S. (2004) \textit{Simulating {H}amiltonian dynamics},
  vol.~14.
\newblock Cambridge University Press.

\bibitem[{Leimkuhler and Skeel(1994)}]{leimkuhler1994symplectic}
Leimkuhler, B.~J. and Skeel, R.~D. (1994) Symplectic numerical integrators in
  constrained {H}amiltonian systems.
\newblock \textit{Journal of Computational Physics}, \textbf{112}, 117--125.

\bibitem[{Leli{\`e}vre et~al.(2019)Leli{\`e}vre, Rousset and
  Stoltz}]{lelievre2019hybrid}
Leli{\`e}vre, T., Rousset, M. and Stoltz, G. (2019) {H}ybrid {M}onte {C}arlo
  methods for sampling probability measures on submanifolds.
\newblock \textit{Numerische Mathematik}.

\bibitem[{Lindgren et~al.(2011)Lindgren, Rue and
  Lindstr{\"o}m}]{lindgren2011explicit}
Lindgren, F., Rue, H. and Lindstr{\"o}m, J. (2011) An explicit link between
  {G}aussian fields and {G}aussian {M}arkov random fields: the stochastic
  partial differential equation approach.
\newblock \textit{Journal of the Royal Statistical Society: Series B
  (Statistical Methodology)}, \textbf{73}, 423--498.

\bibitem[{Livingstone(2015)}]{livingstone2015geometric}
Livingstone, S. (2015) Geometric ergodicity of the random walk {M}etropolis
  with position-dependent proposal covariance.
\newblock \textit{arXiv preprints}.

\bibitem[{Livingstone and Zanella(2019)}]{livingstone2019barker}
Livingstone, S. and Zanella, G. (2019) The {B}arker proposal: combining
  robustness and efficiency in gradient-based {MCMC}.
\newblock \textit{arXiv preprint arXiv:1908.11812}.

\bibitem[{Lotka(1925)}]{lotka1925elements}
Lotka, A. (1925) \textit{Elements of physical biology}.
\newblock Williams and Wilkins.

\bibitem[{Metropolis et~al.(1953)Metropolis, Rosenbluth, Rosenbluth, Teller and
  Teller}]{metropolis1953equation}
Metropolis, N., Rosenbluth, A.~W., Rosenbluth, M.~N., Teller, A.~H. and Teller,
  E. (1953) Equation of state calculations by fast computing machines.
\newblock \textit{The Journal of Chemical Physics}, \textbf{21}, 1087--1092.

\bibitem[{Nagumo et~al.(1962)Nagumo, Arimoto and Yoshizawa}]{nagumo1962active}
Nagumo, J., Arimoto, S. and Yoshizawa, S. (1962) An active pulse transmission
  line simulating nerve axon.
\newblock \textit{Proceedings of the IRE}, \textbf{50}, 2061--2070.

\bibitem[{Neal(2011)}]{neal2012mcmc}
Neal, R.~M. (2011) \textit{MCMC Using Hamiltonian Dynamics}, chap. Chapter 5.
\newblock CRC Press.
\newblock
  \urlprefix\url{https://www.routledgehandbooks.com/doi/10.1201/b10905-7}.

\bibitem[{Petra et~al.(2014)Petra, Martin, Stadler and
  Ghattas}]{petra2014computational}
Petra, N., Martin, J., Stadler, G. and Ghattas, O. (2014) A computational
  framework for infinite-dimensional {B}ayesian inverse problems, {P}art {II}:
  Stochastic {N}ewton {MCMC} with application to ice sheet flow inverse
  problems.
\newblock \textit{SIAM Journal on Scientific Computing}, \textbf{36},
  A1525--A1555.

\bibitem[{Pospischil et~al.(2008)Pospischil, Toledo-Rodriguez, Monier,
  Piwkowska, Bal, Fr{\'e}gnac, Markram and Destexhe}]{pospischil2008minimal}
Pospischil, M., Toledo-Rodriguez, M., Monier, C., Piwkowska, Z., Bal, T.,
  Fr{\'e}gnac, Y., Markram, H. and Destexhe, A. (2008) Minimal hodgkin--huxley
  type models for different classes of cortical and thalamic neurons.
\newblock \textit{Biological cybernetics}, \textbf{99}, 427--441.

\bibitem[{Raue et~al.(2013)Raue, Kreutz, Theis and Timmer}]{raue2013joining}
Raue, A., Kreutz, C., Theis, F.~J. and Timmer, J. (2013) Joining forces of
  {B}ayesian and frequentist methodology: a study for inference in the presence
  of non-identifiability.
\newblock \textit{Philosophical Transactions of the Royal Society A:
  Mathematical, Physical and Engineering Sciences}, \textbf{371}, 20110544.
\newblock
  \urlprefix\url{https://royalsocietypublishing.org/doi/abs/10.1098/rsta.2011.0544}.

\bibitem[{Reich(1996)}]{reich1996symplectic}
Reich, S. (1996) Symplectic integration of constrained {H}amiltonian systems by
  composition methods.
\newblock \textit{{SIAM} Journal on Numerical Analysis}, \textbf{33}, 475--491.

\bibitem[{Roberts et~al.(1997)Roberts, Gelman and Gilks}]{roberts1997weak}
Roberts, G.~O., Gelman, A. and Gilks, W.~R. (1997) Weak convergence and optimal
  scaling of random walk {M}etropolis algorithms.
\newblock \textit{The Annals of Applied Probability}, \textbf{7}, 110--120.

\bibitem[{Roberts and Rosenthal(2001)}]{roberts2001optimal}
Roberts, G.~O. and Rosenthal, J.~S. (2001) Optimal scaling for various
  {M}etropolis--{H}astings algorithms.
\newblock \textit{Statistical science}, \textbf{16}, 351--367.

\bibitem[{Roberts and Stramer(2002)}]{roberts2002langevin}
Roberts, G.~O. and Stramer, O. (2002) Langevin diffusions and
  {M}etropolis--{H}astings algorithms.
\newblock \textit{Methodology and computing in applied probability},
  \textbf{4}, 337--357.

\bibitem[{Rothenberg(1971)}]{rothenberg1971identification}
Rothenberg, T.~J. (1971) Identification in parametric models.
\newblock \textit{Econometrica: Journal of the Econometric Society}, 577--591.

\bibitem[{Rousset et~al.(2010)Rousset, Stoltz and
  Leli{\`e}vre}]{rousset2010free}
Rousset, M., Stoltz, G. and Leli{\`e}vre, T. (2010) \textit{Free Energy
  Computations: A Mathematical Perspective}.
\newblock Imperial College Press.

\bibitem[{Schillings et~al.(2019)Schillings, Sprungk and
  Wacker}]{schillings2019convergence}
Schillings, C., Sprungk, B. and Wacker, P. (2019) On the convergence of the
  {L}aplace approximation and noise level robustness of {L}aplace-based {M}onte
  {C}arlo methods for {B}ayesian inverse problems.
\newblock \textit{arXiv preprints}.

\bibitem[{Sierra et~al.(2014)Sierra, M{\"u}ller and
  Trumbore}]{sierra2014modeling}
Sierra, C., M{\"u}ller, M. and Trumbore, S.~E. (2014) Modeling radiocarbon
  dynamics in soils: {SoilR} version 1.1.
\newblock \textit{Geoscientific Model Development}, \textbf{7}, 1919--1931.

\bibitem[{Stuart(2010)}]{stuart2010inverse}
Stuart, A.~M. (2010) Inverse problems: a {B}ayesian perspective.
\newblock \textit{Acta numerica}, \textbf{19}, 451--559.

\bibitem[{Vehtari et~al.(2019)Vehtari, Gelman, Simpson, Carpenter and
  B{\"u}rkner}]{vehtari2019rank}
Vehtari, A., Gelman, A., Simpson, D., Carpenter, B. and B{\"u}rkner, P.-C.
  (2019) Rank-normalization, folding, and localization: An improved r for
  assessing convergence of {MCMC}.
\newblock \textit{arXiv e-prints}.

\bibitem[{{Virtanen} et~al.(2020){Virtanen}, {Gommers}, {Oliphant},
  {Haberland}, {Reddy}, {Cournapeau}, {Burovski}, {Peterson}, {Weckesser},
  {Bright}, {van der Walt}, {Brett}, {Wilson}, {Jarrod Millman}, {Mayorov},
  {Nelson}, {Jones}, {Kern}, {Larson}, {Carey}, {Polat}, {Feng}, {Moore}, {Vand
  erPlas}, {Laxalde}, {Perktold}, {Cimrman}, {Henriksen}, {Quintero}, {Harris},
  {Archibald}, {Ribeiro}, {Pedregosa}, {van Mulbregt} and {SciPy 1.0
  Contributors}}]{scipy2020scipy}
{Virtanen}, P., {Gommers}, R., {Oliphant}, T.~E., {Haberland}, M., {Reddy}, T.,
  {Cournapeau}, D., {Burovski}, E., {Peterson}, P., {Weckesser}, W., {Bright},
  J., {van der Walt}, S.~J., {Brett}, M., {Wilson}, J., {Jarrod Millman}, K.,
  {Mayorov}, N., {Nelson}, A. R.~J., {Jones}, E., {Kern}, R., {Larson}, E.,
  {Carey}, C., {Polat}, {\.I}., {Feng}, Y., {Moore}, E.~W., {Vand erPlas}, J.,
  {Laxalde}, D., {Perktold}, J., {Cimrman}, R., {Henriksen}, I., {Quintero},
  E.~A., {Harris}, C.~R., {Archibald}, A.~M., {Ribeiro}, A.~H., {Pedregosa},
  F., {van Mulbregt}, P. and {SciPy 1.0 Contributors} (2020) {SciPy} 1.0:
  Fundamental algorithms for scientific computing in {P}ython.
\newblock \textit{Nature Methods}, \textbf{17}, 261--272.

\bibitem[{Volterra(1926)}]{volterra1926fluctuations}
Volterra, V. (1926) Fluctuations in the abundance of a species considered
  mathematically.
\newblock \textit{Nature}, \textbf{118}, 558--560.

\bibitem[{Weckstrom(2010)}]{weckstrom2010intracellular}
Weckstrom, M. (2010) {I}ntracellular recording.
\newblock \textit{Scholarpedia}, \textbf{5}, 2224.
\newblock Revision \#121415.

\bibitem[{Xifara et~al.(2014)Xifara, Sherlock, Livingstone, Byrne and
  Girolami}]{xifara2014langevin}
Xifara, T., Sherlock, C., Livingstone, S., Byrne, S. and Girolami, M. (2014)
  Langevin diffusions and the {M}etropolis-adjusted {L}angevin algorithm.
\newblock \textit{Statistics \& Probability Letters}, \textbf{91}, 14--19.

\bibitem[{Zappa et~al.(2018)Zappa, Holmes-Cerfon and Goodman}]{zappa2018monte}
Zappa, E., Holmes-Cerfon, M. and Goodman, J. (2018) {M}onte {C}arlo on
  manifolds: sampling densities and integrating functions.
\newblock \textit{Communications on Pure and Applied Mathematics}, \textbf{71},
  2609--2647.

\end{thebibliography}

\appendix
\section{Proof of  Theorem \ref{thm.equivalent_dynamics}}
\label{sec.thm.equivalent_dynamics}
From the momentum condition in the constrained Hamiltonian dynamics and differentiating with respect to time we have that
\begin{equation*}
\vct{v}_{\vct{\eta}} = -\tfrac{1}{\sigma} \mtx{F} \vct{v}_{\vct{\theta}}
  \implies
  \dot{\vct{v}}_{\vct{\eta}} = -\tfrac{1}{\sigma} \mtx{F} \dot{\vct{v}}_{\vct{\theta}}.
\end{equation*}
Substituting $\dot{\vct{v}}_{\vct{\theta}} = -\vct{\theta} + \mtx{F}\tr\vct{\lambda}$ and $\dot{\vct{v}}_{\vct{\eta}} = -\vct{\eta} + \sigma \vct{\lambda}$ and solving for $\vct{\lambda}$ gives
\begin{equation*}
\vct{\lambda} = -(\sigma^2 \idmtx + \mtx{F}\mtx{F}\tr)^{-1}(\mtx{F}\vct{\theta} + \sigma \vct{\eta}) = -(\sigma^2 \idmtx + \mtx{F}\mtx{F}\tr)^{-1} (\vct{y} - \vct{f}),
\end{equation*}
with the last equality coming from the constraint equation $\mtx{F}\vct{\theta} + \vct{f} +\sigma\vct{\eta} - \vct{y} = \vct{0}$.

This gives that
\begin{align}
  \dot{\vct{v}}_{\vct{\theta}}
  &= -\vct{\theta} + \mtx{F}(\sigma^2 \idmtx + \mtx{F} \mtx{F}\tr)^{-1} (\vct{y} - \vct{f}) \\
  &= -\vct{\theta} + \tfrac{1}{\sigma^2}(\idmtx + \tfrac{1}{\sigma^2}\mtx{F}\tr \mtx{F})^{-1}\mtx{F}\tr (\vct{y} - \vct{f}) \\
  &= -(\vct{\theta} - \vct{\mu})
\end{align}
with the equality between the first and second lines arising from the push-through identity, and between the second and third lines from the definition of $\vct{\mu}$ above.

Differentiating $\dot{\vct{\theta}} = \vct{v}_{\vct{\theta}}$ with respect to time and substituting the above gives that
\begin{equation*}
\ddot{\vct{\theta}} = -(\vct{\theta} - \vct{\mu})
\end{equation*}
which is equivalent to the second-order dynamics of $\vct{\theta}$ in the original Hamiltonian system for the choice of metric $\mtx{M} = \mtx{\Sigma}^{-1}$.
\section{Proof of  Theorem \ref{thm.equivalent_dyna.general}}
\label{sec.thm.equivalent_dyna.general}

Our proof uses the following standard result from classical mechanics
\begin{lemma}
\label{lem:diffeo_preserves}
If $\gamma : \Theta \to \mathcal{M}$ is a diffeomorphism, then the \it{canonical transformation} $\Gamma : \mathcal{T}^*\Theta \to \mathcal{T}^*\mathcal{M};\; (\vct{\theta}, \vct{p}) \mapsto (\vct{q}, \vct{v})$ implicitly defined by
\begin{equation*}
\vct{q} = \gamma(\vct{\theta}), \quad \jacob\gamma(\vct{\theta})\tr \vct{v} = \vct{p}
\end{equation*}
preserves the form of Hamiltonian's equations that is
\begin{equation*}
(\dot{\vct{q}}, \dot{\vct{v}}) = (\grad_2, -\grad_1)\bar{H}(\vct{q}, \vct{v}) \implies (\dot{\vct{\theta}}, \dot{\vct{p}}) =(\grad_2, -\grad_1)H(\vct{\theta}, \vct{p})
\end{equation*}
for a Hamiltonian function $\bar{H} : \mathcal{T}^*\mathcal{M} \to \mathbb{R}$ and corresponding Hamiltonian function $H: \mathcal{T}^*\Theta \to \mathbb{R}$ defined by $H(\vct{\theta}, \vct{p}) = \bar{H}\circ\Gamma(\vct{\theta}, \vct{p})$.
\end{lemma}
If we define
\begin{equation*}
  \gamma(\vct{\theta}) = (\vct{\theta}, (\vct{y} - F(\vct{\theta})) / \sigma(\vct{\theta}))
\end{equation*}
then $\mathcal{M}$ is the image of $\Theta$ under $\gamma$, i.e. $\gamma(\Theta) = \mathcal{M}$. If we assume $F$ and $\sigma$ are smooth then the corestriction of $\gamma$ onto $\mathcal{M}$ is a diffeomorphism from $\Theta$ to $\mathcal{M}$ and we have that $\mathcal{M}$ is diffeomorphic to $\Theta$. Differentiating $\gamma$ we have that
\begin{equation*}
 \jacob \gamma(\vct{\theta}) = \Big(\idmtx, -\tfrac{1}{\sigma(\vct{\theta})}\big(\jacob F(\vct{\theta}) + \tfrac{1}{\sigma(\vct{\theta})}(\vct{y} - F(\vct{\theta}))\jacob\sigma(\vct{\theta}) \big)\Big)
\end{equation*}
As $(\vct{v}_{\vct{\theta}}, \vct{v}_{\vct{\eta}}) \in \mathcal{T}^*_{\mathcal{M}}(\vct{\theta}, \vct{\eta}) \implies \jacob c(\vct{\theta}, \vct{\eta}) (\vct{v}_{\vct{\theta}}, \vct{v}_{\vct{\eta}}) = 0$ and $\vct{\eta} = \tfrac{1}{\sigma(\vct{\theta})}(\vct{y} - F(\vct{\theta}))$ we have
\begin{equation*}
\vct{v}_{\vct{\eta}} = -\tfrac{1}{\sigma(\vct{\theta})}(\jacob F(\vct{\theta})  + \tfrac{1}{\sigma(\vct{\theta})}(\vct{y}-F(\vct{\theta})) \jacob\sigma(\vct{\theta}))\vct{v}_\theta
\end{equation*}
and so $(\vct{v}_{\vct{\theta}}, \vct{v}_{\vct{\eta}}) = \jacob\gamma(\vct{\theta})\vct{v}_{\vct{\theta}}$. For a canonical transformation $\Gamma : \mathcal{T}^*\Theta \to \mathcal{T}^*\mathcal{M}$; $(\vct{\theta}, \vct{p}) \mapsto (\vct{q}, \vct{v})$ implicitly defined by
\begin{equation*}
\vct{q} =(\vct{\theta}, \vct{\eta}) = \gamma(\vct{\theta}), \quad \jacob\gamma(\vct{\theta})\tr \vct{v} = \jacob\gamma(\vct{\theta})\tr (\vct{v}_{\vct{\theta}}, \vct{v}_{\vct{\eta}}) = \vct{p}
\end{equation*}
we therefore have that
\begin{equation*}
  \vct{p} = \jacob\gamma(\vct{\theta})\tr (\vct{v}_{\vct{\theta}}, \vct{v}_{\vct{\eta}}) = \jacob\gamma(\vct{\theta})\tr \jacob\gamma(\vct{\theta}) \vct{v}_{\vct{\theta}} \implies \vct{v}_{\vct{\theta}} = (\jacob\gamma(\vct{\theta})\tr \jacob\gamma(\vct{\theta}))^{-1} \vct{p}
\end{equation*}
and so the following explicit definition for $\Gamma$
\begin{equation*}
  ((\vct{\theta}, \vct{\eta}), (\vct{v}_{\vct{\theta}}, \vct{v}_{\vct{\eta}})) = \Gamma(\vct{\theta}, p) = (\gamma(\vct{\theta}), \jacob\gamma(\vct{\theta})(\jacob\gamma(\vct{\theta})\tr\jacob\gamma(\vct{\theta}))^{-1}\vct{p}).
\end{equation*}
By Lemma \ref{lem:diffeo_preserves} the constrained Hamiltonian dynamics described by
\[
  (\dot{\vct{q}}, \dot{\vct{v}}) = (\grad_2, -\grad_1)\bar{H}(\vct{q}, \vct{v})
\]
on $\mathcal{T}^*\mathcal{M}$ are equivalent to the Hamiltonian dynamics $(\dot{\vct{\theta}}, \dot{\vct{p}}) = (\grad_2, -\grad_1)H(\vct{\theta}, \vct{p})$ on $\mathcal{T}^*\Theta$ with Hamiltonian function $H$ defined by
\begin{align}
  H(\vct{\theta}, \vct{p}) &= \bar{H}(\Gamma(\vct{\theta}, \vct{p}))\\
  &=
  \tfrac{1}{2}\| \vct{\theta} \|^2 + \tfrac{1}{2\sigma(\vct{\theta})^2}(\vct{y}-F(\vct{\theta}))\tr(\vct{y}-F(\vct{\theta})) + \tfrac{1}{2}\log G(\gamma(\vct{\theta}))\\
  &\phantom{=}\,+ \tfrac{1}{2} \vct{p}\tr(\jacob\gamma(\vct{\theta})\tr\jacob\gamma(\vct{\theta}))^{-1} \jacob\gamma(\vct{\theta})\tr\jacob\gamma(\vct{\theta})(\jacob\gamma(\vct{\theta})\tr\jacob\gamma(\vct{\theta}))^{-1}\vct{p}
\end{align}
By the matrix determinant lemma we have that
\begin{align}
  \left|G(\vct{\theta}, \vct{\eta})\right| &= \left|(\jacob F(\vct{\theta}) + \vct{\eta} \jacob\sigma(\vct{\theta}))(\jacob F(\vct{\theta})+ \vct{\eta} \jacob\sigma(\vct{\theta}))\tr + \sigma(\vct{\theta})^2\idmtx\right|\\
  &= \left| \tfrac{1}{\sigma(\vct{\theta})^2}(\jacob F(\vct{\theta})+ \vct{\eta} \jacob\sigma(\vct{\theta}))\tr(\jacob F(\vct{\theta}) + \vct{\eta} \jacob\sigma(\vct{\theta})) + \idmtx \right| | \sigma(\vct{\theta})^2\idmtx|
\end{align}
therefore we have that
\begin{align*}
  H(\vct{\theta}, \vct{p}) &=
  \tfrac{1}{2\sigma(\vct{\theta})^2}(\vct{y}-F(\vct{\theta}))\tr(\vct{y}-F(\vct{\theta})) + d_{\mathcal{Y}}\log\sigma(\vct{\theta}) + \tfrac{1}{2}\| \vct{\theta} \|^2 \\
  &\phantom{=}\qquad+ \tfrac{1}{2}\vct{p}\mtx{M}(\vct{\theta})^{-1}\vct{p} + \tfrac{1}{2}\log|\mtx{M}(\vct{\theta})|
\end{align*}
with $\mtx{M}$ defined for $\vct{\eta}(\vct{\theta}) = \tfrac{1}{\sigma(\vct{\theta})}(\vct{y} - F(\vct{\theta}))$ by
\begin{align*}
  M(\vct{\theta})
  &=
  \jacob\gamma(\vct{\theta})\tr\jacob\gamma(\vct{\theta})
  \\
  &=
  \idmtx + \tfrac{1}{\sigma(\vct{\theta})^2} (\jacob F(\vct{\theta}) + \vct{\eta(\vct{\theta})}\jacob\sigma(\vct{\theta}))\tr(\jacob F(\vct{\theta}) +\vct{\eta}(\vct{\theta})\jacob\sigma(\vct{\theta})).
\end{align*}

\makeatletter\@input{a_supp.tex}\makeatother

\end{document}


\section{Position-dependent MCMC methods in the vanishing noise regime}
\label{app.position-dependent-methods-vanishing-noise}

\begin{figure}[t]
  \centering
  \includegraphics[width=\linewidth]{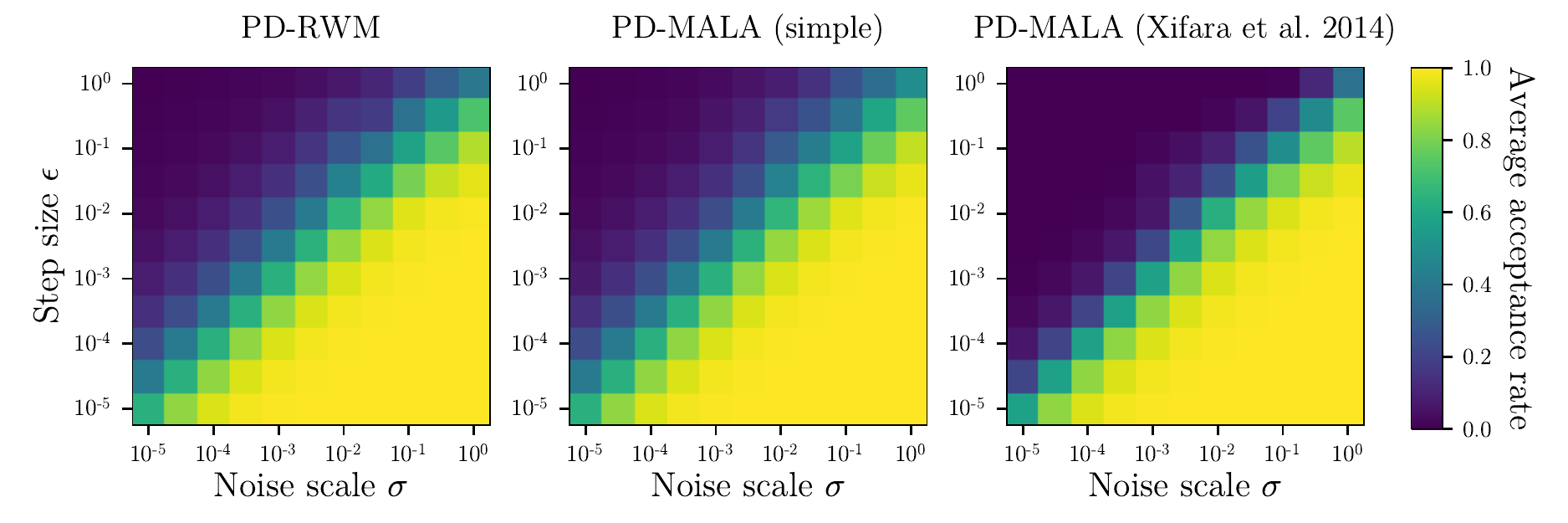}
  \caption{Average acceptance rate for chains simulated using different position-dependent \acs{mcmc} methods (using a Fisher information based metric) targetting the posterior $\pi^\sigma$ of toy example for varying noise scales $\sigma$ and (integrator) step sizes $\epsilon$. The values displayed are means of the acceptance probabilities over four chains of 1000 samples initialised at equispaced points around the limiting manifold $\cS$.
  }
  \label{fig.loop_position_dependent_experimental_results}
\end{figure}

In Section \ref{sec.vanishing_noise} in the main text, we compared the performance of various existing \acs{mcmc} methods in terms of how the average acceptance rate varies as a function of the observation noise scale $\sigma$ and integrator step size $\epsilon$ in our toy two-dimensional model. There we compared the performance of both `simple' schemes such as \acs{rwm}, \acs{mala} and \acs{hmc} with a fixed (isotropic) proposal covariance or metric, and the more complex \acs{rmhmc}, which accounts for the locally varying geometry of the posterior by simulating Hamiltonian dynamics on a Riemannian manifold with position dependent metric, with this requiring using an integrator with implicit steps involving solving a non-linear system of equations. For all these methods we found that the step size $\epsilon$ needed to be proportional to $\sigma$ for the acceptance rate to not vanish as $\sigma \to 0$.

In the case of \acs{rmhmc} we suggested the requirement of reducing $\epsilon$ with $\sigma$ to maintain a non-zero acceptance rate was due to the need to maintain the stability of the fixed-point solver used in the implicit integrator. As there are \acs{mcmc} schemes which use position-dependent preconditioners to account for locally varying geometry in the target distribution but do not require solving a non-linear system of equations on each step as in \acs{rmhmc}, an obvious question is whether such simpler schemes suffer from the same degradation in performance as $\sigma \to 0$. If not, it would seem that such methods would provide a less implementationally complex approach to get comparable behaviour to the \acs{chmc} approach described in the paper.

As an initial concrete example, \citet{girolami2011riemann} suggests a simple position-dependent \acs{mala} variant, which ignores terms involving derivatives of the metric function $\mtx{M}$ (accounting for manifolds with varying curvature), with Gaussian proposals generated according to
\begin{equation}\label{eq.simple-preconditioned-mala}
  \vct{\theta}' \,|\, \vct{\theta} \sim
  \normal\big(\vct{\theta} - \tfrac{\epsilon}{2}\mtx{M}(\vct{\theta})^{-1} \Phi(\vct{\theta}), \epsilon^2 \mtx{M}^{-1}(\vct{\theta}) \big),
\end{equation}
and accepted or rejected in a \emph{Rosenbluth--Teller-Metropolis--Hastings} acceptance step \citep{metropolis1953equation,hastings1970monte}.

In \citet{xifara2014langevin}, the authors note that the diffusion that \eqref{eq.simple-preconditioned-mala} corresponds to a Euler--Maruyama discretization of does not have the target distribution $\pi(\dr\vct{\theta}) \propto \exp(-\Phi(\vct{\theta})) \dr\vct{\theta}$ as its stationary distribution. They show that by adding a correction to the drift term, a diffusion with the correct invariant distribution can be derived, at the cost of requiring evaluation of derivatives of the metric function $\mtx{M}$. Again using an Euler--Maruyama discretisation, the resulting scheme generates Gaussian proposals from
\begin{equation}\label{eq.xifara-preconditioned-mala}
  \vct{\theta}' \,|\, \vct{\theta} \sim
  \normal\Big(\vct{\theta} - \tfrac{\epsilon}{2}\mtx{M}(\vct{\theta})^{-1} \big(\Phi(\vct{\theta}) + \Xi(\mtx{M})(\vct{\theta}) \big), \epsilon^2 \mtx{M}^{-1}(\vct{\theta}) \Big),
\end{equation}
with $\Xi_i(\mtx{M})(\vct{\theta}) = \sum_{j,k=1}^{d_\Theta}\partial_k\mtx{M}_{ij}(\vct{\theta})\mtx{M}^{-1}_{jk}(\vct{\theta}), ~ i \in 1{:}d_\Theta$.

An alternative approach analysed by \citet{livingstone2015geometric} is to omit the drift term altogether, and instead use $\mtx{M}$ as a position-dependent proposal covariance in a \acs{rwm} scheme, that is generate Gaussian proposals from
\begin{equation}\label{eq.preconditioned-rwm}
  \vct{\theta}' \,|\, \vct{\theta} \sim
  \normal\big(\vct{\theta}, \epsilon^2 \mtx{M}^{-1}(\vct{\theta}) \big).
\end{equation}

In all three cases, as for the metric in \acs{rmhmc}, the pre-conditioner or proposal covariance function $\mtx{M}$, can be chosen as any positive-definite matrix valued function. One possible choice is the expected Fisher information plus the negative Hessian of the log prior density, which for a model with fixed observation noise scale $\sigma$ and standard normal prior corresponds to
\begin{equation}\label{eq.fisher-information-based-metric-fixed-sigma}
  M(\vct{\theta}) = \idmtx + \tfrac{1}{\sigma^2} \jacob F(\vct{\theta})\tr\jacob F(\vct{\theta}).
\end{equation}
For large $\sigma$ the identity term will dominate and the position-dependent \ac{rwm} and \acs{mala} schemes will behave increasingly similar to \ac{rwm} and \acs{mala} schemes with a fixed identity metric (proposal covariance), while for small $\sigma$ the Fisher information term will dominate, resulting in proposals which adjust for the differing scales in the tangential and normal directions to the limiting manifold $\lbrace \vct{\theta} : F(\vct{\theta}) = \vct{y} \rbrace$.

In Figure \ref{fig.loop_position_dependent_experimental_results} we show comparable heatmaps of the average acceptance rate as a function of the observation noise scale $\sigma$ and step size $\epsilon$ to those in Figure \ref{fig.loop_experimental_results} in the main paper, for these three \acs{mcmc} scheme with position-dependent proposals when applied to our toy two-dimensional model, using the metric function defined in \eqref{eq.fisher-information-based-metric-fixed-sigma}. We see a similar pattern of performance as observed for the \acs{rwm} and \acs{mala} schemes with fixed identity metric and \acs{rmhmc} with this same choice of metric, with all three schemes requiring for the step size $\epsilon$ to be proportional to $\sigma$ to maintain a non-zero acceptance rate (though the constants of proportionality are larger than for the schemes with fixed metric).

We therefore see that these simpler position-dependent schemes do not provide a remedy to the issue of poor scaling in the small noise regime. We also note that, while these position-dependent schemes are implementationally simpler than \acs{rmhmc} and \acs{chmc} due to the lack of any requirement to solve non-linear systems of equations, for the choice of metric in \eqref{eq.fisher-information-based-metric-fixed-sigma}, they retain the same overall per-step $\min(d_{\mathcal{Y}}d_\Theta^2, d_{\mathcal{Y}}^2d_\Theta)$ complexity due to the requirement to evaluate the determinant of (and solve linear systems in) the metric function $\mtx{M}$.

\section{Computational cost of constrained leapfrog integrator}
\label{app.computational_cost}

Typically the operations dominating the computational cost of each constrained leapfrog integrator step (Algorithm \ref{alg.constrained_integrator}) will be: evaluation of the constraint Jacobian $\jacob\constr$, solving linear systems of the form $\jacob\constr(\vct{q})\jacob\constr(\vct{q}')\tr\vct{x} = \vct{b}$ and $\gram{\vct{q}}\vct{x} = \vct{b}$ and evaluating the Gram matrix log-determinant $\log\det\gram{\vct{q}}$ and its derivative.

The exact number of operations performed will vary on each integrator step depending on the number of iterations taken for the projection solver to reach convergence. We found in practice however both the Newton and symmetric Newton solvers, when they did converge, tended to do so within a small number ($\sim$ 5 -- 10) of iterations and that this was reflected across all of the different models used in the experiments. This suggests that the average number of iterations for convergence is not strongly affected by either the dimensions of the unknowns $d_\Theta$ or observations $d_{\mathcal{Y}}$ and so can be assumed to be roughly constant with respect to these variables.

For a model of the form in \eqref{eq.general_model}, the constraint function $\constr(\vct{\theta},\vct{\eta}) = \forw(\vct{\theta}) + \sigma(\vct{\theta})\vct{\eta} - \vct{y}$ has a Jacobian of the form
\begin{equation}
  \jacob\constr(\vct{\theta},\vct{\eta}) = (
    \jacob\forw(\vct{\theta}) + \vct{\eta}\jacob\sigma(\vct{\theta}),
    \sigma(\vct{\theta}) \idmtx_{d_{\mathcal{Y}}}
  ).
\end{equation}
By a standard result from algorithmic differentiation \citep{griewank1993some}, the Jacobian of a function $\mathbb{R}^{m} \to \mathbb{R}^n$ can always be evaluated at a cost proportional to $\min(m, n)$ times the cost of evaluating the function itself. Evaluation of $\jacob\forw$ will therefore cost $\mathcal{O}(\min(d_\Theta, d_{\mathcal{Y}}))$ times the cost of evaluating the forward function $F$ while evaluation of $\jacob\sigma$ will be comparable to evaluating (the scalar valued) $\sigma$ itself.

The projection steps require solving linear systems with matrix terms of the form
\begin{equation}\label{eq.general_model_jacobian_product}
  \jacob\constr(\vct{\theta},\vct{\eta})\jacob\constr(\vct{\theta}',\vct{\eta}')\tr =\\
    (\jacob\forw(\vct{\theta}) + \vct{\eta}\jacob\sigma(\vct{\theta}))(\jacob\forw(\vct{\theta}') + \vct{\eta}'\jacob\sigma(\vct{\theta}'))\tr +
    \sigma(\vct{\theta})\sigma(\vct{\theta}') \idmtx_{d_{\mathcal{Y}}}.
\end{equation}
corresponding to the Gram matrix when $(\vct{\theta},\vct{\eta}) = (\vct{\theta}',\vct{\eta}')$. For \emph{underdetermined models} where $d_{\Theta} > d_{\mathcal{Y}}$ direct computation of the matrix product will have a $\mathcal{O}(d_{\Theta}d_{\mathcal{Y}}^2)$ cost and solving a linear system in the resulting $d_{\mathcal{Y}}\times d_{\mathcal{Y}}$ matrix will have a $\mathcal{O}(d_{\mathcal{Y}}^3)$. For \emph{overdetermined models} with $d_{\Theta} < d_{\mathcal{Y}}$, by recognising that the above corresponds to a rank $d_{\Theta}$ correction to a diagonal matrix, we can exploit the Woodbury matrix identity to solve the linear systems at $\mathcal{O}(d_{\Theta}^2d_{\mathcal{Y}})$ cost.

By a similar argument, evaluating the determinant of the Gram matrix $\det\gram{\vct{q}}$, can in the undetermined setting be performed directly at $\mathcal{O}(d_{\mathcal{Y}}^3)$ cost plus a $\mathcal{O}(d_{\Theta}d_{\mathcal{Y}}^2)$ to compute the Gram matrix itself, and in the overdetermined setting, by exploiting the matrix determinant lemma at a $\mathcal{O}(d_{\Theta}^2d_{\mathcal{Y}})$ cost. As the log-determinant is a scalar valued function computing the derivative of $\log\det\gram{\vct{q}}$ with respect to $\vct{q}$ by reverse-model algorithmic differentiation will cost no more than a constant multiple of the cost of evaluating $\log\det\gram{\vct{q}}$ itself.

Overall we therefore have that the linear algebra operations involved in each constrained integrator step have a $\mathcal{O}(\min(d_\Theta d_\cY^2, d_\cY d_\Theta^2))$ complexity, with each constraint Jacobian evaluation additionally having a cost of $\mathcal{O}(\min(d_\Theta, d_{\mathcal{Y}}))$ times the (model dependent) cost of evaluating the forward function $\forw$ (assuming evaluation of the noise scale $\sigma$ is negligible in comparison to the cost of evaluating $\forw$).

\section{Additional figures for simulated data experiments}
\label{app.simulated-data-additional-results}

\begin{figure}[H]
  {
  \centering
  \includegraphics[width=0.9\textwidth]{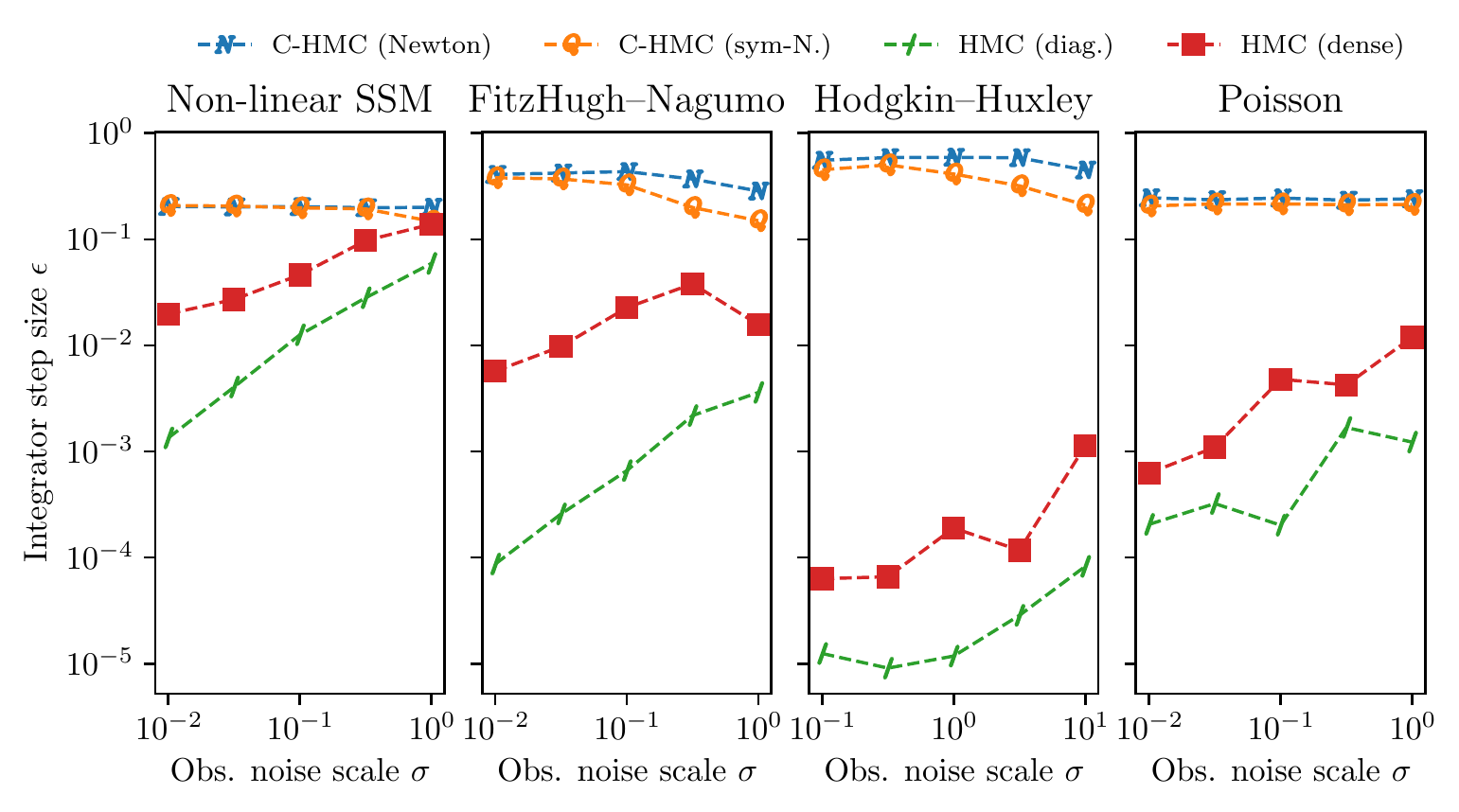}
  }
  \caption{Adapted integrator step size $\epsilon$ for varying $\sigma$ for non-linear \acs{ssm}, FitzHugh--Nagumo and Hodgkin--Huxley \acs{ode} and Poisson \acs{pde} simulated data models.}
  \label{fig.simulated_data_model_step_size_results}
\end{figure}

\begin{figure}[H]
  {
  \centering
  \includegraphics[width=0.95\textwidth]{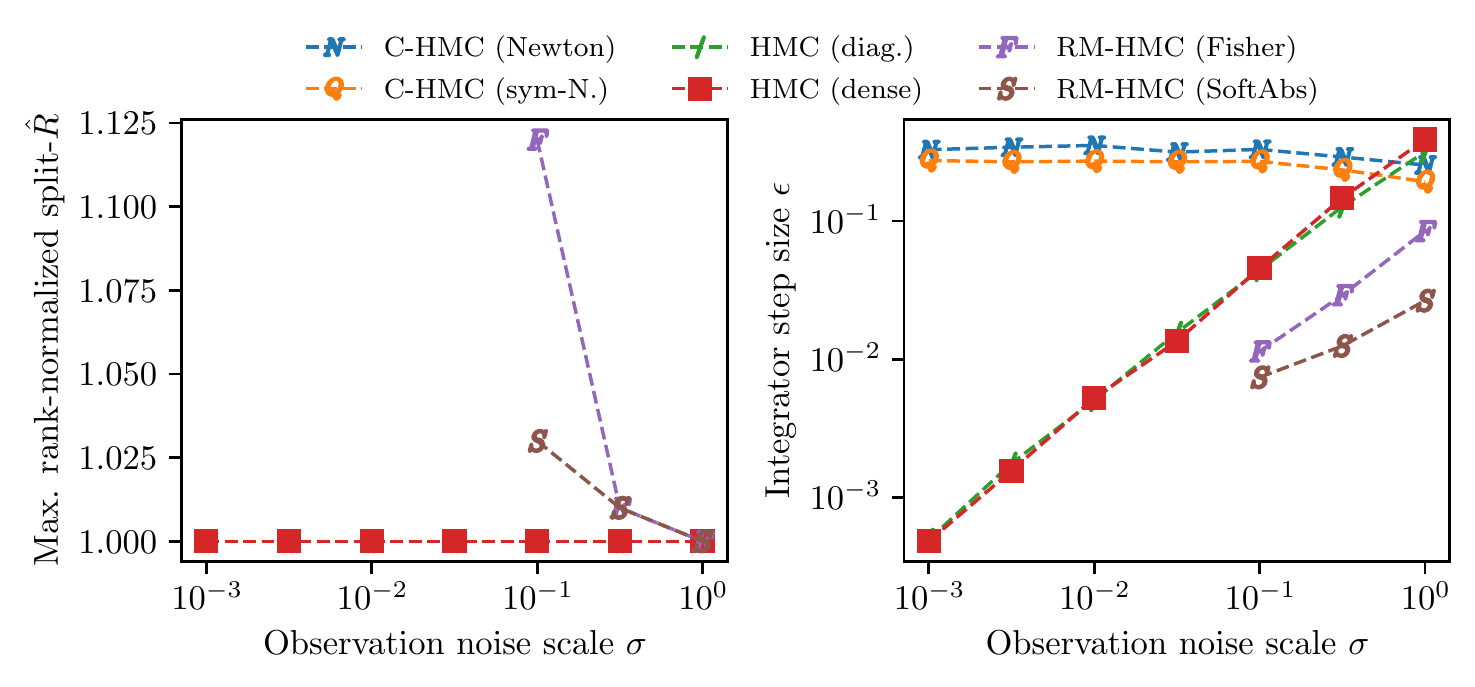}
  }
  \caption{Toy two-dimensional model. Maximum (across $\theta_0$ and $\theta_1$) rank-normalized split-$\hat{R}$ convergence diagnostic and integrator step size $\epsilon$ for varying $\sigma$.}
  \label{fig.loop_rhat_step_size_results}
\end{figure}

\begin{figure}[H]
  \centering
  \includegraphics[width=0.9\textwidth]{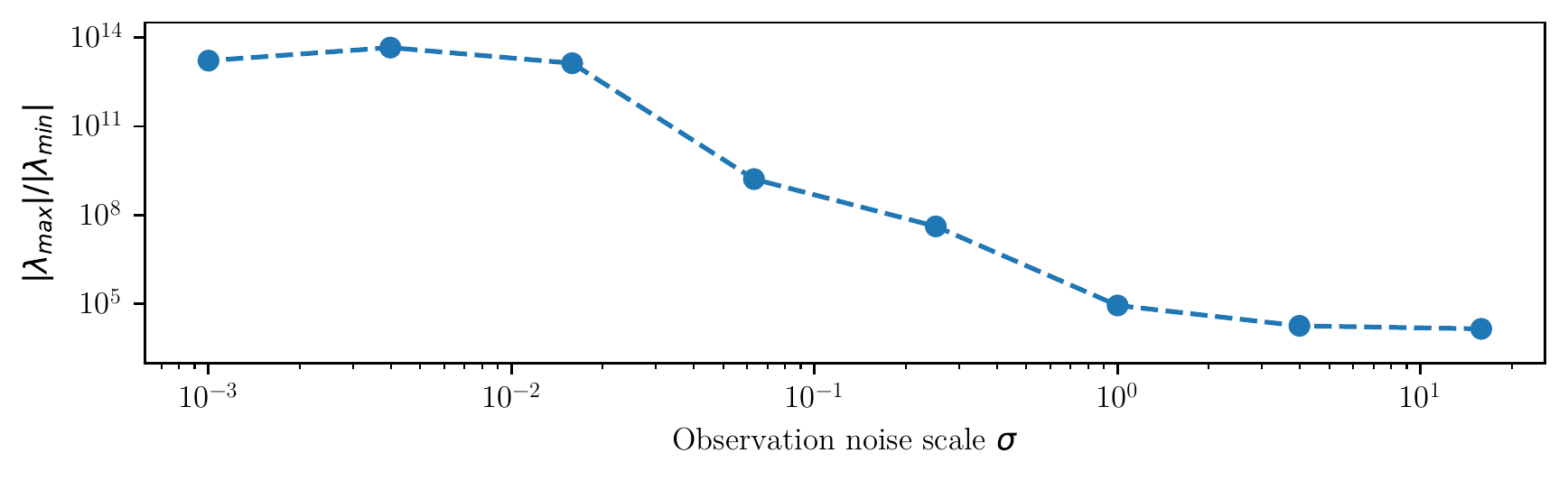}
  \caption{Nonlinear \acs{ssm} model. Condition number of the Hessian of the negative log-density evaluated at a MAP estimate for varying $\sigma$.  $\lambda_{\text{max}}$ and $\lambda_{\text{min}}$ denote the largest and smallest eigenvalues (by magnitude) respectively.}
  \label{fig.nonlinear_ssm_hessian_evals}
\end{figure}

\begin{figure}[H]
  \centering
  \includegraphics[width=0.45\textwidth]{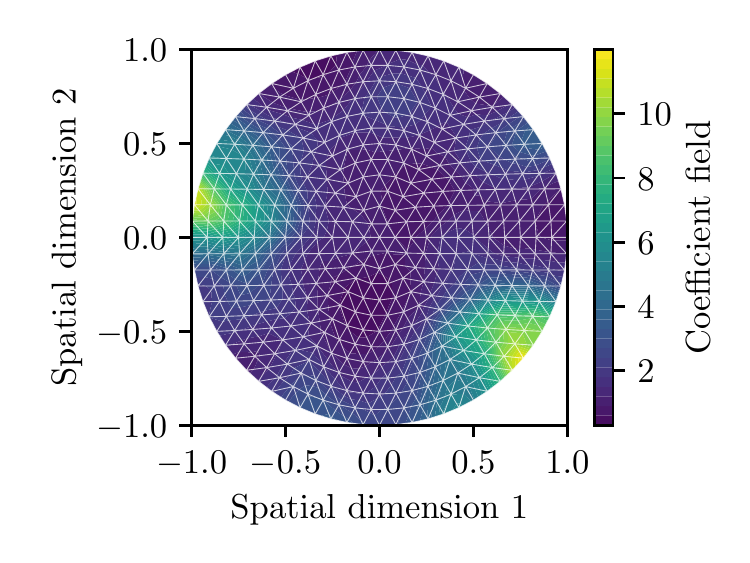}%
  \includegraphics[width=0.45\textwidth]{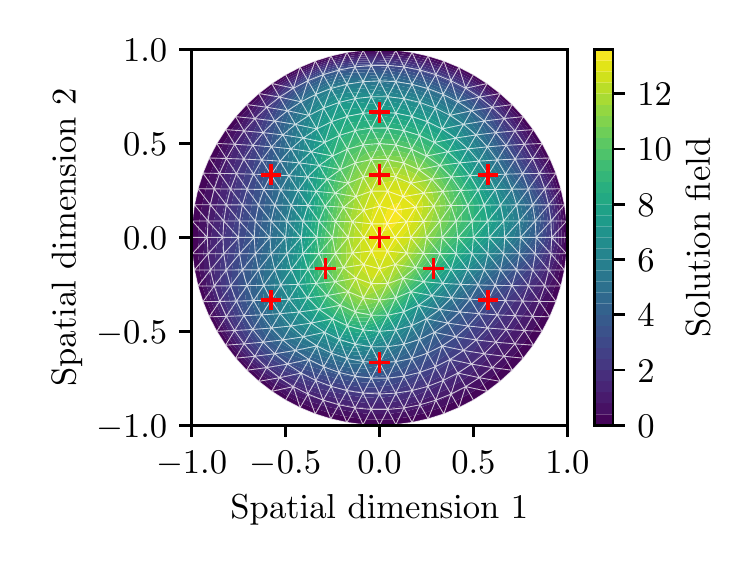}
  \caption{Poisson \acs{pde} model. Left: True coefficient field; right: solution field with observed locations shown by red crosses and mesh by white triangulation.}
  \label{fig.pde_coefficient_and_solution_fields}
\end{figure}

\begin{figure}[H]
  \centering
  \includegraphics[width=0.6\textwidth]{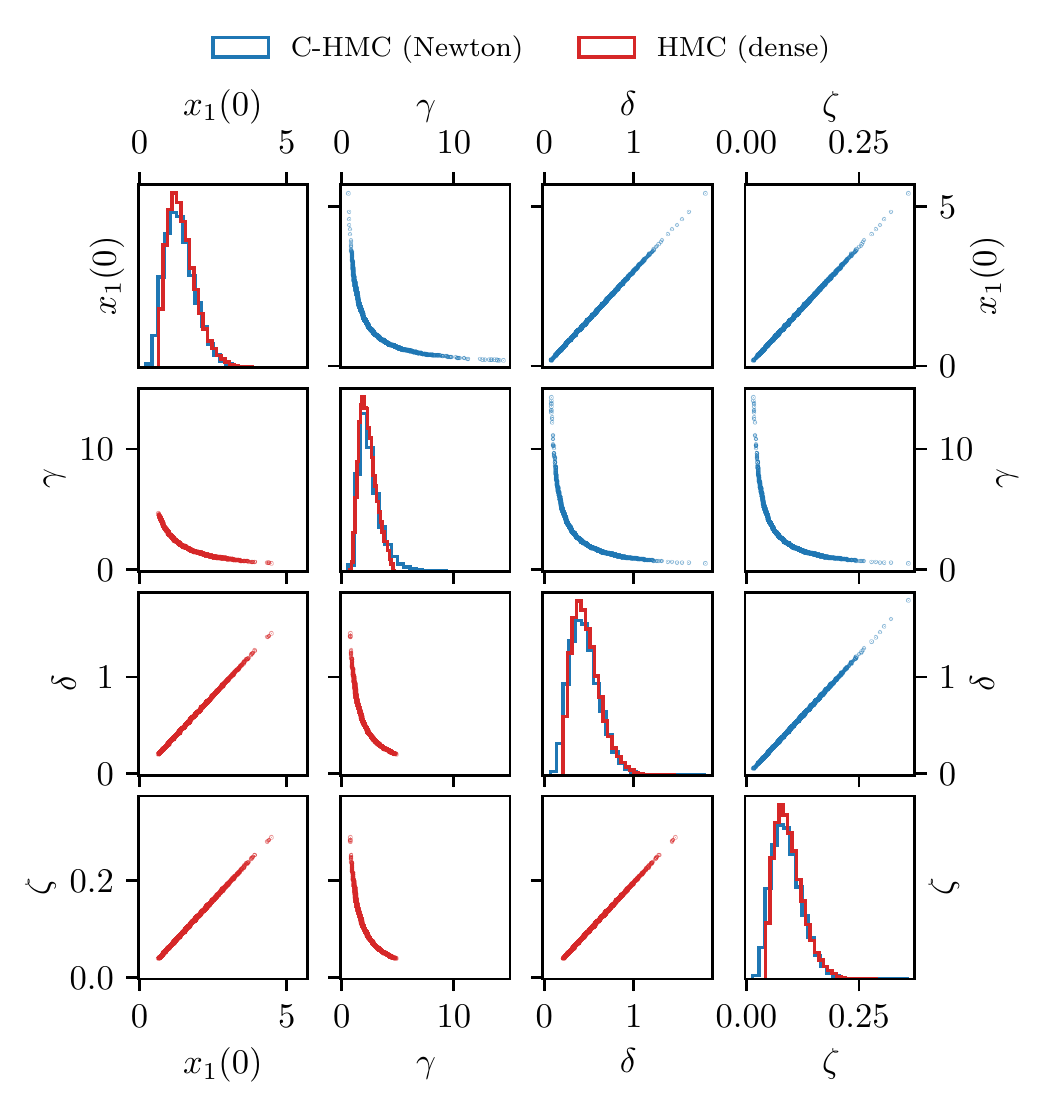}%
  \\
  \includegraphics[width=0.6\textwidth]{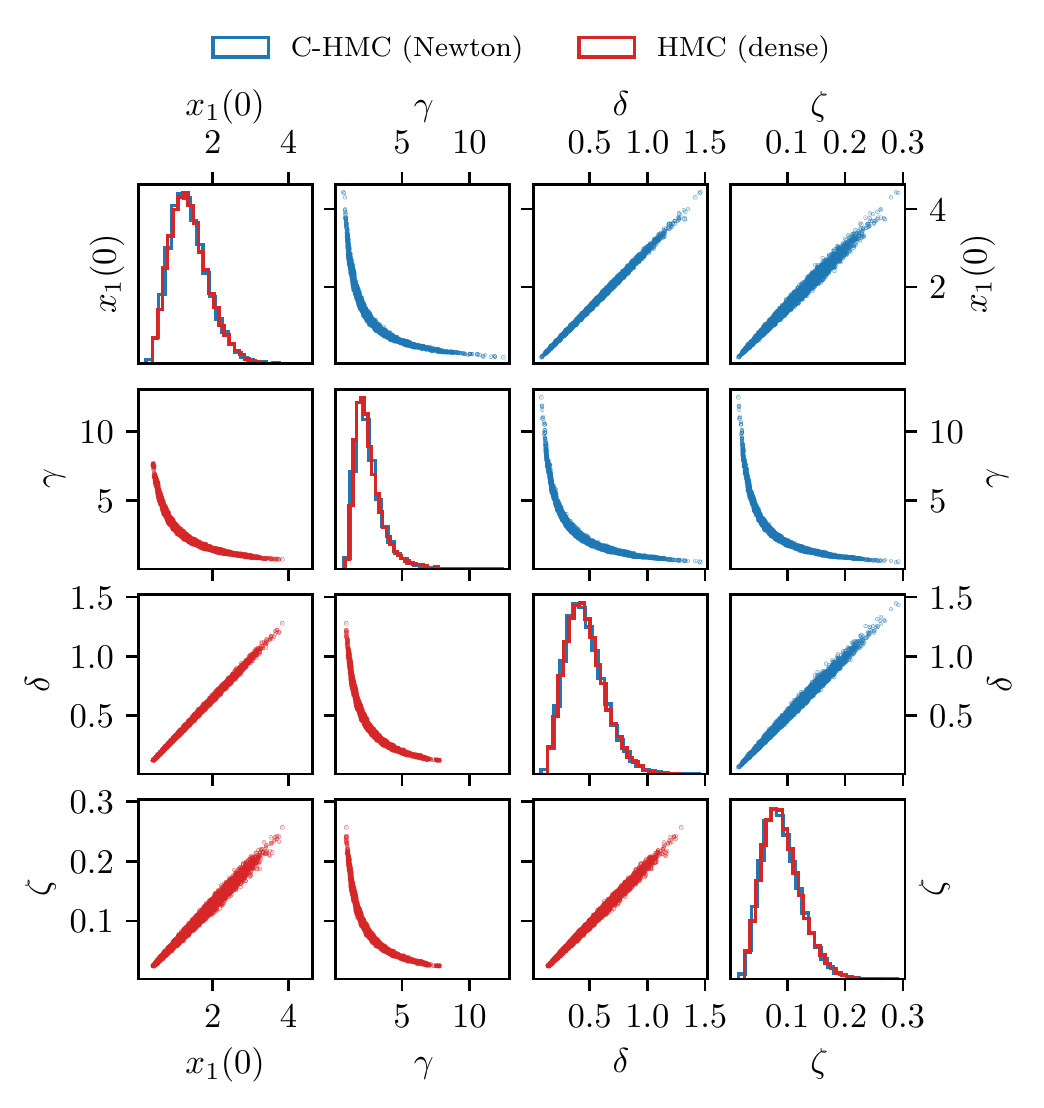}%
  \caption{FitzHugh--Nagumo \acs{ode} model. Pairwise scatter plots and histograms of non-identifiable parameters $x_{1}(0), \gamma, \delta, \zeta$ for true observation scale $\sigma=0.01$ (top) and $\sigma=0.1$ (bottom) for \acs{hmc} (with dense metric, red) and \acs{chmc} (with Newton projection solver, blue) chains. The \acs{hmc} chains have failed to explore the narrower regions of the posterior for the pairs of variables with a non-linear relationship.}
  \label{fig.fhn_pair_plots_sigma}
\end{figure}

\begin{figure}[H]
  \centering
  \includegraphics[width=0.9\textwidth]{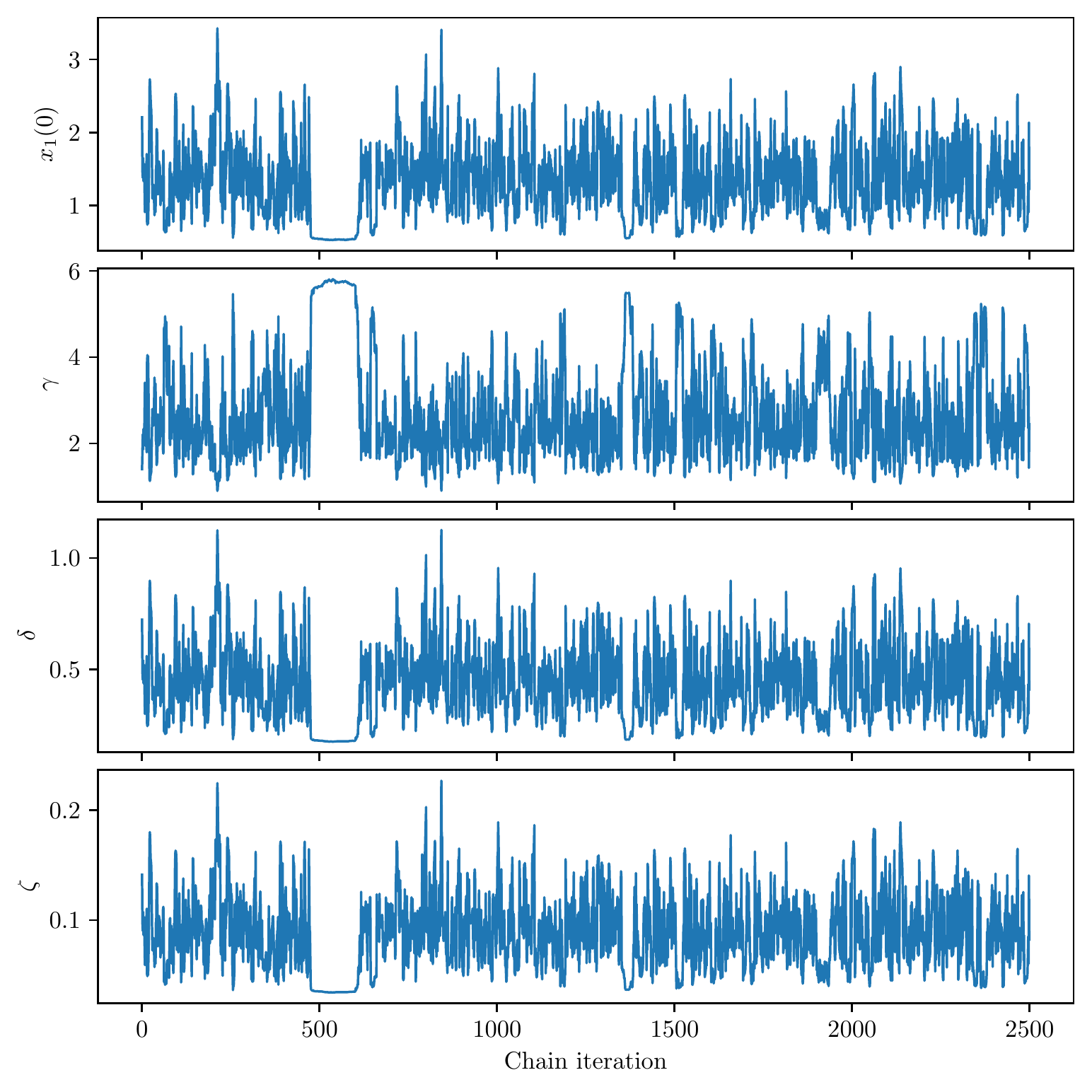}%
  \caption{FitzHugh--Nagumo \acs{ode} model. Example trace plots of non-identifiable parameters $x_{1}(0), \gamma, \delta, \zeta$ for true observation scale $\sigma=10^{-1.5}$ for \acs{hmc} chain (with dense metric) illustrating `sticking' behaviour.}
  \label{fig.fitzhugh_nagumo_hmc_sticking_example_trace}
\end{figure}

\section{Exploiting Markovian structure in state space models}
\label{sec.exploiting-markovian-structure-in-ssms}

A \acs{ssm} of the form
\begin{align*}
  \vct{x}_1 &= G_1(\vct{\phi}, \vct{\nu}_1), \\
  \vct{x}_{t} &=
  G_t(\vct{\phi}, \vct{\nu}_t)(\vct{x}_{t-1}) \quad t \in 2{:}T, \\
  \vct{y}_{t} &=
  H_t(\vct{\phi})(\vct{x}_t) + \sigma_t(\vct{\phi}) \vct{\eta}_t \quad t \in 1{:}T,
\end{align*}
can be considered as a special case of the general model class described by \eqref{eq.general_model}, with $\vct{\theta} = (\vct{\phi}, \vct{\nu})$, $\vct{\nu} = (\vct{\nu}_1, \dots , \vct{\nu}_T)$, $\vct{\eta} = (\vct{\eta}_1, \dots, \vct{\eta}_T)$ and $\vct{y} = (\vct{y}_1, \dots, \vct{y}_T)$. If $\vct{\phi} \in \reals^{d_{\Phi}}$, $\vct{\nu}_t \in \reals^{d_x}$ and $\vct{y}_t \in \reals^{d_{y}}$ for all $t \in 1{:}T$ then $d_{\Theta} = d_{\Phi} + Td_{x}$ and $d_{\mathcal{Y}} = Td_{y}$. Using the lifting described in Section \ref{sec.state_space_augmentation} in the main paper, with constraint function
\begin{align*}
  C(\vct{\phi},\vct{\nu},\vct{\eta}) &= (C_1, \dots, C_T)(\vct{\theta}, \vct{\nu}, \vct{\eta}),\\
  C_t(\vct{\phi}, \vct{\nu}, \vct{\eta}) &=
  H_t(\vct{\phi}) \circ \big(\bigcirc_{s=1}^t G_s(\vct{\phi}, \vct{\nu}_t) \big)
  + \sigma_t(\vct{\phi}) \vct{\eta}_t
  - \vct{y}_t
  \quad t \in 1{:}T,
\end{align*}
evaluating the $Td_y \times (d_\Phi + Td_x + Td_y)$ Jacobian of the constraint function will have an $\mathcal{O}(T^2)$ cost (as evaluating each $C_t$ and its Jacobian $\jacob C_t$ via reverse-mode automatic differentiation will have an $\mathcal{O}(t)$ cost) and
evaluating the determinant of the $Td_y \times Td_y$ Gram matrix or solving a linear system in it will have an $\mathcal{O}(T^3)$ cost. This would suggest that our proposed approach will be of limited applicability in \acp{ssm} unless the number of observation times $T$ is small.

In some cases however we can exploit the additional structure in \acp{ssm} to significantly reduce the computational costs. Specifically if the observation functions $(H_t(\vct{\phi}))_{t=1}^T$ are invertible for all values of the parameters $\vct{\phi}$ then can reduce the cost of evaluating the constraint Jacobian and Gram matrix to $\mathcal{O}(T)$. In this case, we have that the system states $\vct{x}_{1{:}T}$ at each time step can be recovered from the observations $\vct{y}_{1:T}$, parameters $\vct{\phi}$ and observation noise vectors $\vct{\eta}_{1{:}T}$ using
\[
  \vct{x}_t =
  \bar{\vct{x}}_t(\vct{\phi},\vct{\eta}_t) = (H_t(\vct{\phi}))^{-1}(\vct{y}_t - \sigma_t(\vct{\phi}) \vct{\eta}_t)
  \quad t \in 1{:}T,
\]
with we explicitly differentiating between the values of the states $\vct{x}_t$ and functions of the parameters and noise vectors mapping to those values $\bar{\vct{x}}_t$ here for clarity.

We can then define an alternative constraint function
\begin{align*}
  \bar{C}(\vct{\phi}, \vct{\nu}, \vct{\eta}) &= (\bar{C}_1, \dots, \bar{C}_T)(\vct{\phi}, \vct{\nu}, \vct{\eta}),\\
  \bar{C}_1(\vct{\phi}, \vct{\nu}, \vct{\eta}) &=
  G_1(\vct{\phi}, \vct{\nu}_1) - \bar{\vct{x}}_1(\vct{\phi},\vct{\eta}_1),
  \\
  \bar{C}_t(\vct{\phi}, \vct{\nu}, \vct{\eta}) &=
  G_t(\vct{\phi}, \vct{\nu}_t)(\bar{\vct{x}}_{t-1}(\vct{\phi},\vct{\eta}_{t-1})) - \bar{\vct{x}}_t(\vct{\phi},\vct{\eta}_t)
  \quad t \in 2{:}T,
  \
\end{align*}
with the property that the zero-level sets of $\bar{C}$ and $C$ coincide; that is the constraint equations $C(\vct{\phi}, \vct{\nu}, \vct{\eta}) = \vct{0}$ and $\bar{C}(\vct{\phi}, \vct{\nu}, \vct{\eta}) = \vct{0}$ describe the same manifold $\mathcal{M}$.

Importantly however we can evaluate the Jacobian of the constraint function $\bar{C}$ and perform computations with the associated Gram matrix at much lower cost than for the corresponding computations for the original constraint function $C$. Firstly, as each $\bar{C}_t$ can be evaluated at cost that is independent of $t$ (compared to the $\mathcal{O}(t)$ evaluation cost for separately evaluating each $C_t$), $\jacob \bar{C}_t(\vct{\phi}, \vct{\nu},\vct{\eta})$ can be evaluated at $\mathcal{O}(1)$ cost (with respect to $t$) using reverse-mode automatic differentiation and the overall constraint Jacobian at $\mathcal{O}(T)$ cost.

Secondly, we have that $\bar{C}_t$ depends only on $\vct{\phi}$, $\vct{\nu}_t$, $\vct{\eta}_t$ and $\vct{\eta}_{t-1}$ for all $t \in 2{:}T$ and $\bar{C}_1$ depends only on $\vct{\phi}$, $\vct{\nu}_1$ and $\vct{\eta}_1$. Therefore, $\jacob_2 \bar{C}(\vct{\phi},\vct{\nu},\vct{\eta})$ is block diagonal with $T$ blocks of size $d_y \times d_x$ (and zeros elsewhere) and $\jacob_3 \bar{C}(\vct{\phi},\vct{\nu},\vct{\eta})$ is block tridiagonal, with $T$ blocks of size $d_y \times d_y$ along the main diagonal and a further $T - 1$ blocks of size $d_y \times d_y$ in the first lower diagonal (and zeros elsewhere). The corresponding Gram matrix
\begin{align}
  \bar{G}(\vct{\phi},\vct{\nu},\vct{\eta})
  =
  \jacob \bar{C}(\vct{\phi},\vct{\nu},\vct{\eta})
  \jacob\bar{C}(\vct{\phi},\vct{\nu},\vct{\eta})\tr
  =
  \sum_{i \in 1{:}3}
  \jacob_i \bar{C}(\vct{\phi},\vct{\nu},\vct{\eta})
  \jacob_i \bar{C}(\vct{\phi},\vct{\nu},\vct{\eta})\tr
\end{align}
can therefore be decomposed as the sum of a rank-$d_\Phi$ matrix
\[
  \jacob_1 \bar{C}(\vct{\phi},\vct{\nu},\vct{\eta})
  \jacob_1 \bar{C}(\vct{\phi},\vct{\nu},\vct{\eta})\tr,
\]
and the block tri-diagonal matrix
\[
  \jacob_2 \bar{C}(\vct{\phi},\vct{\nu},\vct{\eta})
  \jacob_2 \bar{C}(\vct{\phi},\vct{\nu},\vct{\eta})\tr +
  \jacob_3 \bar{C}(\vct{\phi},\vct{\nu},\vct{\eta})
  \jacob_3 \bar{C}(\vct{\phi},\vct{\nu},\vct{\eta})\tr,
\]
with $T$ blocks of size $d_y \times d_y$ along the main diagonal, and $T-1$ blocks of size $d_y \times d_y$ on the lower and upper diagonals. Linear systems in the block tridiagonal matrix and its determinant can both be evaluated at $\mathcal{O}(T)$ cost, therefore by further using the matrix determinant lemma and Woodbury identity to account for the rank-$d_\Phi$ term, we can then compute the determinant of and solve linear systems in the Gram matrix at an $\mathcal{O}(T)$ cost.

\section{Hodgkin--Huxley current stimulus model details}
\label{sec.hh_full}

For the numerical experiments in Section \ref{sec.hh} we use a \emph{Hodgkin--Huxley} \citep{hodgkin1952quantitative} conductance-based model of neuronal dynamics subject to injected current stimulus, in particular following the particular model formulation described in \citet{pospischil2008minimal}. The dynamics are described by the following system of \acp{ode}
\begin{align*}
  \partial_1 V(t, \vct{\theta}) &=
  \tfrac{1}{C_m}\big(
    \current{S}(t)
    -
    {
      \textstyle
      \sum_{
        \text{ch} \in
        \lbrace \text{Na},\text{K},\text{M},\text{L} \rbrace
      }
      \current{ch}
      (V(t, \vct{\theta}), (m, n, h, p)(t, \vct{\theta}), \vct{\theta})
    }
  \big), \\
  \partial_1 m(t, \vct{\theta}) &=
  \alpha_m(V(t, \vct{\theta}), \vct{\theta})(1-m(t, \vct{\theta}))
  - \beta_m(V(t, \vct{\theta}), \vct{\theta}) m(t, \vct{\theta}), \\
  \partial_1 n(t, \vct{\theta}) &=
  \alpha_n(V(t, \vct{\theta}), \vct{\theta})(1-n(t, \vct{\theta}))
  - \beta_n(V(t, \vct{\theta}), \vct{\theta}) n(t, \vct{\theta}), \\
  \partial_1 h(t, \vct{\theta}) &=
  \alpha_h(V(t, \vct{\theta}), \vct{\theta})(1-h(t, V, \vct{\theta}))
  - \beta_h(V(t, \vct{\theta}), \vct{\theta}) h(t, \vct{\theta}), \\
  \partial_1 p(t, \vct{\theta}) &=
  \big(p_\infty(V(t, \vct{\theta}), \vct{\theta}) - p(t, \vct{\theta})\big)
  / \tau_p(V(t, \vct{\theta}), \vct{\theta}),
\end{align*}
where $V$ is the real-valued \emph{membrane voltage} and $m$, $n$, $h$ and $p$ are ion channel gating variables, which takes values in $[0, 1]$.

The \acs{ode} for the membrane voltage $V$ includes terms for the \emph{stimulus} current $\current{S}$, here defined as a square wave
\begin{equation}
\current{S}(t) =
\begin{cases}
    3.248 ,& \text{if } 10\leq t \leq 110\\
    0,              & \text{otherwise}\\
\end{cases}
\end{equation}
and various ionic channel current components, namely the sodium ionic current $\current{Na}$, the potassium ionic current $\current{K}$, the `delayed-rectifier' potassium ionic current $\current{M}$ \citep{traub_miles_1991} and a catch-all leakage current $\current{L}$, with these defined by
\begin{align*}
  \current{Na}(V, (m, n, h, p), \vct{\theta}) &= \gbar{Na} m^3 h (V - 53), &
  \current{K}(V, (m, n, h, p), \vct{\theta}) &= \gbar{K} n^4 ( V + 107), \\
  \current{M}(V, (m, n, h, p), \vct{\theta}) &= \gbar{M} p (V + 107), &
  \current{L}(V, (m, n, h, p), \vct{\theta}) &= g_\text{L} (V + 75).
\end{align*}
The dynamics of the gating variables $\lbrace m, n, h \rbrace$ are governed by rate functions
\begin{align*}
  \alpha_n(V, \vct{\theta}) &=
  \frac{0.032 \, (V - V_T - 15) }{ \exp(-(V - V_T - 15) / 5) - 1},
  &
  \beta_n(V, \vct{\theta}) &=
  0.5 \exp(-(V - V_T - 10) / 40),
  \\
  \alpha_m(V, \vct{\theta}) &=
  \frac{0.32 (V - V_T - 13) }{\exp(-(V  - V_T -13) /4) - 1},
  &
  \beta_m(V, \vct{\theta}) &=
  \frac{0.28 (V - V_T - 40) }{\exp(-(V - V_T - 40)/ 5) - 1},
  \\
  \alpha_h(V, \vct{\theta}) &=
  0.128 \exp(-(V - V_T - 17) / 18),
  &
  \beta_h(V, \vct{\theta}) &=
  \frac{4 }{ \exp(-(V  - V_T -40) / 5) + 1},
\end{align*}
which can be used to define \emph{steady state} and \emph{time constant} functions
\begin{equation*}
  x_\infty(V, \vct{\theta}) =
  \frac
    {\alpha_x(V, \vct{\theta})}
    {\alpha_x(V,\vct{\theta}) + \beta_x(V, \vct{\theta})},
  \quad
  \tau_x(V, \vct{\theta}) =
  \frac
    {1}
    {\alpha_x(V,\vct{\theta}) + \beta_x(V, \vct{\theta})},
  \quad
  x \in \lbrace m, n, h \rbrace,
\end{equation*}
corresponding respectively to the steady state and time constant of the solutions to governing \acp{ode} for fixed $V$, for which the \acp{ode} are linear.
In the case of $p$ the steady-state and time-constant functions are defined directly as
\begin{align*}
  p_\infty(V, \vct{\theta}) &=
  \big(
    1 + \exp\big(-(V + 35) / 10\big)
  \big)^{-1}
  ,\\
  \tau_p(V, \vct{\theta}) &=
  k_{\tau_p} \big(3.3 \exp((V + 35)/ 20) + \exp(-(V + 35) / 20)\big)^{-1}.
\end{align*}
The initial state of the \acp{ode} system is then defined as
\begin{equation*}
  V(0, \vct{\theta}) = -75, \quad
  x(0, \vct{\theta}) = x_{\infty}(-75, \vct{\theta})
  \quad \forall x \in \lbrace m, n, h, p \rbrace.
\end{equation*}
The \acs{ode} system is numerically solved using a Lie--Trotter splitting based approach that exploits the property that the system is \emph{conditionally linear} \citep{chen2020structure}: the \acp{ode} for the gating variables $\lbrace m, n, h, p \rbrace$ for a fixed membrane voltage $V$ are linear as is the \acs{ode} for the membrane voltage $V$ for fixed values of the gating variables $\lbrace m, n, h, p \rbrace$. A integrator timestep of $\Delta = 0.1$ is used.
The $d_{\mathcal{Y}} = 1000$ observations are modelled as being generated according to
\begin{equation}
  y_i = V(10 + i\Delta, \vct{\theta}) + \sigma \eta_i,
  \quad
  \eta_i \sim \normal(0, 1),
  \quad
  \forall i \in 1{:}1000.
\end{equation}
that is, observations of the membrane voltage only, at discrete times and subject to independent additive Gaussian noise.

The $d_\Theta = 7$ parameters, \(
  \vct{\theta} = (\gbar{Na}, \gbar{K}, \gbar{M}, g_{\text{L}}, V_T, k_{\tau_p}, \sigma),
\)
are given priors
\begin{equation}
\begin{aligned}[t]
  \gbar{Na} &\sim \lognormal(4, 1),\\
  \gbar{K} &\sim \lognormal(2, 1),\\
  \gbar{M} &\sim \lognormal(-3, 1),\\
  g_{\text{L}} &\sim \lognormal(-3, 1),
\end{aligned}
\quad
\begin{aligned}[t]
  V_T &\sim \normal(-60, 10),\\
  k_{\tau_p} &\sim \lognormal(8, 1),\\
  \sigma &\sim \lognormal(0, 1).
\end{aligned}
\end{equation}

\section{Details for ordinary differential equation models fitted to real data}
\label{sec.real-data-model-details}

\subsection{Lotka--Volterra model}
\label{subsec.lotka-volterra-details}

A classic model of the dynamics of an ecosystem in which two species interact, one as predator and one as prey, due to \citet{lotka1925elements} and \citet{volterra1926fluctuations}. The system dynamics are described by the \acp{ode}
\begin{equation}
  \label{eq.lotka_volterra}
  \partial_1 x_1(t, \vct{\theta}) =
  (\alpha - \beta x_2(t, \vct{\theta})) x_1(t, \vct{\theta}),
  \qquad
  \partial_1 x_2(t, \vct{\theta}) =
  (\delta - \gamma x_1(t, \vct{\theta})) x_2(t, \vct{\theta}),
\end{equation}
with initial conditions
\begin{equation}
  x_1(t_0, \vct{\theta}) = x_{1,0},
  \qquad
  x_2(t_0, \vct{\theta}) = x_{2,0},
\end{equation}
where $x_1$ describes the prey population and $x_2$ the predator population. The model was fitted to a dataset of annual measurements from 1900 to 1920 of snowshoe hare and Canadian Lynx populations, based on pelt numbers collected by the Hudson's Bay Company \citep{hewitt1921conservation, howard2009modeling}. Following the approach of \citet{carpenter2018predator}, a log-normal observation model is assumed with the $d_{\mathcal{Y}} = 42$ observations $y_{1{:}42}$ of the prey $x_1$ and predator $x_2$ populations for the years $t_j = 1900 + j,~\forall j \in 0{:}20$ assumed to be generated according to
\begin{equation}\label{eq.lotka_volterra_observation}
  \log y_{2j + i} = \log x_i(t_j, \vct{\theta}) + \sigma_i \eta_{2j+i},
  \quad
  \eta_{2j+i} \sim \normal(0, 1)
  \quad
  \forall i \in \lbrace 1, 2\rbrace, j \in 0{:}20.
\end{equation}
The \acs{ode} system is numerically solved using a fourth-order Runge--Kutta method with fixed time step $\Delta =0.1\,\text{days}$. Also following \citet{carpenter2018predator}, the $d_\Theta = 8$ parameters, $\vct{\theta} = (\alpha, \beta, \gamma, \delta, x_{1,0}, x_{2,0}, \sigma_1, \sigma_2)$, are given priors
\begin{equation}
  \begin{aligned}[t]
    \alpha &\sim \mathsf{TruncatedNormal}(1, 0.5, 0, \infty), \\
    \beta &\sim \mathsf{TruncatedNormal}(0.05, 0.05, 0, \infty), \\
    \gamma &\sim \mathsf{TruncatedNormal}(1, 0.5, 0, \infty),
  \end{aligned}
  ~~
  \begin{aligned}[t]
    \delta &\sim \mathsf{TruncatedNormal}(0.05, 0.05, 0, \infty), \\
    x_{i,0} & \sim \lognormal(\log(10), 1) ~~ \forall i \in \lbrace 1, 2\rbrace, \\
    \text{and } \sigma_i &\sim \lognormal(-1, 1) ~~ \forall i \in \lbrace 1, 2\rbrace.
  \end{aligned}
\end{equation}

\subsection{Soil carbon two-pool model}

A two-pool model for carbon flow within soil, based on a \emph{Stan} case study \citep{carpenter2014soil}. The system is described by the linear \acp{ode}
\begin{equation}\label{eq.soil_incubation}
  \partial_1 x_1(t, \vct{\theta}) = -k_1 x_1(t, \vct{\theta}) + \alpha_{12} k_2 x_2(t, \vct{\theta}),
  \quad
  \partial_1 x_2(t, \vct{\theta}) -k_2 x_2(t, \vct{\theta}) + \alpha_{21} k_1 x_1(t, \vct{\theta}),
\end{equation}
with initial conditions
\begin{equation}
 x_1(t_0, \vct{\theta}) = \gamma \, C_0,
 \qquad
 x_2(t_0, \vct{\theta}) = (1 - \gamma) \, C_0.
\end{equation}
The observed data correspond to measurements of the carbon cumulatively evolved by the system over time, with an assumption of independent normal noise in the observations of unknown standard deviation, with the observation model for a set of measurements at $d_{\mathcal{Y}}$ times $t_{1{:}d_{\mathcal{Y}}}$ then defined by
\begin{equation}
  y_i = (C_0 - x_1(t_i, \vct{\theta}) - x_2(t_i, \vct{\theta})) + \sigma \eta_i
  \quad
  \forall i \in 1{:}d_{\mathcal{Y}}.
\end{equation}
Data from two soil incubation experiments are taken from the \emph{R} package \emph{SoilR} \citep{sierra2014modeling}. The two datasets, \emph{HN-T35} and \emph{AK-T25} use soil samples from two different sites and incubated at different temperatures, and  both have measurements at $d_{\mathcal{Y}} = 25$ times. The \acs{ode} system can be exactly solved in this case. The $d_{\Theta} = 7$ parameters, $\vct{\theta} = (k_1, k_2, \alpha_{12}, \alpha_{21}, \gamma, C_0, \sigma)$ are given priors
\begin{equation}
  \begin{aligned}[t]
    k_1 &\sim \halfnormal(1),\\
    k_2 &\sim \mathsf{TruncatedNormal}(0, 1, 0, k_1),\\
    \alpha_{12} &\sim \uniform(0, 1),\\
    \alpha_{21} &\sim \uniform(0, 1 - \alpha_{12}),
  \end{aligned}
  \qquad
  \begin{aligned}[t]
    \gamma &\sim \uniform(0,1),\\
    C_0 &\sim \lognormal(1, 2),\\
    \text{and } \sigma &\sim \halfnormal(1).
  \end{aligned}
\end{equation}

\subsection{Hodgkin--Huxley voltage clamp model}

A conductance-based model of action potential generation in neurons, as proposed in \citet{hodgkin1952quantitative}. Compared to the model described in Section \ref{sec.hh_full}, which corresponded to the free dynamics of a neuron subject to a current stimulus, in accordance with the experimental protocol of \citet{hodgkin1952quantitative} the model here is used to simulate the dynamics of a giant squid axon under a \emph{voltage clamp}, with the membrane voltage held at a specified level using a feedback circuit. In addition to clamping the voltage, in the experiments of \citet{hodgkin1952quantitative}, the dynamics of specific ionic channels were isolated by manipulating the ions present in the extracellular fluid during experiments, with this allowing individual measurements to be made of the potassium and sodium channel conductances.

The conductance of a particular set of ion channels is modelled as being determined by one or more \emph{gating variables} with temporally varying activations in $[0, 1]$, with the overall conductance then a polynomial expression in the gating variable values and an overall maximum conductance coefficient. Each of the gating variables is assumed to have dynamics governed by the linear (for fixed parameters $\vct{\theta}$ and membrane voltage $V$) \acs{ode}
\begin{equation}
  \partial_1 x(t, V, \vct{\theta}) =
  \alpha_x(V, \vct{\theta}) \big(1 - x(t, V, \vct{\theta})\big) -
  \beta_x(V, \vct{\theta}) x(t, V, \vct{\theta})
\end{equation}
where $\alpha_x$ and $\beta_x$ are channel-specific parametrized and voltage-dependent rate functions. For a fixed membrane voltage $V$ this \acs{ode} has the exact solution
\begin{equation}
  x(t, V, \vct{\theta}) = x(0, V, \vct{\theta}) + \big(x_{\infty}(V, \vct{\theta}) - x(0, V, \vct{\theta})\big) \big( 1 - \exp(-t / \tau_x(V, \vct{\theta}))\big)
\end{equation}
where $x_\infty$ and $\tau_x$, respectively the \emph{steady state} and \emph{time constant}, are defined
\begin{equation}
  x_\infty(V, \vct{\theta}) = \frac{\alpha_x(V, \vct{\theta})}{\alpha_x(V,\vct{\theta}) + \beta_x(V, \vct{\theta})},
  \quad
  \tau_x(V, \vct{\theta}) = \frac{1}{\alpha_x(V,\vct{\theta}) + \beta_x(V, \vct{\theta})}.
\end{equation}
We assume in all cases the initial condition for a gating variable is the steady state value for a membrane voltage $V = 0$, corresponding to the resting potential, i.e.
\begin{equation}
  x(0, V, \vct{\theta}) = x_\infty(0, \vct{\theta}).
\end{equation}
We fit the models to a digitization of the original voltage clamp experimental data from \citet{daly2015hodgkin}, specifically sequences of
sodium and potassium channel conductances for varying depolarizations (fixed values of the membrane voltage).

\subsubsection{Potassium channel conductances}

For the potassium channel conductances, the model uses a single gating variable $n$ with corresponding parametrized rate functions
\begin{equation}
  \alpha_n(V, \vct{\theta}) = \frac{\kan{1} \, (V + \kan{2}) }{\exp\big((V + \kan{2})/\kan{3}\big) - 1},
  \qquad
  \beta_n(V, \vct{\theta}) = \kbn{1} \, \exp(V / \kbn{2}).
\end{equation}
The potassium conductances are then assumed to be defined as $\bar{g}_{\text{K}}n^4$, with the observed noisy conductance data therefore assumed to be generated according to
\begin{equation}
  y_{i} = \bar{g}_{\text{K}} n(t_i, V_i, \vct{\theta})^4 + \sigma \eta_{i},
  \quad
  \eta_i \sim \normal(0, 1)
  \quad
  \forall i \in 1{:}d_{\mathcal{Y}}
\end{equation}
where $(t_i, V_i)_{i=1}^{d_\mathcal{Y}}$ are the set of $d_{\mathcal{Y}} = 136$ measurement times and applied membrane voltages (depolarizations) used in the experiments.

The $d_\Theta = 7$ parameters $\vct{\theta} = (\kan{1}, \kan{2}, \kan{3}, \kbn{1}, \kbn{2}, \bar{g}_{\text{K}}, \sigma)$ are given priors
\begin{equation}
  \begin{aligned}[t]
    \kan{1} &\sim \lognormal(-3, 1),\\
    \kan{2} &\sim \lognormal(2, 1),\\
    \kan{3} &\sim \lognormal(2, 1),\\
    \kbn{1} &\sim \lognormal(-3, 1),
  \end{aligned}
  \qquad
  \begin{aligned}[t]
    \kbn{2} &\sim \lognormal(2, 1),\\
    \bar{g}_{\text{K}} &\sim \lognormal(2, 1),\\
    \text{and } \sigma &\sim \lognormal(0, 1).
  \end{aligned}
\end{equation}

\subsubsection{Sodium channel conductances}

For the sodium channel conductances, the model uses two gating variables $m$ and $h$ with corresponding parametrized rate functions
\begin{equation}
  \alpha_m(V, \vct{\theta}) = \frac{\kam{1} \, (V + \kam{2}) }{ \exp ((V + \kam{2})/ \kam{3}) - 1},
  \quad
  \beta_m(V, \vct{\theta}) = \kbm{1} \, \exp( V / \kbm{2}),
\end{equation}
\begin{equation}
  \alpha_h(V, \vct{\theta}) = \kah{1} \, \exp(V / \kah{2}),
  \quad
  \beta_h(V, \vct{\theta}) = \frac{1}{ \exp ((V + \kbh{1}) /\kbh{2}) - 1}.
\end{equation}

The sodium conductances are then assumed to be defined as $\bar{g}_{\text{Na}}m^3 h$, with the observed noisy conductance data therefore assumed to be generated according to
\begin{equation}
  y_{i} = \bar{g}_{\text{Na}} m(t_i, V_i, \vct{\theta})^3 h(t_i, V_i, \vct{\theta}) + \sigma \eta_{i},
  \quad
  \eta_i \sim \normal(0, 1)
  \quad
  \forall i \in 1{:}d_{\mathcal{Y}}
\end{equation}
where $(t_i, V_i)_{i=1}^{d_\mathcal{Y}}$ are the set of $d_{\mathcal{Y}} = 142$ measurement times and applied membrane voltages (depolarizations) used in the experiments.

The $d_\Theta = 11$ parameters \[\vct{\theta} = (\kam{1}, \kam{2}, \kam{3}, \kbm{1}, \kbm{2}, \kah{1}, \kah{2}, \kbh{1}, \kbh{2}, \bar{g}_{\text{Na}}, \sigma)\] are given priors
\begin{equation}
  \begin{aligned}[t]
    \kam{1} &\sim \lognormal(-3, 1),\\
    \kam{2} &\sim \lognormal(2, 1),\\
    \kam{3} &\sim \lognormal(2, 1),\\
    \kbm{1} &\sim \lognormal(0, 1),
  \end{aligned}
  \quad
  \begin{aligned}[t]
    \kbm{2} &\sim \lognormal(2, 1),\\
    \kah{1} &\sim \lognormal(-3, 1),\\
    \kah{2} &\sim \lognormal(2, 1),\\
    \kbh{1} &\sim \lognormal(2, 1),
  \end{aligned}
  \quad
  \begin{aligned}[t]
    \kbh{2} &\sim \lognormal(1, 1),\\
    \bar{g}_{\text{Na}} &\sim \lognormal(2, 1),\\
    \text{and } \sigma &\sim \lognormal(0, 1).
  \end{aligned}
\end{equation}

\newpage

\section{Posterior pair plots for models fitted to real data}
\label{sec.real-data-posterior-pair-plots}

\begin{figure}[H]
  \centering
  \includegraphics[width=\linewidth]{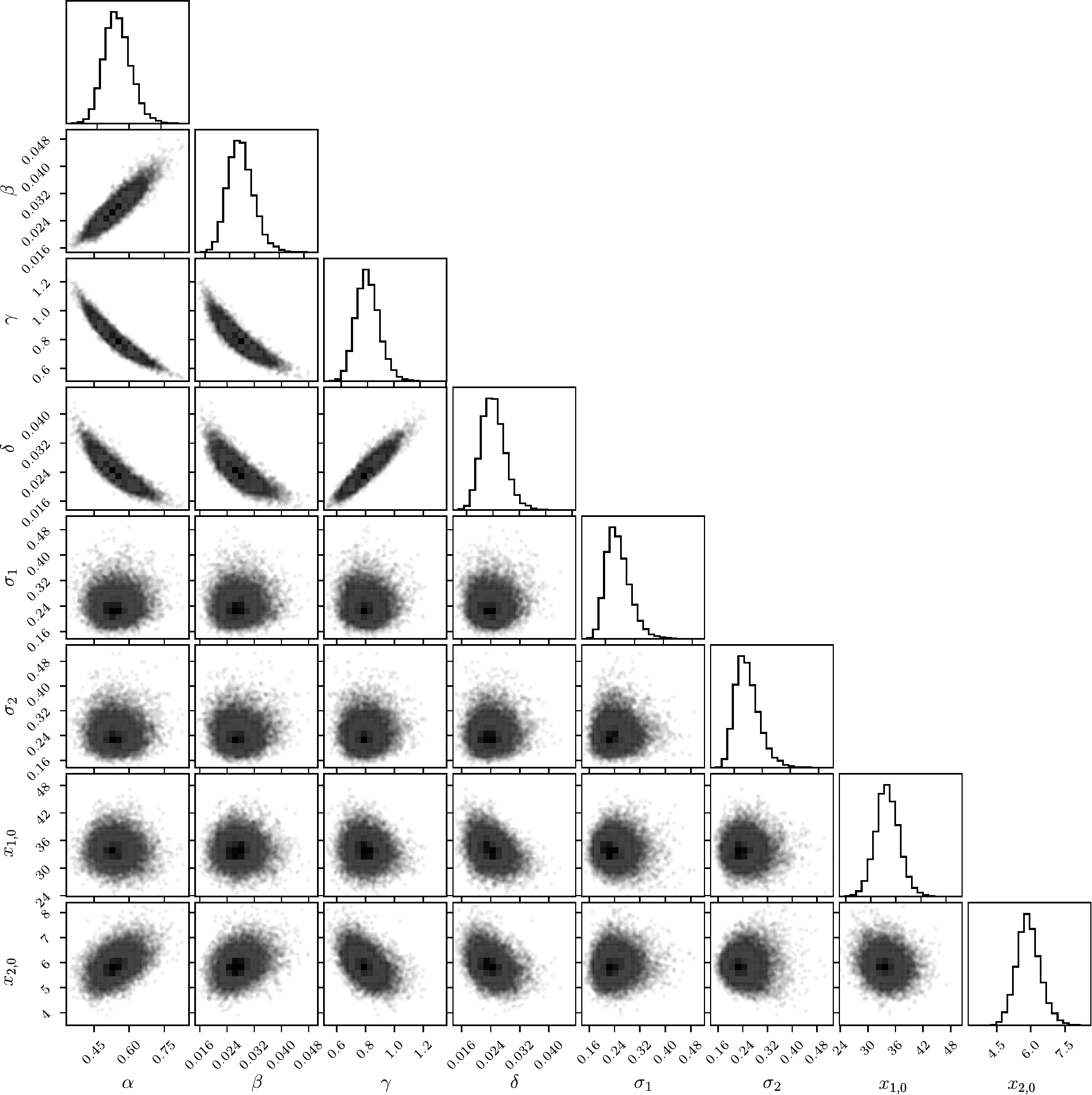}%
  \caption{\emph{Lotka--Volterra}: Parameter space}
  \label{fig.lotka_volterra_posterior_pair_plot}
\end{figure}
\begin{figure}[H]
  \centering
  \includegraphics[width=\linewidth]{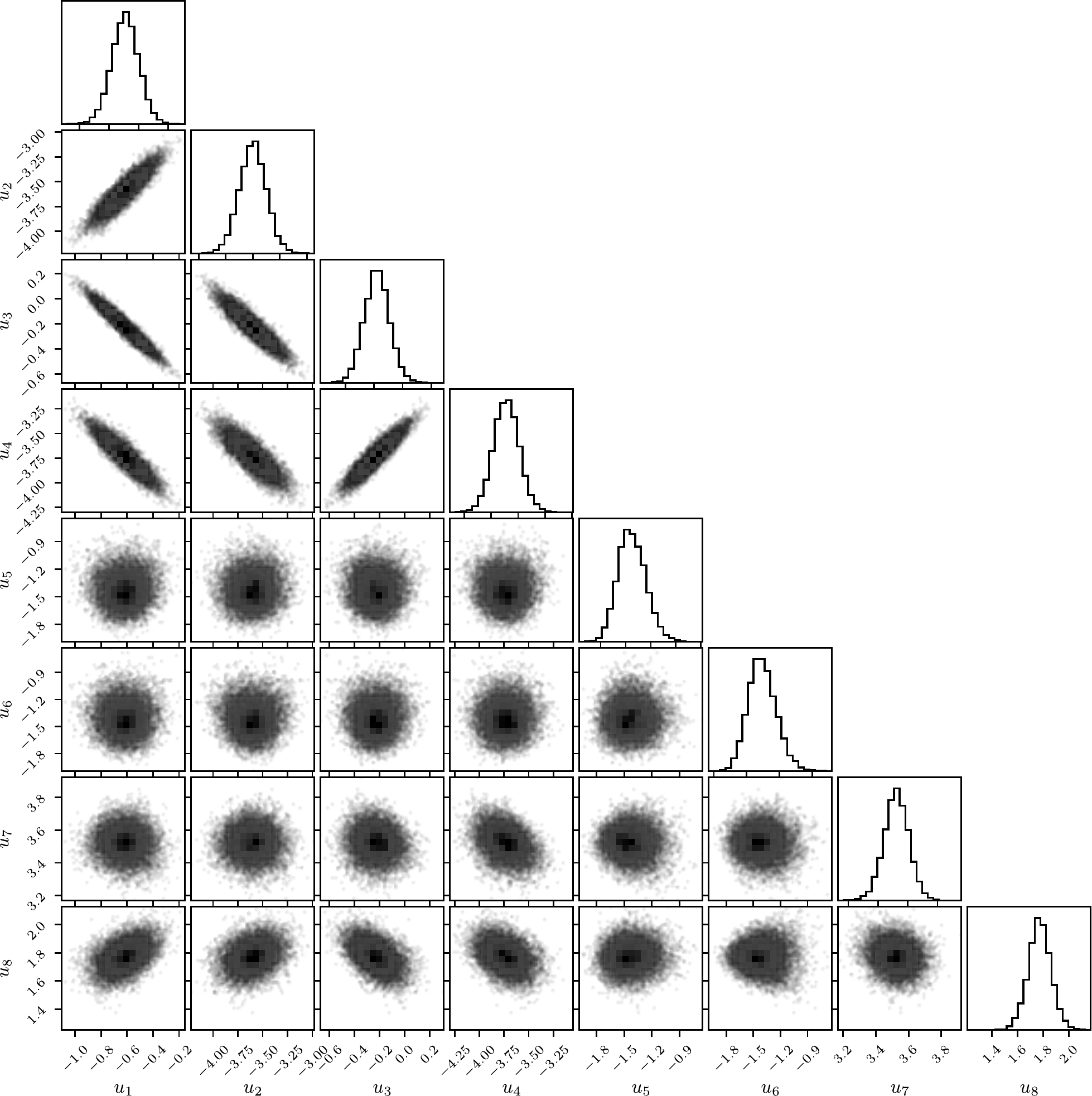}%
  \caption{\emph{Lotka--Volterra}: Unconstrained space}
  \label{fig.lotka_volterra_unconstrained_posterior_pair_plot}
\end{figure}
\begin{figure}[H]
  \centering
  \includegraphics[width=\linewidth]{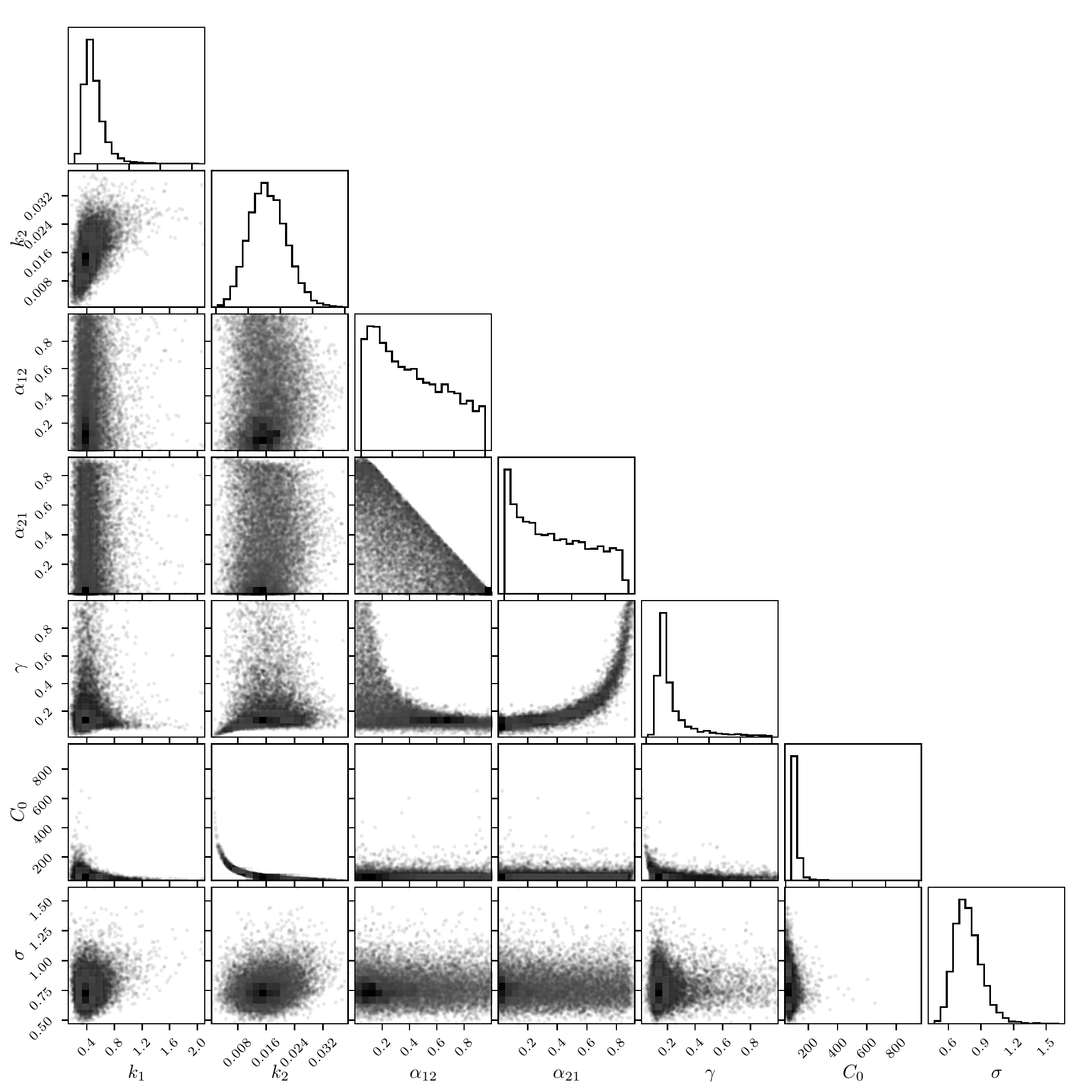}%
  \caption{\emph{Soil incubation (AK-T25)}: Parameter space}
  \label{fig.soil_incubation_ak-t25_posterior_pair_plot}
\end{figure}
\begin{figure}[H]
  \centering
  \includegraphics[width=\linewidth]{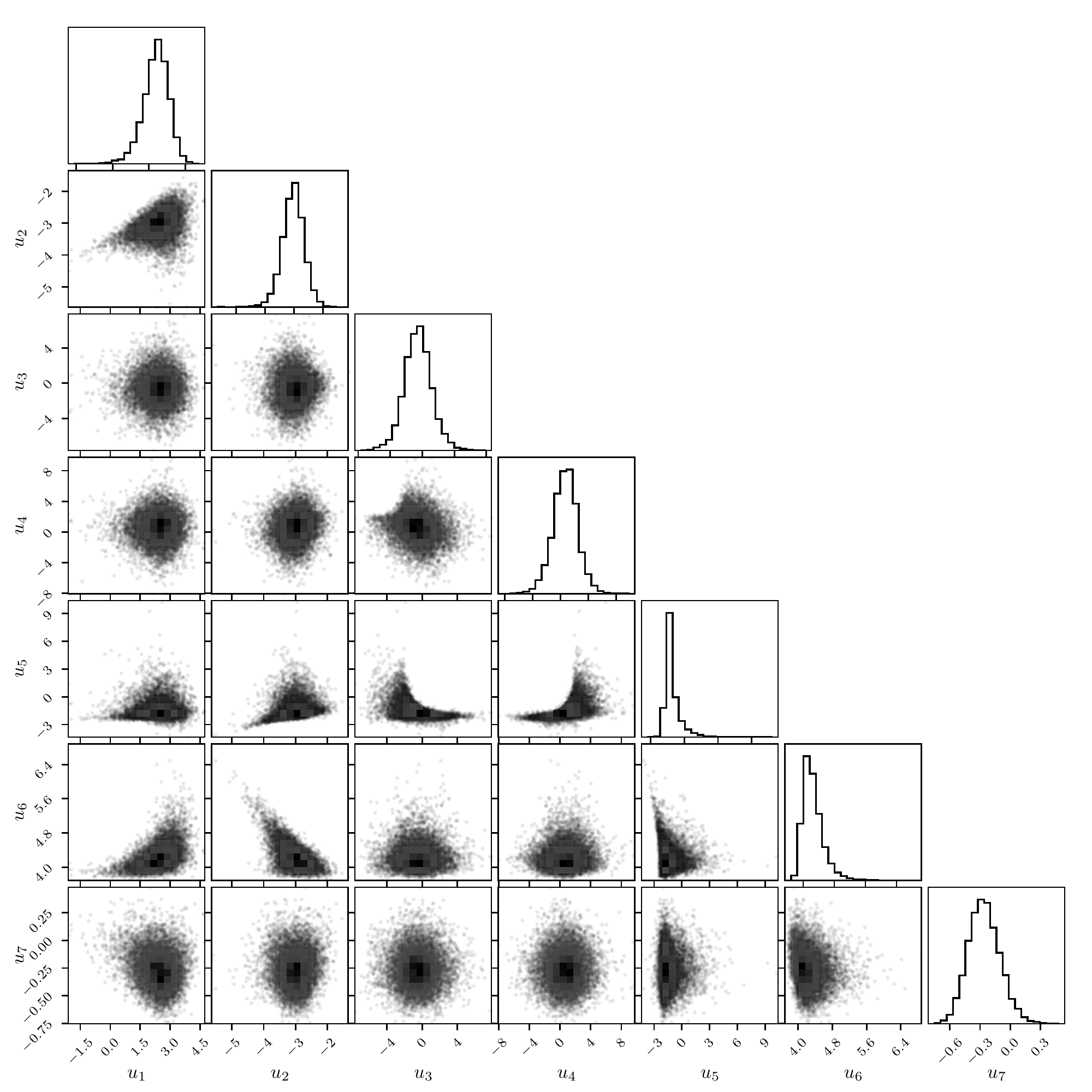}%
  \caption{\emph{Soil incubation (AK-T25)}: Unconstrained space}
  \label{fig.soil_incubation_ak-t25_unconstrained_posterior_pair_plot}
\end{figure}
\begin{figure}[H]
  \centering
  \includegraphics[width=\linewidth]{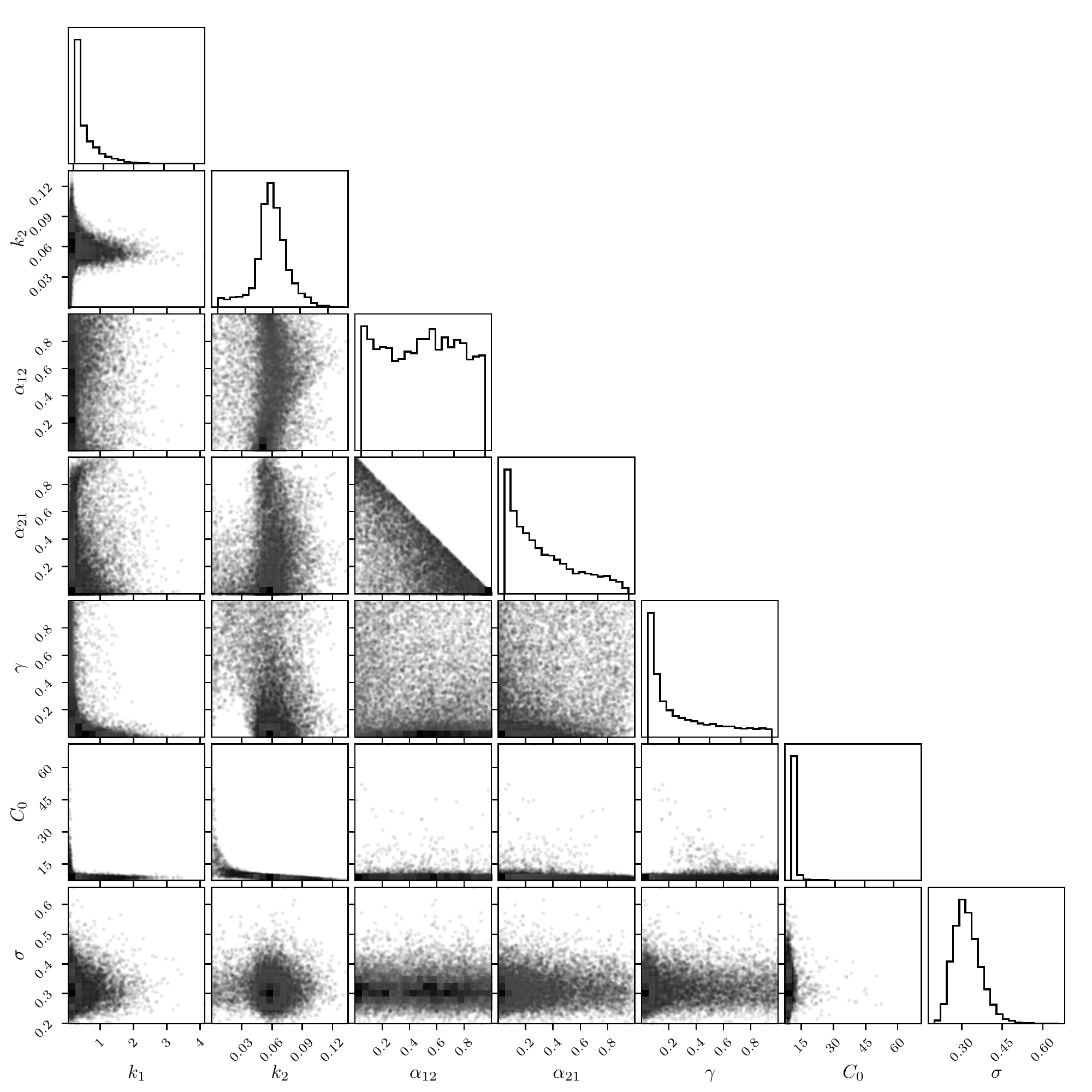}%
  \caption{\emph{Soil incubation (HN-T35)}: Parameter space}
  \label{fig.soil_incubation_hn-t35_posterior_pair_plot}
\end{figure}
\begin{figure}[H]
  \centering
  \includegraphics[width=\linewidth]{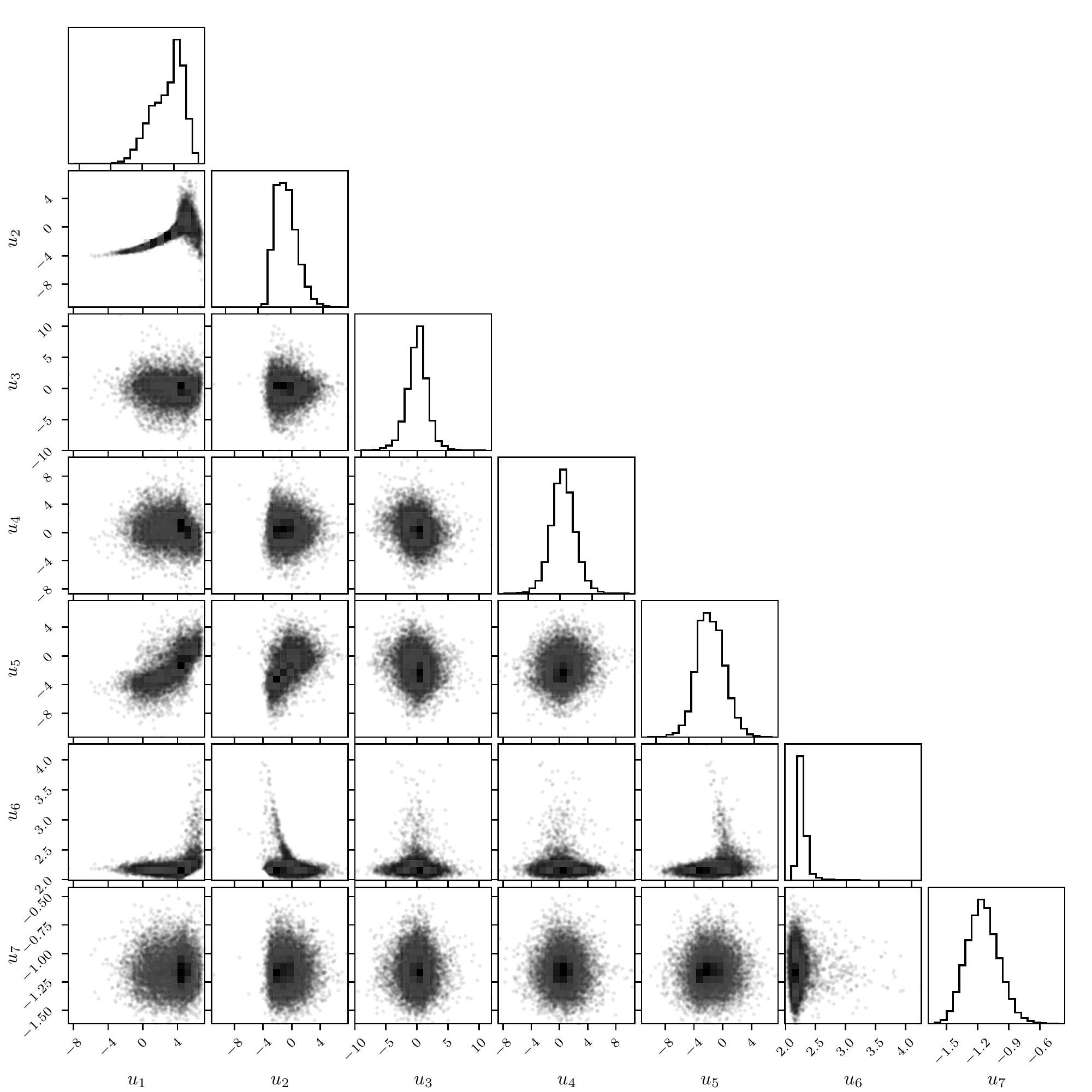}%
  \caption{\emph{Soil incubation (HN-T35)}: Unconstrained space}
  \label{fig.soil_incubation_hn-t35_unconstrained_posterior_pair_plot}
\end{figure}
\begin{figure}[H]
  \centering
  \includegraphics[width=\linewidth]{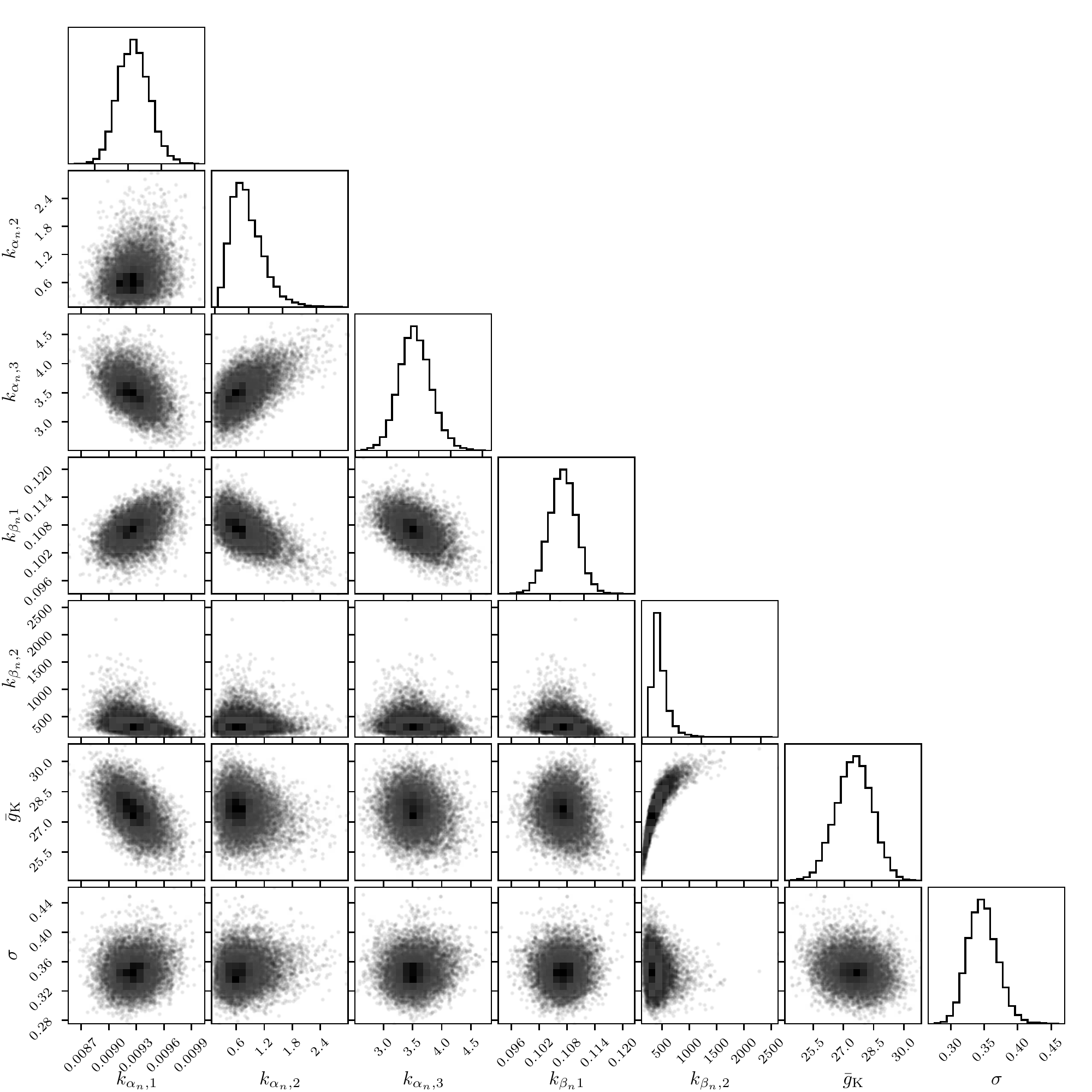}%
  \caption{\emph{Hodgkin--Huxley voltage clamp (potassium)}: Parameter space}
  \label{fig.hh_voltage_clamp_potassium_posterior_pair_plot}
\end{figure}
\begin{figure}[H]
  \centering
  \includegraphics[width=\linewidth]{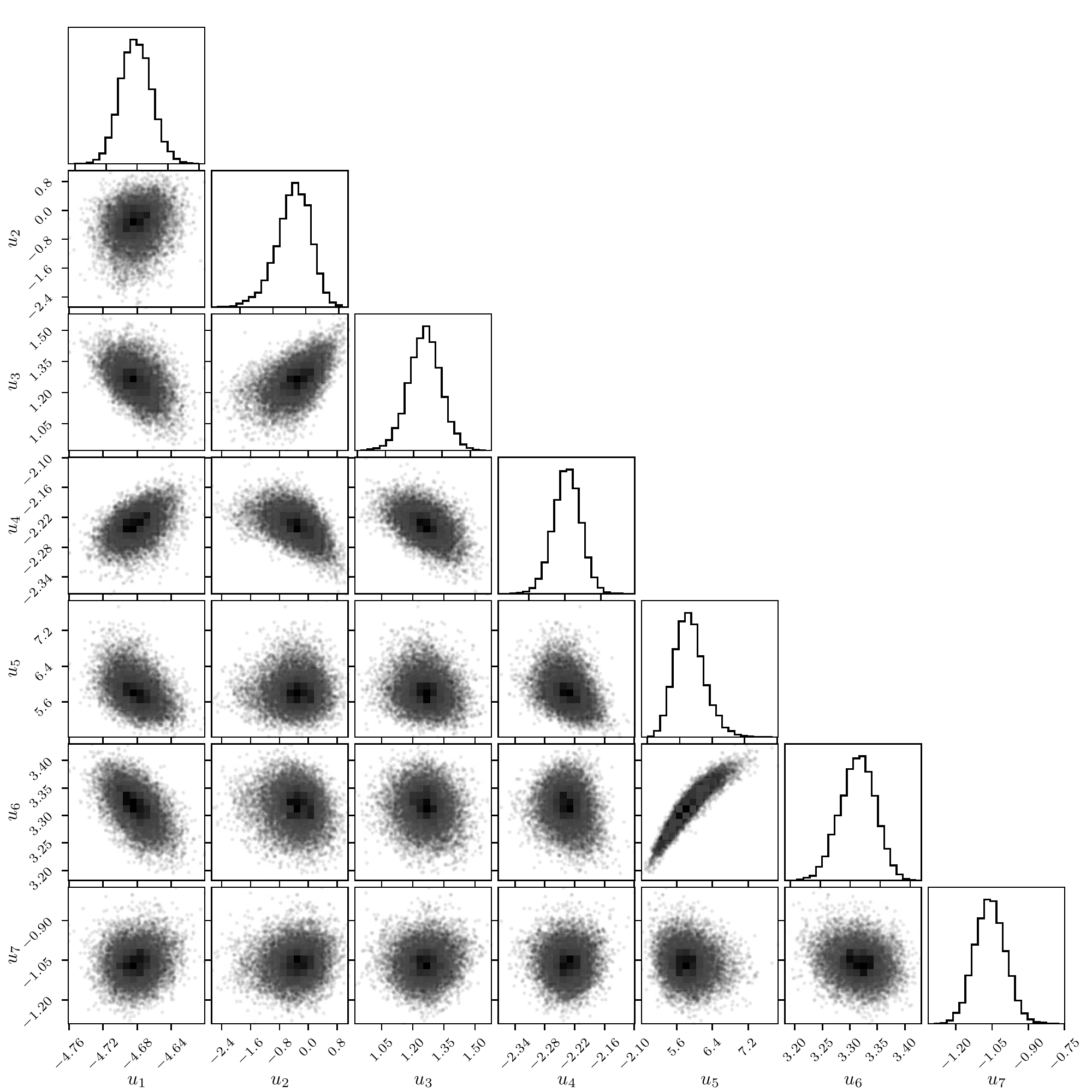}%
  \caption{\emph{Hodgkin--Huxley voltage clamp (potassium)}: Unconstrained space}
  \label{fig.hh_voltage_clamp_potassium_unconstrained_posterior_pair_plot}
\end{figure}
\begin{figure}[H]
  \centering
  \includegraphics[width=\linewidth]{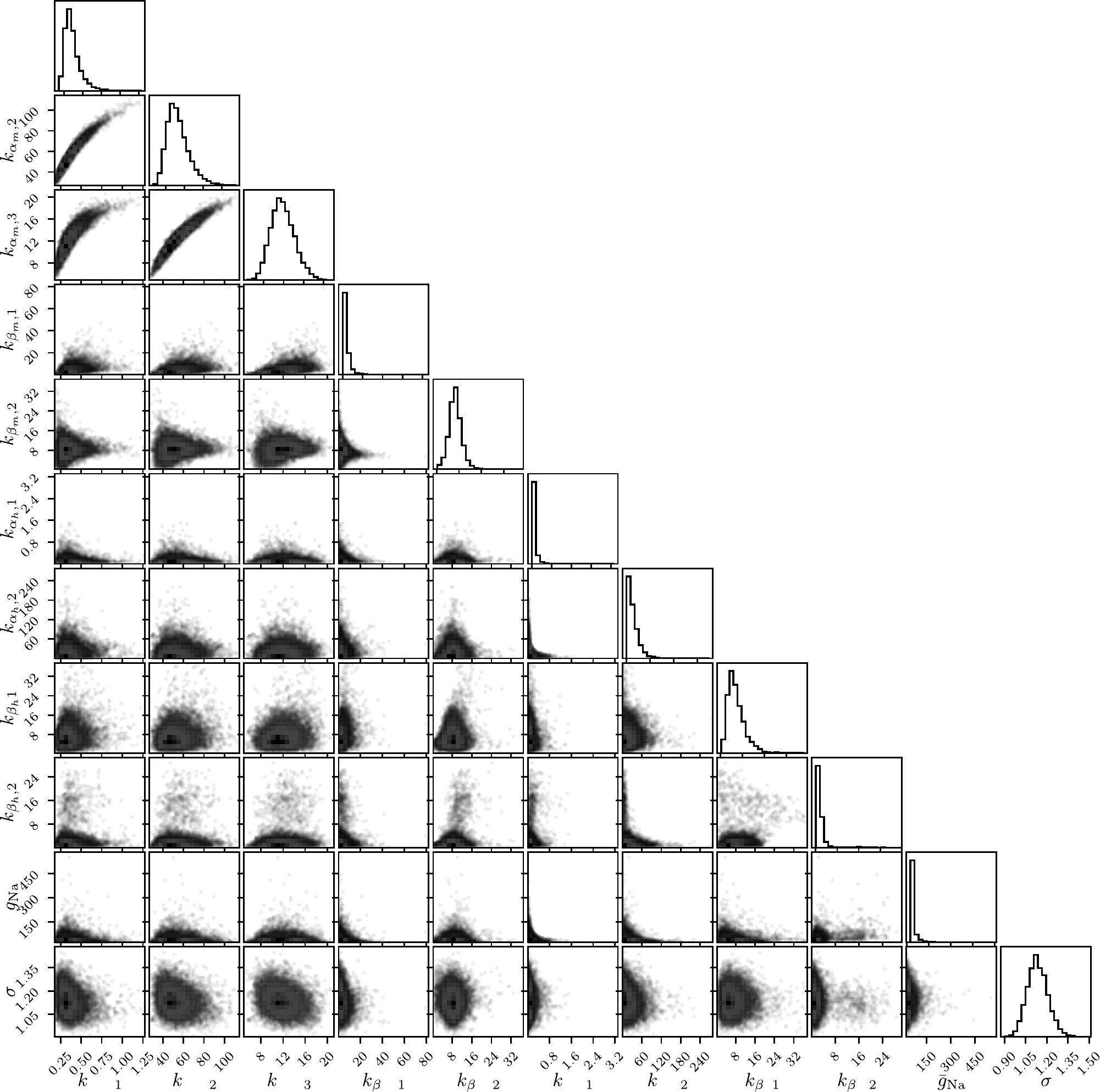}%
  \caption{\emph{Hodgkin--Huxley voltage clamp (sodium)}: Parameter space}
  \label{fig.hh_voltage_clamp_sodium_posterior_pair_plot}
\end{figure}
\begin{figure}[H]
  \centering
  \includegraphics[width=\linewidth]{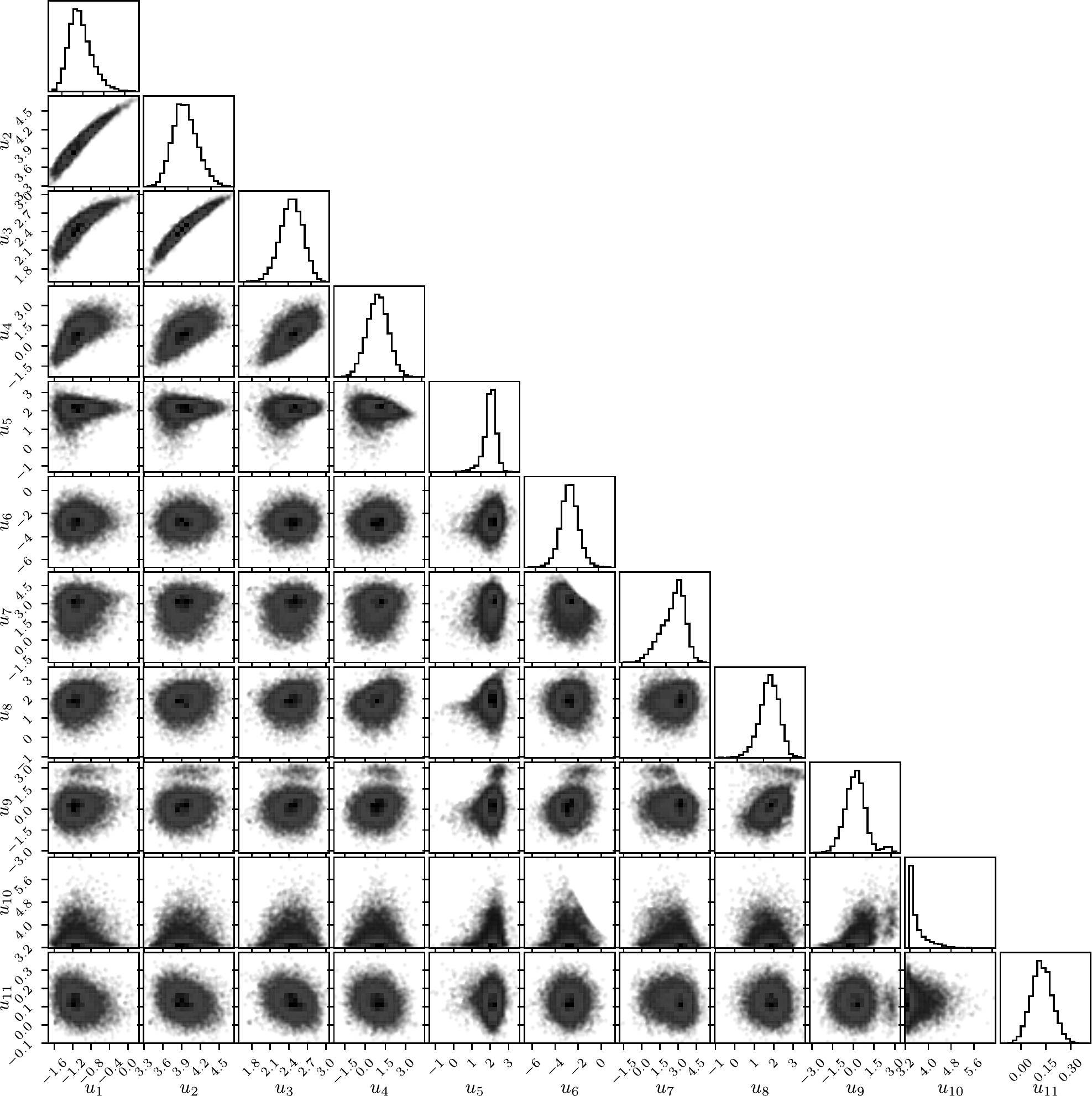}%
  \caption{\emph{Hodgkin--Huxley voltage clamp (sodium)}: Unconstrained space}
  \label{fig.hh_voltage_clamp_sodium_alt_unconstrained_posterior_pair_plot}
\end{figure}

{
\bibliographystyle{rss}

}

\makeatletter\@input{c_ms.tex}\makeatother